\documentclass[11pt]{article}
\usepackage[table]{xcolor}				
\usepackage{colortbl}
\usepackage{booktabs}					
\usepackage[pdftex,pdfborder={0 0 0}]{hyperref}	
\usepackage{graphicx}
\usepackage{graphics}
\usepackage{dsfont}
\usepackage{epsfig}
\usepackage{amsmath,amssymb,amsthm,amscd}
\usepackage{float}
\usepackage[parfill]{parskip}
\usepackage{verbatim} 
\usepackage{youngtab}
\usepackage{srcltx}

\allowdisplaybreaks

\newcommand{\xdownarrow}[1]{%
  {\left\downarrow\vbox to #1{}\right.\kern-\nulldelimiterspace}
}

\usepackage{tikz}
\usetikzlibrary{decorations.pathmorphing}
\usetikzlibrary{arrows,decorations.markings}
\usetikzlibrary{calc}

\newcommand{\ei}{\epsilon_1}
\newcommand{\eii}{\epsilon_2}

\newcommand{\base}{b}

\newcommand{\zbetasing}{Z_{\beta}^{\mathrm{single}}}
\newcommand{\zbetaambig}{Z_{\beta}^{\mathrm{ambig}}}
\newcommand{\phibetaambig}{\phi_{\beta}^{\mathrm{ambig}}}
\newcommand{\num}{\phi}

\newcommand{\SLtz}{{\rm SL}(2,\mathbb{Z})}

\newcommand{\G}{\mathfrak{g}}
\newcommand{\eeight}{\mathfrak{e}_8}

\newcommand{\nm}{n_{-}}
\newcommand{\np}{n_{+}}

\newcommand{\BC}{C_k}
\newcommand{\ckB}{{\check B}}

\newcommand{\Tr}{\textrm{Tr}}

\newcommand{\be}{\begin{equation}}
\newcommand{\ee}{\end{equation}}

\numberwithin{equation}{section}

\numberwithin{equation}{section}
\numberwithin{table}{section}

\def\beq{\begin{eqnarray}}
\def\eeq{\end{eqnarray}}
\def\ba{\begin{eqnarray}}
\def\ea{\end{eqnarray}}

\newcommand{\IZ}{\mathbb{Z}}
\newcommand{\IC}{\mathbb{C}}
\newcommand{\IP}{\mathbb{P}}
\newcommand{\IN}{\mathbb{N}}

\newcommand{\IQ}{\mathbb{Q}}
\newcommand{\IH}{\mathbb{H}}
\newcommand{\IF}{\mathbb{F}}

\newcommand{\Ms}{\mathrm{M}}
\newcommand{\Es}{\mathrm{E}}

\newcommand{\Kthree}{\mathrm{K3}}
\newcommand{\halfKthree}{\frac{1}{2}\mathrm{K3}}

\newcommand{\nn}{\nonumber}

\newcommand{\cN}{{\cal N}}

\newcommand{\cB}{{\cal B}}

\newcommand{\cH}{{\cal H}}
\newcommand{\cA}{{\cal A}}

\newcommand{\cO}{{\cal O}}
\newcommand{\cC}{{\cal C}}

\newcommand{\bmu}{\boldsymbol{\mu}}

\def\mb{\mathbb}
\def\mc{\mathcal}
\def\md{\mathbf}
\def\mf{\mathfrak}

\def\({\left(}
\def\){\right)}

\newcommand{\ie}{{\rm e}}
\newcommand{\ri}{{\mathsf{i}\,}}

\newcommand{\Ztop}{\mathrm{Z_{top}}}

\begin{document}
\begin{titlepage}
{}~ \hfill\vbox{ \hbox{} }\break

\rightline{USTC-ICTS-16-22}
\rightline{LPTENS 17/01}

\vskip 1 cm


\begin{center}
		\Large \bf  Refined BPS invariants of 6d SCFTs \\from anomalies and modularity
\end{center}

\vskip 0.8 cm

\centerline{ Jie Gu${}^{a}$, Min-xin Huang${}^{b}$, Amir-Kian Kashani-Poor${}^{a}$, Albrecht Klemm${}^{c}$}

\vskip 0.2in
\begin{center}{\footnotesize
		\begin{tabular}{c}
			${}^{\, a}${\em Laboratoire de Physique Th\'{e}orique de l'\'{E}cole Normale Sup\'{e}rieure,} \\
			$\phantom{{}^{\, b}}${\em CNRS, PSL Research University, Sorbonne Universit\'{e}s, UPMC, 75005 Paris, France} \\[2ex]
			${}^{\, b}${\em Interdisciplinary Center for Theoretical Study,} \\
			$\phantom{{}^{\, a}}${\em University of Science and Technology of China, Hefei, Anhui 230026, China}\\[2ex]
			${}^{\, c}${\em Bethe Center for Theoretical Physics (BCTP),} \\
			$\phantom{{}^{\, c}}${\em Physikalisches Institut, Universit\"{a}t Bonn, 53115 Bonn, Germany}
		\end{tabular}
}\end{center}

\setcounter{footnote}{0}
\renewcommand{\thefootnote}{\arabic{footnote}}
\vskip 60pt
\begin{abstract}
	F-theory compactifications on appropriate local elliptic Calabi-Yau manifolds engineer six dimensional superconformal 
	field theories and their mass deformations. The partition function $\Ztop$ of the refined topological string on these geometries 
	captures the particle BPS spectrum of this class of theories compactified on a circle. Organizing $\Ztop$ in terms of contributions 
	$Z_\beta$ at base degree $\beta$ of the elliptic fibration, we find that these, up to a multiplier system, are meromorphic Jacobi 
	forms of weight zero with modular parameter the K\"ahler class of the elliptic fiber and elliptic parameters the couplings and 
	mass parameters. The indices with regard to the multiple elliptic parameters are fixed by the refined holomorphic anomaly equations, 
	which we show to be completely determined from knowledge of the chiral anomaly of the corresponding SCFT. We express $Z_\beta$ as a quotient of weak Jacobi forms, 
	with a universal denominator inspired by its pole structure as suggested by the  form of $\Ztop$ in terms of 5d BPS numbers. 
	The numerator is determined by modularity up to a finite number of coefficients, which we prove to be fixed uniquely by imposing
	 vanishing conditions on 5d BPS numbers as boundary conditions. We demonstrate the feasibility of our approach  
	 with many examples, in particular solving the E-string and M-string theories including mass deformations, as well as 
	 theories constructed as chains of these. We make contact with previous work by showing that spurious singularities are cancelled 
	when the partition function is written in the form advocated here. Finally, we use the BPS invariants of the E-string thus 
	 obtained to test a generalization of the G\"{o}ttsche-Nakajima-Yoshioka $K$-theoretic blowup equation, as inspired by the Grassi-Hatsuda-Mari\~{n}o conjecture, to generic 
	 local Calabi-Yau threefolds.

\end{abstract}

{
	\let\thefootnote\relax
	\footnotetext{jie.gu@lpt.ens.fr, minxin@ustc.edu.cn, kashani@lpt.ens.fr, aklemm@th.physik.uni-bonn.de}
}

\end{titlepage}
\vfill \eject


\newpage

\baselineskip=16pt

\tableofcontents

\section{Introduction}
\label{Introduction} 

How well and how generally we can compute the topological string partition function $\Ztop$ serves as a benchmark for how well we understand topological string theory. The computational tools available depend sensitively on the class of geometries on which we consider the theory. The most computable class of geometries to date are local toric Calabi-Yau 3-folds. The key methods that exist for the computation of $\Ztop$ on these geometries rely on localization~\cite{Klemm:1999gm,CKK}, large N-dualities involving matrix 
models~\cite{Bouchard:2007ys} or 3d Chern-Simons theories giving rise to the topological vertex~\cite{Aganagic:2003db,IKV}, as well as the modular approach~\cite{Haghighat:2008gw,Huang:2010,Huang:2011qx} 
based on the holomorphic anomaly equations~\cite{BCOV,Huang:2010}. In this paper, following \cite{Huang:2015sta}, we will use modular methods in conjunction with vanishing conditions on 5d BPS invariants $N_{j_-j_+}^\kappa$  
to compute the {\it refined} topological string partition function on a class of non-toric geometries, consisting of elliptically fibered local Calabi-Yau manifolds $\check M$.

The $N_{j_-j_+}^\kappa$  are the multiplicities of 5d BPS states that arise upon M-theory compactification on the Calabi-Yau manifold $\check M$. Such BPS states  
were first considered in~\cite{Gopakumar:1998ii,Gopakumar:1998jq} and play a decisive role in our analysis. They fall into spin representations of the 5d little 
group $SU(2)_{+}\times SU(2)_{-}$ and are labelled by classes $\kappa\in H_2(\check M,\mathbb{Z})$ determining the mass of the corresponding BPS particles. 
The $N_{j_-j_+}^\kappa$  determine  the refined partition function $\Ztop$, which depends on K\"ahler parameters of the geometry as fugacities 
for the classes $\kappa$, and the parameters $\epsilon_\pm=\frac{1}{2}(\epsilon_1\pm \epsilon_2)$ which serve as fugacities for the $\pm$ spins $j_-$ and $j_+$.

On general Calabi-Yau manifolds, a deformation invariant BPS index is obtained only upon summing over the right spin quantum number. 
This corresponds to setting  $s^2=-(2 \epsilon_+)^2$ to zero and leads to the conventional topological string partition function 
in which the string coupling constant --- a genus counting parameter --- is identified as $g_s^2=-\epsilon_1 \epsilon_2$~\cite{Gopakumar:1998ii,Gopakumar:1998jq}. 
A geometrical model for computing these 5d BPS invariants was proposed in~\cite{KKV}. If the Calabi-Yau manifold admits a $U(1)$ isometry, as will be the case with the geometries $\check M$ that we consider in this paper, the 
integers $N_{j_-j_+}^\kappa$ are individually invariant~\cite{CKK,NO}. 
On local Calabi-Yau manifolds that engineer $\cN=2$ gauge theories in 4 and 5 dimensions, $\Ztop$ can be identified \cite{Iqbal:2003ix,Iqbal:2003zz,IKV} with the 
Nekrasov partition function \cite{Nekrasov:2002qd}, with $\epsilon_i$ playing the role of the equivariant parameters introduced by Nekrasov in his localization calculation. 

The refined $\Ztop$ plays a central role in the AGT correspondence \cite{Alday:2009aq} and its generalizations, via which the refined partition function $\Ztop$ is related to correlators in Liouville and Toda conformal field theories \cite{Wyllard:2009hg,KashaniPoor:2012wb,Nieri:2013yra,Aganagic:2013tta}. In the $\epsilon_2 \rightarrow 0$ limit, refined $\Ztop$ relates to integrable models underlying the space of vacua of $\cN=2$ supersymmetric gauge theories \cite{Nekrasov:2009rc} . It also features prominently in a recent proposal \cite{Grassi:2014zfa} for a non-perturbative completion of the refined topological string on toric geometries which relies crucially on refinement for the so-called pole cancellation mechanism \cite{Hatsuda:2012hm,Hatsuda:2013oxa}.

In addition to the coupling constants $g_s$ and $s$, refined $\Ztop$ depends on K\"ahler parameters~${\boldsymbol t}$ of the underlying geometry. The modular methods referred to in the opening paragraph yield $\log \Ztop$ as an asymptotic series in $g_s$ and $s$ with coefficients that are exact in $\boldsymbol{t}$, whereas topological vertex computations yield $\Ztop$ as an expansion in $e^{\boldsymbol{t}}$ with coefficients that are exact in $g_s$ and $s$ \cite{Aganagic:2003db,IKV}. For the latter, on appropriate geometries, the sum over some \cite{Iqbal:2004ne} but not all of the K\"ahler parameters can be performed. The methods \cite{Huang:2015sta,HKK2} we will advance in this paper for elliptically fibered manifolds rely on modular considerations and the holomorphic anomaly equation in the form \cite{Witten:1993ed}, and yield $\Ztop$ again as an expansion in  K\"ahler parameters assigned to the base of the elliptic fibration, but now with coefficients that enjoy modular properties making the invariance of $\Ztop$ under fiber monodromies manifest. These coefficients are meromorphic Jacobi forms of vanishing weight and of index determined by the holomorphic anomaly equations, with modular parameter the K\"ahler parameter of the fiber, and elliptic parameters built from the $\epsilon$ parameters and additional K\"ahler parameters identified with masses in the six dimensional setting that we will discuss presently. Based on the pole structure of these forms, as implied by the form of $\Ztop$ in terms of 5d BPS states \cite{Gopakumar:1998ii,Gopakumar:1998jq,Hollowood:2003cv}, we can argue that they must be quotients of weak Jacobi forms, with a universal denominator. Determining the numerator at each base degree then becomes a finite dimensional problem, which can be solved by imposing boundary conditions in the form of vanishing conditions on the refined BPS invariants $N_{j_-j_+}^\kappa$. These methods can also be applied to compact geometries~\cite{Huang:2015sta}, in which case the vanishing conditions are however not sufficient to solve the theory, as the index of the denominator grows too rapidly with the base degree~\cite{HKK2}.

The geometries on which we will consider $\Ztop$ yield via F-theory compactification an intriguing class of chiral 6d supersymmetric field theories. These theories are exotic in that they generally do not admit a Lagrangian description, exhibit strings in their spectrum (which become tensionless in the IR), and defy the expectation based on power counting that they should be trivial in the infrared. Indeed, we will focus on geometries that lead to 6d theories with a superconformal fixed point in the infrared. The topological string on these geometries captures the Kaluza-Klein modes of the tensionless string upon circle compactification. As initiated in \cite{Klemm:1996}, studying these modes can yield insight into the nature of such strings. The elliptic genus of their 2d chiral worldsheet theory is closely related to $\Ztop$, and provides a physical explanation for the transformation properties as Jacobi forms, based on the chiral anomaly of the theory. 

A classification of geometries leading to 6d super conformal field theories upon F-theory compactification, conjectured to be complete, is presented in \cite{Heckman:2013pva,Heckman:2015bfa}, based on earlier work in \cite{Morrison:2012np}. They consist of elliptic fibrations over non-compact complex surfaces $\check B$ which contain tree-like configurations of intersecting $\IP^1$'s. In this work, we will focus on geometries leading to 6d theories without enhanced gauge symmetry; this requires the self-intersection number of these curves to be $-1$ or $-2$. In upcoming work \cite{AGHKZ}, we will take up the more general case.

The cases in which all curves in $\check B$ have self-intersection number $-2$ are covered by the resolution of the quotient $\IC^2/\Gamma$, with $\Gamma$ a discrete subgroup of $SU(2)$. As $\check B$ in this case has trivial canonical class, the elliptic fibration yielding the Calabi-Yau 3-fold is trivial, and the corresponding 6d theories have $(2,0)$ supersymmetry. The $A_n$ series in this class yields superconformal theories which describe a stack of $n+1$ M5 branes. The $A_1$ case, corresponding to the geometry 
$T^2 \times \left[ \cO(-2) \rightarrow \IP^1 \right]$, is called the M-string \cite{Haghighat:2013}. 

The generic case requires a non-trivial elliptic fibration to yield a Calabi-Yau 3-fold over $\check B$, and leads to a 6d theory with $(1,0)$ supersymmetry. The simplest example of this class is called the E-string \cite{Morrison:1996na,Morrison:1996pp,Klemm:1996}, and consists of the elliptic fibration over the non-compact base surface $\cO(-1) \rightarrow \IP^1$. This geometry can also be constructed as the total space of the canonical bundle of the compact elliptic surface $\frac{1}{2}$K3, a nine point blow-up of $\IP^2$. According to the classification results in \cite{Heckman:2013pva,Heckman:2015bfa}, an ADE chain of $(-2)$ curves with a single $(-1)$ curve at one end also engineers a 6d theory without gauge symmetry with a superconformal fixed point. 

Our methods yield closed results at a given base degree from which all refined BPS invariants can easily be extracted. We list some of these invariants for ease of reference for mathematicians approaching them by other means. We use this data to refine the vanishing conditions on these invariants which follow from application of the adjunction formula. For the E-string, this data also allows us to provide some circumstantial evidence that the E-string BPS spectrum is computable via the quantization of an appropriate underlying quantum curve: We show that a suitably generalized consistency condition \cite{Wang:2015wdy} between two perspectives \cite{Grassi:2014zfa,Wang:2015wdy} on the quantization of the mirror curve in the case of toric Calabi-Yau manifolds is satisfied by these invariants. For a special class of toric Calabi-Yau manifolds, the consistency condition is shown in \cite{Grassi:2016nnt} to be the Nekrasov-Shatashvili limit of the G\"{o}ttsche-Nakajima-Yoshioka $K$-theoretic blowup equation \cite{Nakajima:2005fg,Gottsche:2006bm,Nakajima:2011} for the gauge theoretic Nekrasov partition functions. The natural generalization of the blowup equation that we propose, and that is satisfied by the E-string, can be applied to any local Calabi-Yau geometry on which the refined topological string can be formulated.

This paper is organized as follows. In section \ref{strategy}, we outline the general strategy to obtain the 
refined topological string partition on elliptically fibered Calabi-Yau manifolds and introduce the coefficients $Z_\beta$ of the expansion of $\Ztop$ in base K\"ahler parameters. We then review integer BPS invariants and their vanishing from a geometric point of view in section~\ref{IntegerGeometric} 
and present the geometries we will discuss in this paper, centered around the E- and the M-string, in
section~\ref{EMstringgeometries}. We discuss the geometric meaning of the mass parameters in section~\ref{sc:mass}. In section \ref{s:diff_equ_J}, we show that Jacobi forms satisfy a differential equation, which we identify with a generalization of the holomorphic anomaly equations in wave function form to the refined case in subsection~\ref{AnomalyEM}. The differential equation encodes the indices of the various elliptic parameters on which $Z_\beta$ depends. These indices can be derived from the anomaly polynomial of the 2d chiral theory living on the worldsheet of the non-critical strings of the 6d theories, as we demonstrate in section~\ref{sc:anomaly}. In section~\ref{s:ansatz_z_base}, we make a universal ansatz for the denominator of $Z_\beta$ based on the pole structure of refined $\Ztop$ as implied by the form of $\Ztop$ in terms of 5d BPS numbers. Together with the information regarding the indices, this determines $Z_\beta$ up to a finite number of coefficients. In section~\ref{s:sufficiency}, we determine under what circumstances vanishing conditions on BPS invariants suffice to determine these coefficients. We conclude that for all geometries relevant to this paper, they suffice\footnote{At lowest base wrapping degree, they need to be supplemented with one non-vanishing BPS number.}. Our approach thus leads to a complete solution for the class of theories we are considering. We present concrete results for the M-string in section~\ref{Mstring}, for the E-string 
in section~\ref{Estring}, and for the E-M string chain in section~\ref{EMchain}. Further results and examples for refined 
BPS invariants are relegated to appendix~\ref{Modularandenumerative}. In section \ref{s:BPSvanishing}, we extract BPS numbers from our closed expressions for $Z_\beta$ and discuss their structure and their vanishing. In section~\ref{section:domain_wall_method}, we relate our results  for the E- and M-string to the ones 
obtained by~\cite{Haghighat:2014} using a domain wall argument in 
terms of ratios of theta function. Based on theta functions identities, proven in appendix~\ref{appendix:proofs_identities}, we show that additional poles exhibited by these latter expressions are spurious. Finally, in section~\ref{sc:blowup}, we propose a generalization of the blowup equation and find that it is satisfied for the E-string. Some of the data needed for this check is presented in appendix \ref{sc:Sakai}. Some background on the ring of modular and generalized Jacobi forms is provided in appendix~\ref{Appendixmod}. As we summarize in the conclusions in section \ref{Conclusions}, our method has a wide range of applicability. We are confident that it can be adapted to encompass all geometries engineering 6d superconformal field theories \cite{DelZotto:2016pvm,AGHKZ}.

\section{General strategy for the solution of the topological string partition function on elliptically fibered Calabi-Yau manifolds} \label{strategy}

We will compute the refined topological string partition function on elliptic Calabi-Yau spaces 
with at least one zero section recursively as an expansion in the base classes $\beta$. The fundamental objects in this study are the expansion coefficients $Z_\beta$ in these classes, defined via
\be
Z(\boldsymbol{t_b},\tau,\boldsymbol{t_m},\ei,\eii)=\exp(F_\beta(\tau,\boldsymbol{t_m},\ei,\eii)\boldsymbol{Q}^\beta)=Z_{\beta=0}\left(1+ \sum_{\beta\neq 0} Z_\beta(\tau,\boldsymbol{t_m},\ei,\eii) \boldsymbol{Q}^\beta \right) .
\label{betaexpansion}  
\ee
We distinguish between three classes of K\"ahler parameters: those associated to the base, denoted $\boldsymbol{t_b}$, the exponentials $\boldsymbol{Q}^\beta=\exp(2 \pi i \boldsymbol{t}_{\beta})=\exp(2 \pi i \sum \beta_i t_{b_i})$ of which provide our expansion parameters,\footnote{More precisely, the $\boldsymbol{t_b}$ are shifted with regard to the geometric K\"ahler parameters $\tilde t_b$
by $t_{b_i}=\tilde t_{b_i}-\tau (C_{b_i} \cdot K)/2$, with $C_{b_i}$ a curve in the class corresponding to $b_i$ and $K$ the canonical class of the base surface, see \cite{HKK2} for details.} the K\"ahler parameter $\tau$ of the elliptic fiber, with $q=\exp(2\pi i \tau)$, which will play the role of modular parameter, and the remaining K\"ahler parameters $\boldsymbol{t_m}$ associated to mass deformations of the theory (for geometries containing curves of self-intersection number less than -2, there will also be K\"ahler parameters associated to the resolution of singular elliptic fibers, see \cite{DelZotto:2016pvm,AGHKZ}).

For the unrefined case $\ei=-\eii$, an ansatz for $Z_\beta(\boldsymbol{t_b},\tau,\boldsymbol{t_m},\ei,\eii)$ has been presented in~\cite{Huang:2015sta} based on general arguments that is valid both for compact and non-compact elliptically fibered Calabi-Yau spaces. In this paper, we seek to generalize this ansatz to the refined case. The ingredients are the following:
\begin{itemize} 
\item Elliptically fibered Calabi-Yau manifolds $M$ exhibit special auto-equivalences of the derived 
category of coherent sheafs $D^b(M)$ which can be identified with the $S$ and the $T$ transformation of the modular group ${\rm SL}(2,\IZ)$. Via the Fourier-Mukai transform, they induce an action on the $K$-theory charges $K(M)=H^{even}(M)$. Via mirror symmetry, this action can be identified with the monodromy action of the symplectic group ${\rm Sp}(h^3(W),\mathbb{Z})$ on the middle cohomology $H^3(W)$ of the mirror manifold $W$ to $M$. This action can also be computed directly on $W$. One obtains     
\be 
\tau \rightarrow  \tau_\gamma=\frac{a \tau + b}{c \tau + d},\quad  t_{m_i} \rightarrow  t_{m_i}^\gamma= \frac{t_{m_i}}{c\tau +d}, \quad  
t_\beta\rightarrow t_\beta+\mu_\gamma(\beta)  +  {\cal O}(Q_\beta) \ .
\label{transkaehler}
\ee
The projective nature of the first two transformations arises as the K\"ahler parameters 
are ratios of periods. The more non-trivial fact is the transformation of $t_\beta$ up to exponentially suppressed contributions.  
In particular $\mu_\gamma(\beta)$ gives rise to a multiplier system $\epsilon_\gamma(\beta) = \exp(2 \pi i\mu_\gamma(\beta))$. 
It is characterized by $\mu_{S}(\beta)= \mu_{T}(\beta)=- (C_\beta\cdot K)/2$, see  \cite{HKK2} for a derivation. 
\item The transformation properties of $Z(\boldsymbol{t_b},\tau,\boldsymbol{t_m},\ei,-\ei)$ under this action have been determined in~\cite{Witten:1993ed}.
\item Demanding that $Z(\boldsymbol{t_b},\tau,\boldsymbol{t_m},\ei,-\ei)$ (before taking the holomorphic limit!) be {\it invariant} under the monodromy transformations (\ref{transkaehler}) requires also transforming $g_s$ as 
\be
g_s^2\rightarrow (g_s^\gamma)^2 =\frac{g_s^2}{(c\tau +d)^2} \,.
\ee
Combining this with results on the refined holomorphic anomaly equations allows us to extend this transformation property to the refined case, yielding
\be 
\epsilon_{1/2} \rightarrow \epsilon^\gamma_{1/2} = \frac{\epsilon_{1/2}}{c\tau +d}\ .   
\label{transeps}
\ee

\item 
The invariance of $Z(\boldsymbol{t_b},\tau,\boldsymbol{t_m},\ei,\eii)$ under the joint transformations (\ref{transkaehler}) and (\ref{transeps}) implies that in the holomorphic limit, $Z_\beta(\tau,\boldsymbol{t_m},\ei,\eii)$ transform as 
Jacobi forms with the multiplier system $\epsilon_\gamma(\beta)$, modular parameter $\tau$ and elliptic parameters $(\boldsymbol{t_m},\ei,\eii)$.      
\item The pole structure of $Z(\boldsymbol{t_b},\tau,\boldsymbol{t_m},\ei,\eii)$ as suggested by the form of the free energy expressed in terms of 5d BPS numbers (as reviewed in the next section, see formula (\ref{defrefinedinv})) implies that $Z_\beta(\tau,\boldsymbol{t_m},\ei,\eii)$ is a ratio of weak Jacobi forms, and allows us to make a universal ansatz for the denominator, see (\ref{ansatz_z_nb}).
  
\item 
The indices related to the elliptic parameters can be calculated in the
unrefined case from the topological terms that appear in 
the holomorphic anomaly equations of~\cite{Witten:1993ed} as explained 
in~\cite{Huang:2015sta,HKK2}. A derivation of the indices in the refined case should be possible based on the 
refined holomorphic anomaly equations. We provide such a derivation, modulo several constants which we fix by studying examples. We demonstrate that our final result also follows from anomaly considerations,
 following \cite{DelZotto:2016pvm}.\footnote{We thank Michele Del Zotto and Guglielmo Lockhart for explaining their ideas to us.}
 
\item The ring of weak Jacobi forms with given even weight and integer index in a single elliptic parameter $z$ is generated by the forms $A(\tau,z)$ and $B(\tau,z)$, see appendix \ref{Appendixmod}. Lacking an analogous structure theorem for weak Jacobi forms exhibiting multiple elliptic parameters, we conjecture that the numerator of $Z_\beta(\tau,\boldsymbol{t_m},\ei,\eii)$ is an element of the ring generated by $\{A(,\tau,z_i),B(\tau,z_i)\}_{z_i \in \{ \epsilon_{+},\epsilon_{-} \}}$, where $\epsilon_{\pm} = (\epsilon_1\pm \epsilon_2)/2$,  together with an appropriate generating set of Weyl invariant Jacobi forms: these exhibit the mass parameters $\boldsymbol{t_m}$ as tuples of elliptic parameters, and are invariant under the action of the Weyl group of the flavor symmetry group on these tuples. This fixes the numerator up to a finite number of coefficients.

\item Vanishing conditions on 5d BPS invariants suffice to uniquely fix the numerator, and therewith $Z_\beta$. We thus obtain explicit expressions for $Z(\boldsymbol{t_b},\tau,\boldsymbol{t_m},\ei,\eii)$ which, aside from passing the stringent test of integrality for all 5d BPS invariants encompassed, match all results available in the literature computed by other means. 
\end{itemize}

\section{Geometry} \label{Geometry} 
We begin this section by recalling some aspects of the geometric invariants associated to the topological string and its refinement.
We then introduce the elliptically fibered geometries that will be the subject of this paper.

\subsection{Integer BPS numbers and their vanishing} 
\label{IntegerGeometric} 

The free energy $F(t,g_s)$ of the topological string on a Calabi-Yau 
manifold $M$ was initially defined from the world-sheet point of view as a sum over connected word-sheet instanton contributions, via the expansion
\be 
\label{gw}
F(\boldsymbol{t},g_s)= \sum_{g=0}^\infty\sum_{\kappa \in H_2(M,\mathbb{Z})} g_s^{2g-2} r_g^\kappa \boldsymbol{Q}^\kappa \,, \quad\quad r^\kappa_g\in \IQ \,.   
\ee 
The rational invariants $r^\kappa_g$ are called Gromov-Witten invariants. Mathematically, they can be defined as follows: Let  
${\cal M}_{g,\kappa}(M)$ be the moduli stack of holomorphic embedding 
maps $X:\Sigma_g\rightarrow M$ such that the image $[X(\Sigma_g)]$ lies in the curve 
class $\kappa$. Then  $r^\kappa_g=\int_{{\cal M}_{g,\kappa}(M)} {\bf 1}$ 
is the degree of the virtual fundamental class. As the virtual 
dimension ${\rm dim}_{vir}({\cal M}_{g,\kappa}(M))=\int_{\kappa} c_1(M)+ (g-1)(3-{\rm dim}_\mathbb{C}(M))$ 
is zero on Calabi-Yau 3-folds, the determination of $r^\kappa_g$ reduces to a point counting problem with 
rational weights, generically with a non-zero answer.

An important insight into the structure of the topological string free energy was provided by string-string duality. 
Comparing the expansion (\ref{gw}) with a BPS saturated heterotic one 
loop computation, Gopakumar and Vafa~\cite{Gopakumar:1998ii,Gopakumar:1998jq} conjectured
that the $g_s$ expansion takes the general form 
\be 
\label{bpsindices}
F(\boldsymbol{t},g_s)=\sum_{m=1\atop g=0}^\infty\sum_{\kappa\in H_2(M,\mathbb{Z})} 
\frac{I^\kappa_g}{m} \left(2 \sin \left( \frac{g_s m}{2}\right)\right)^{2 g -2} \boldsymbol{Q}^{m \kappa}  \,,\quad \quad I^\kappa_g\in \mathbb{Z}\,. 
\ee 
The coefficients $I^\kappa_g$ of this expansion are integers. They can be identified with the counting parameters in the BPS index  
\be 
{\rm Tr}_{{\cal H}_{\rm BPS}} (-1)^{2 j^3_+}u^{2 j^3_-}  \boldsymbol{Q}^{H} = \sum_{\kappa \in H_2(M,\mathbb{Z})} 
\sum_{g=0}^{\infty} I^{\kappa}_{g} (u^{\frac{1}{2}} + u^{-\frac{1}{2}})^{2 g} \boldsymbol{Q}^{\kappa} \,,
\label{unrefinedindex}
\ee
defined via a trace over the Hilbert space ${\cal H}_{BPS}$ of $5d$ BPS states in the compactification of M-theory on the Calabi-Yau manifold $M$.\footnote{A universal factor $2[(0,0)]\oplus [(\frac{1}{2},0)]$ is factored out of the Hilbert space to obtain $\cH_{BPS}$ \cite{Gopakumar:1998ii,Gopakumar:1998jq}.}  The BPS states can be organized into representations of the $5d$ Poincar\'e  group; their mass is proportional to their charge $\kappa$, and each component $\cH_\kappa$ of ${\cal H}_{BPS}$ can be decomposed into irreducible representations of the little group $SU(2)_{-}\times SU(2)_{+}$,
\be \label{intrN}
\cH_\kappa = \bigoplus_{j_-,j_+} N^\kappa_{j_- j_+} [(j_-,j_+)] \,.
\ee
Generically, this decomposition is not invariant upon motion in moduli space. The invariant index \eqref{unrefinedindex} is obtained upon summing over the spin quantum numbers  of $SU(2)_{+}$ 
with alternating signs. Using ideas 
of~\cite{Gopakumar:1998jq}, this index was geometrically interpreted 
in~\cite{KKV} based on the $SU(2)_{-}\times SU(2)_{+}$ Lefschetz decomposition 
of the moduli space ${\cal M}_g^\kappa$ of a $D0$-$D2$ brane system. ${\cal M}_g^\kappa$ 
was realized in~\cite{KKV} as the Jacobian fibration over the geometric deformation 
space of a family of  curves of maximal genus $g$ in the class $\kappa$ wrapped 
by the $D2$ brane. The $D0$ brane number counts the degeneration of the 
Jacobian and is hence related to the 
geometric genus of the curve.

An important ingredient in our determination of the topological string partition function will be the vanishing of the invariants $I^\kappa_g$ at fixed class $\kappa$ for sufficiently large $g$. For geometries $\check M$ which are the total space of the canonical bundle of a compact surface $S$, as is the case for the E-string and the M-string geometry which we will review below, this vanishing can be argued for very simply: the only curves that contribute to the partition function lie inside the compact surface. Given a smooth representative $C_\kappa$ of a class $\kappa \in H_2(S,\mathbb{Z})$, the adjunction formula permits us to express the canonical class $K_C$ of $C_\kappa$ in terms of its normal bundle in the surface, and the canonical class $K$ of the surface $S$. Via the relation between the degree and the arithmetic genus $p_a(\kappa)$ of the curve, $2 p_a(\kappa)-2 = \deg K_C$, this yields
\be 
2p_a(\kappa)-2 = C_\kappa^2 + K\cdot C_\kappa  \,,
\label{adjunction}  
\ee
providing an upper bound  
\be 
g_{max}(\kappa) =\frac{C_\kappa^2 + K\cdot C_\kappa}{2}+1
\label{gmax}
\ee 
on the geometric genus of any representative of the class $\kappa$. Note that a smooth representative of a class $\kappa \in H_2(S,\mathbb{Z})$ does not necessarily exist. More generally, we have to resort to Castelnuovo theory, 
which studies the maximal arithmetic genus of curves of given degree in projective space, 
see e.g. \cite{MR685427}. Either way, at fixed $\kappa$, there exists a bound beyond which $I^\kappa_{g}$ vanishes, simply because the 
moduli space ${\cal M}_g^\kappa$ is empty. This is in contrast to the Gromov-Witten invariants 
$r^\kappa_g$, which due to multi-covering contributions do not vanish for 
fixed $\kappa $ even at arbitrarily large genus. At large degree, the bound scales 
as $g_{max} \sim \kappa^2$. We will use the term {\it Castelnuovo bound} generically to refer to a bound beyond which BPS invariants vanish.

Mathematically, the BPS data of the $D0$-$D2$ brane system is best described by stable pair or PT invariants~\cite{PT1}. 
If the curve is smooth  at the bound and the obstructions vanish, the dimension of the 
deformation space $\mathbb{P}H^0({\cal O}(C_\kappa))$ follows from the Riemann-Roch theorem
\be 
\chi({\cal O}(C_\kappa ))=\sum_{i=0}^2 h^i({\cal O}(C_\kappa))= \frac{C_\kappa^2 - K\cdot C_\kappa}{2}+1    \,,
\ee
and the $I^\kappa_{g_{max}(\kappa)}$ evaluate to 
\be
I^\kappa_{g_{max}(\kappa)}=(-1)^{\chi({\cal O}(C_\kappa ))}\chi(\mathbb{P}^{\chi({\cal O}(C_\kappa ))-1})\ .
\label{leadingunrefined}
\ee
The invariants $I^\kappa_{g}$ at $g< g_{max}(\kappa)$ generically evaluate to much larger numbers due to the increasing 
degeneration of the Jacobian.

When the Calabi-Yau manifold admits a $\mathbb{C}^*$ isometry, one can also give a 
geometrical interpretation to states with definite + Lefschetz 
number and the BPS numbers $N_{j_- j_+}^\base\in \mathbb{N}$ introduced in \eqref{intrN} which keep track 
of the degeneracy of states with definite $SU(2)_{-}$ and the $SU(2)_{+}$ spin.  The corresponding index is 
\be 
{\rm Tr}_{{\cal H}_{\rm BPS}} u^{2 j^3_-} v^{2 j^3_+} \boldsymbol{Q}^{H} = \sum_{\kappa \in H_2(M,\mathbb{Z})} \sum_{j_-,j_+\in\frac{1}{2} \mathbb{N}} 
N^\kappa_{j_- j_+} [j_-]_u [j_+]_v  \boldsymbol{Q}^\kappa \ . 
\label{refinedindex} 
\ee
We have here assigned to every irreducible $SU(2)$ representation labeled by $j\in \mathbb{N}/2$ a Laurent polynomial, 
\be 
[j]_x=\sum_{k=-j}^{j} x^{2k},  
\ee
where the summation index $k$ is increased in increments of $1$ starting at $-j$. These BPS numbers are computed by the refined topological string, whose free energy takes the form \cite{Hollowood:2003cv}
\be 
F(\boldsymbol{t},\ei,\eii)=\sum_{m=1\atop j_-,j_+ \in \mathbb{N}/2}^\infty\sum_{\kappa\in H_2(M,\mathbb{Z})} 
\frac{N^\kappa_{j_-j_+} (-1)^{2 (j_-+j_+)}}{m} \frac{[j_-]_{u^m} [j_+]_{v^m}}{(x^\frac{m}{2}- 
	x^{-\frac{m}{2}}) (y^\frac{m}{2}- y^{-\frac{m}{2}})} \boldsymbol{Q}^{m \kappa}   \ .
\label{defrefinedinv} 
\ee 
We have introduced the variables
\be\label{eq:Nek-parameters}
x=\exp(\ri\epsilon_1)=u v,  \quad y =\exp(\ri\epsilon_2)= \frac{v}{u},   \quad 
u=\exp(\ri\epsilon_-)=\sqrt{\frac{x}{y}}, \quad v=\exp(\ri\epsilon_+)=\sqrt{x y}\ , \\ 
\ee
with 
\be 
\epsilon_-=\frac{1}{2}( \epsilon_1 - \epsilon_2),\qquad  \epsilon_+=\frac{1}{2}( \epsilon_1 + \epsilon_2) \,.
\ee
For geometries $\check M$ that geometrically engineer $\cN=2$ gauge theories in 5 dimensions, \eqref{defrefinedinv} coincides with the $K$-theoretic instanton partition function of Nekrasov \cite{Nekrasov:2002qd}, and  $\epsilon_{1,2}$ map to equivariant parameters in the localization computation performed in the gauge theoretic setting.

To facilitate the transition between the refined and the unrefined free energy, it is also convenient to introduce the variables
\be\label{eq:gs-s}
g_s^2 = - \epsilon_1 \epsilon_2 \,, \quad s = - (\epsilon_1 + \epsilon_2) \,.
\ee
Comparing (\ref{unrefinedindex}) and (\ref{refinedindex}) allows us to relate the BPS number $N^\kappa_{j_-j_+}$ to the invariants $I_g^\kappa$. In terms of the tensor 
representations $T_g=\left( 2 [0] +\left[\frac{1}{2} \right]\right)^{\otimes g}$, where the 
bracket $[j]$ indicates the irreducible $SU(2)$ representation of spin $j$, we obtain
\be 
\sum_{j_-,j_+\in\frac{1}{2} \mathbb{N}} (-1)^{2 j_+} (2 j_++1) N^\kappa_{j_- j_+}[j_-]=\sum_{g \in \mathbb{N}} I^\kappa_g T_g \ .
\label{sumoverrightspin}
\ee
This relation implies that if at fixed $\kappa$, the maximal genus for which $I_g^\kappa \neq 0$ is $g_{max}(\kappa)$, the maximum left spin for which the number $N^\kappa_{j_-j_+} \neq 0$ is  
\be \label{leftbound}
2 j_-^{max}(\kappa) = g_{max}(\kappa)=\frac{C_\kappa^2 + K\cdot C_\kappa}{2}+1 \,.
\ee

To describe a generic bound on the right spin, we will need to review some aspects of PT theory of stable pairs~\cite{PT1,PT2} and its relation 
to KKV theory \cite{KKV} and refined KKV theory as outlined in \cite{CKK}. As above, we will assume that there is a smooth 
curve $C_\kappa$ in the class $\kappa\in H_2(S,\mathbb{Z})$. Following the notation in~\cite{CKK}, we denote 
by  $P_n(M,\kappa)$ the moduli space of stable pairs $(F,s)$, where $F$ is a free sheaf of pure dimension 1 
generated by the section $s\in H^0(F)$ outside $d$ points, with $\textrm{ch}_2(F)=\kappa$ and 
holomorphic Euler characteristic $\chi(F)=n=1-p_a +d$. The PT invariant 
$P_{n,\kappa}$ is defined as the degree of the 
virtual fundamental class of $P_n(M,\kappa)$. These invariants were related to $I_g^\kappa$ in \cite{PT2}. Following \cite{CKK},
we will review a generic model for $P_n(M,\kappa)$ based on which the $SU(2)_-\times SU(2)_+$ decomposition of the Hilbert space $\cH_\beta$ of 5d BPS states can be computed. This will permit us to arrive at a bound for the maximal right spin $j_+^{max}(\kappa)$ of non-vanishing BPS numbers $N_{j_{-}^{max} j_{+}}^\kappa$. Based on experience, this also serves as a bound below $j_-^{max}$.

The basic identity relating the geometric $SU(2)$ Lefschetz decomposition to the $SU(2)_-\times SU(2)_+$ decomposition of $\cH_\beta$ is motivated by the KVV approach \cite{KKV,CKK}. It is given, at sufficiently small Euler characteristic as we explain presently, by
\be \label{basic_identity}
H^*(\cC^{[k]}) = \left(\theta^{p_a-k}\cH_\kappa \right)_{SU(2)_\Delta} \oplus H^*(\cC^{[k-2]})  \,.
\ee
$\cC^{[k]}$ here is the relative Hilbert scheme parameterizing curves $C$ of class $\kappa$ with $k$ distinguished points. In the generic cases we will consider, such Hilbert schemes will be projective spaces, and the corresponding cohomology carries the standard Lefschetz decomposition. $\theta$ is an $SU(2)_-$ lowering operator, i.e. it acts as
\be
\theta([(j_-,j_+)]) =\left( \theta[j_-] \right)\otimes [j_+] = [(j_--1,j_+)] \,,
\ee
with the understanding that $[j]=0$ for $j<0$. $(\cdot)_{SU(2)_\Delta}$ is the map from $SU(2)_- \times SU(2)_+$ representations to the diagonal $SU(2)_\Delta$ representation. Corrections to \eqref{basic_identity} arise from reducible curves. At a given class $\kappa$, the minimal Euler characteristic of a reducible curve with components of class $\kappa_1 + \kappa_2 = \kappa$ will be $\chi_1 + \chi_2 = 1 - p_a(\kappa_1) + 1 - p_a(\kappa_2)$. At given $\kappa$, the minimal Euler characteristic at which corrections due to reducible curves can arise is thus given by
\be
{\rm min}\big\{1-p_a(\kappa_1) + 1-p_a(\kappa_2)|\kappa_1+\kappa_2=\kappa \big\} .
\label{reducible}
\ee
Below this bound, \eqref{basic_identity} holds unmodified.

We will argue for a bound on the $SU(2)_+$ number at given $\kappa$ for geometries $\check M$ which are the total space of the canonical bundle over a Fano surface $S$. In this case, for small $d$  
bounded by a constraint linear in $\kappa$ \cite{PT2}, the moduli space of stable pairs on $\check M$ coincides with that of $S$, $P_{1-p_a+d}(\check M,\kappa)=P_{1-p_a+d}(S,\kappa)$. The latter in turn is isomorphic to the appropriate relative Hilbert scheme of curves of class $\kappa$ with $d$ distinguished points (cf. proposition B.8 of \cite{PT2}),
\be
P_{1-p_a +d}(S,\kappa) \simeq \cC^{[d]} \,.
\ee
For $C_\kappa$ a curve of class $\kappa$, a model for this Hilbert scheme is given by $\mathbb{P}^{(C_\kappa^2-C_\kappa \cdot K)/2-d}$ over $S^{[d]}$.

Returning to \eqref{basic_identity}, we see that the contributions to highest left spin can be read off at $d=0$. We obtain
\be
H^*(\cC) = H^*(\IP^{(C_\kappa^2-C_\kappa \cdot K)/2}) = [ \frac{C_\kappa^2-C_\kappa \cdot K}{4}] = N^\kappa_{p_a,j_+^{max}} [j_+^{max}] \,,
\ee
i.e.
\be 
2 j^{max}_+(\kappa)= \frac{C_\kappa^2-C_\kappa \cdot K}{2}\ .
\label{rightbound}
\ee
We reiterate that we have derived $j_-^{max}$ as the right spin at highest left spin, but expect it to present more generally a bound on right spin at given $\kappa$.
%
This generic bound is quadratic in $\kappa$, just as the bound on $j_-$. It is implicit in the 
formalism of \cite{CKK}, but not spelled out explicitly. Instead, more 
complicated calculations are performed there for $d>0$, showing an asymptotic 
pattern of the refined BPS numbers that we also observe in our examples.
The generic bound is satisfied for the large $\kappa$ asymptotics in all known 
geometric examples, in particular for the $E$-string for which $S$ is only semi-Fano. 

Let us finally comment on geometries bases on chains of $(-1)$ and $(-2)$ curves. We can consider each such curve as the base of an elliptic surface. 
For curve classes lying in a single such surface, the bounds \eqref{leftbound} and \eqref{rightbound} should apply unaltered. 
More general classes will require a more detailed analysis. In these cases, the leading contributions must come from reducible 
curves which have been excluded by the condition (\ref{reducible}), so the precise bound will have to include these. We 
still expect it to remain quadratic in the curve class. This expectation is confirmed by our explicit computations, see section \ref{s:BPSvanishing}.

%
\subsection{Geometries underlying the E- and M-string and generalizations}
\label{EMstringgeometries} 
The geometries we will consider in this paper consist of elliptically fibered Calabi-Yau manifolds $\check M$ over non-compact bases $\ckB$, such that 
\begin{itemize}
\item all compact curves $C_i$ in $\ckB$ are contractible, and
\item the elliptic fiber does not degenerate over all of $C_i$ for any $i$.
\end{itemize}
The first condition is required for the 6d theories engineered by compactifying F-theory on $\check M$ to have a superconformal fixed point \cite{Heckman:2013pva,Heckman:2015bfa}, and is a prerequisite, as we explain below, for the boundary conditions supplied by vanishing conditions on GV invariants to completely fix our ansatz. The second condition implies that the 6d theories do not exhibit gauge symmetry. We will treat the case with gauge symmetry in a forthcoming publication \cite{AGHKZ}.

The simplest examples that we consider contain a single compact curve $C$ in the base $\ckB$. These geometries can be obtained via decompactification limits of elliptic fibrations over Hirzebruch surfaces. Recall that a Hirzebruch surface $\IF_k$ is a $\IP^1$ fibration over a base $\IP^1$. Two irreducible effective divisors $C_k$ and $F$ can be chosen to span $H_2(\IF_k,\IZ)$. $F$ corresponds to a fiber of $\IF_k$, hence $F \cdot F=0$, and $C_k$ is a section of the fibration, thus $C_k \cdot F =1$, with $C_k \cdot C_k = -k$. The canonical class of $\IF_k$ in terms of these two divisors is given by $-K_{\IF_k} = 2C_k +(2+k)F$.

Calabi-Yau elliptic fibrations over the Hirzebruch surfaces $\IF_k$, $k=0,1,2$, can be constructed as hypersurfaces in an appropriate ambient toric variety. The constructions considered in $\cite{Hosono:1993qy}$ yield geometries with $h^{1,1}=3$, corresponding to the lifts of $C_k$ and $F$, and the elliptic fiber $E$. Taking the defining equation of the hypersurface to be in Weierstrass form,
\be 
y^2= 4 x^3 - f_4 x -g_6  \,, 
\label{weierstrass} 
\ee
we must choose $f_4$ and $g_6$ as sections of particular line bundles over the base surface $B$,
\be
f_4 \in -4 K_B \,, \quad g_6 \in -6 K_B \,,
\ee
to ensure that the total space of this fibration is a Calabi-Yau manifold. Here, $K_B$ is the canonical class of $B$. The elliptic fibration is singular along the discriminant locus of the fibration, defined as the zero locus of the discriminant 
\be 
\Delta_{12} = f_4^3-27 g_6^2 \,.
\label{disc}
\ee
The type of singularity along each component of the discriminant locus depends on the vanishing order of $\Delta$ together with that of $f$ and $g$ on this component, and has been classified by Kodaira.

The minimal vanishing order $\gamma$ of a section of the line bundle $-n K_B$ along a curve $C_k$ with negative self-intersection number $-k$ can be computed by decomposing $-nK_B = \gamma C_k + A$, with $\gamma$ the smallest positive integer such that $C_k \cdot A \geq 0$. Equivalently, we can consider a decomposition $-nK_B = \gamma' C_k + A'$, where now, $\gamma' \in \IQ$ and we impose $A' \cdot C_k=0$. We have $\gamma' = \gamma + \frac{A \cdot C_k}{C_k^2}$, and by minimality of $\gamma$, $-1 <\frac{A \cdot C_k}{C_k^2} \le 0$; hence the minimal vanishing order is given by the smallest integer larger or equal to $\gamma'$. Either way, based on the canonical class of the Hirzebruch surface $\IF_k$ cited above, we can argue that for $k=0,1,2$, sections of $-4K_B, -6K_B$ or $\Delta$ generically do not vanish identically over $C_k$.
Thus, $\Delta|_{C_k}$ will generically vanish only at the intersection of $C_k$ with $-12 K_B$, hence at $-12 K_B \cdot C_k = 12(2-k)$ points. As $f |_{C_k}$ and $g |_{C_k}$ will generically not vanish at these points, the elliptic fibration above $C_k$ (recall that $C_k$ is a section of $B$) will generically exhibit $12(2-k)$ isolated $I_1$ singularities, hence correspond to a $\Kthree$ surface, $\halfKthree$ surface, and the trivial product $\IP^1 \times T^2$ respectively.

We decouple gravity by decompactifying these geometries along the direction of the fiber $F$ of the Hirzebruch surface. The fibration structure of these geometries is summarized in diagram \eqref{fibrationstructure}.
\be    \label{fibrationstructure}                            
\begin{array}{rr} {\cal E} \rightarrow M & \\
	\big\downarrow &\!\!\! \!\!\!\!  \pi\\ [3 pt]
	F=\mathbb{P}^1  \rightarrow  B=\mathbb{F}_k& \\
	\big\downarrow &\!\!\!\!\!\!\! \pi'\\ [3 pt]
	\BC=\mathbb{P}^1\end{array} 
\qquad \rightarrow \qquad  
\begin{array}{rr} {\cal E} \rightarrow \check M & \\
	\big\downarrow &\!\!\! \!\!\!\! \check \pi\\ [3 pt]
	{\cal O}(-k) \rightarrow \check B& \\
	\big\downarrow &\!\!\!\!\!\!\! \check \pi'\\ [3 pt]
	\BC=\mathbb{P}^1\end{array}             \qquad .    
\ee
We can engineer more general 6d superconformal theories by replacing the base $\BC$ of the non-compact base $\check B$ of the elliptic fibration by a string of intersecting $\IP^1$'s. Setting $\ckB = \IC^2/\Gamma$, with $\Gamma$ chosen among the discrete subgroups of $SU(2)$, yields theories with $(2,0)$ supersymmetry, as the elliptic fibration here is trivial. The string of compact curves in $\ckB$ and their intersections is encoded in the Dynkin diagram corresponding to the subgroups $\Gamma$; recall that these enjoy an ADE classification. In particular, all of these curves have the topology of $\IP^1$ and have normal bundle $-2H$. We refer to them as $(-2)$ curves. The massless M-string is the simplest member of this class of geometries, with $\Gamma= \IZ_2 \sim A_1$. 

The papers \cite{Heckman:2013pva,Heckman:2015bfa} discuss which F-theory compactifications on elliptic fibrations over non-compact bases can lead to 6d SCFTs with generically only (1,0) supersymmetry. The only additional SCFTs without gauge symmetry contained in this classification arise upon including a single $(-1)$ curve at the end of a string of $(-2)$ curves; in the case $\Gamma = \IZ_{n+1}$, this corresponds to an E-M$^n$ string chain bound state.

The topological data that we will need to determine the topological string partition function on these geometries consists of the intersection numbers of these curves, as well as their intersection product with the canonical line bundle of $\ckB$. The latter is once again simply determined by the adjunction formula (\ref{adjunction}), this time applied to $\check B$. As all compact curves $C$ in $\ckB$ have genus 0, we obtain
\be
K_{\check B} \cdot C = -2 - C \cdot C \,. \label{KC_from_adjunction}
\ee

\subsection{Turning on mass parameters}
\label{sc:mass}

Above, we have identified the elliptic fibration over the total space of the bundle $\cO(-1) \rightarrow \IP^1$ as local $\halfKthree$. Let us briefly recall the geometry of the $\halfKthree$ surface. It can be obtained as the blow up of $\mathbb{P}^2$  
at nine points. One can construct a series of del Pezzos surfaces $B_k$ by blowing up 
$k$ generic points in $\mathbb{P}^2$. The canonical class is given by
\be 
-K=3H -\sum_{i=1}^k e_i\ .  
\ee
Here, $H$ is the hyperplane class of $\mathbb{P}^2$ and $e_i$, $i=1,\ldots,k$, denote 
the exceptional $\mathbb{P}^1$ curves resulting from the blowups. The intersection numbers between these divisors are
\be
H^2=1 \,,\quad e_i\cdot H=0\,,\quad e_i^2=-1\,,\quad i=1,\ldots,k \,.
\ee 
Let $\Lambda_k=H_2(B_m,\mathbb{Z})$ and   
\be 
\Lambda'_k=\{ x \in \Lambda_k| x \cdot K =0\} \ .
\ee
The intersection product induces an inner product on this lattice. One can readily show that $\Lambda'_k$ can be equipped with 
a root (or co-root) basis whose negative 
Cartan matrix is $-,-,A_1,A_1\times A_2, A_4,D_5,E_6,E_7,E_8$ for 
$k=0,\ldots,8$. The simple roots are always of the form $\beta_{ij}=e_i-e_j$ or 
$\gamma_{ijk}=H-e_i-e_j-e_k$, where $i,j,k$ are taken to have distinct values.

\begin{figure} 
\begin{center} 
	\resizebox{6cm}{!}{\begin{tikzpicture}{xscale=1cm,yscale=1cm}
		\coordinate[label=below:${\bf 1}$,label=above:$0$](A) at (5,0);
		\coordinate[label=below:${\bf 2}$,label=above:$1$](B) at (-2,0);
		\coordinate[label=below:${\bf 4}$,label=above:$2$](C) at (-1,0);
		\coordinate[label=below:${\bf 6}$, label={[label distance=.5mm]45:$3$}](D) at (0,0);
		\coordinate[label=below:${\bf 5}$,label=above:$4$](E) at (1,0);
		\coordinate[label=below:${\bf 4}$,label=above:$5$](F) at (2,0);
		\coordinate[label=below:${\bf 3}$,label=above:$6$](G) at (3,0);
		\coordinate[label=below:${\bf 2}$,label=above:$7$](I) at (4,0);
		\coordinate[label=left:${\bf 3}$,label=above:$8$](H) at (0,1);
		\draw (B)--(C)--(D)--(E)--(F)--(G)--(I);
		\draw (D)--(H);
		\draw (4,0) --(4.9,0);
		\draw (A) circle (.1);
		\fill (B) circle (.1);
		\fill (C) circle (.1);
		\fill (D) circle (.1);
		\fill (E) circle (.1);
		\fill (F) circle (.1);
		\fill (G) circle (.1);
		\fill (H) circle (.1);
		\fill (I) circle (.1);
		\end{tikzpicture}} 
\end{center}
\caption{The root lattice of affine $E_8$.}\label{rootlattice}
\end{figure}
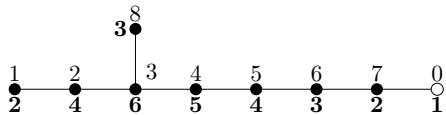

For example, in the case $k=8$, we obtain the root lattice of $E_8$, as depicted in figure \ref{rootlattice}. We set $\alpha_k=e_k-e_{k+1}$, $k=1,\ldots,7$, and $\alpha_8=\gamma_{123}$. The labels correspond to the upper indices in the figure (the lower indices are the Dynkin labels). The full set of positive roots is then all $\beta_{ij}$, of which there are 28, all $\gamma_{ijk}$ (56), 
all $\delta_{ij}=2 H -\sum_{k\neq i,j} e_k$ (28)  
and $\epsilon_{i}=3 H -\sum_{k\neq i} e_k$ (8), yielding $120 = \frac{248-8}{2}$.

It follows from the intersection numbers that $K^2=9-k$. The $\halfKthree$ surface is obtained by blowing up the unique base point of the elliptic pencil that models $B_8$. It is elliptically fibered over this exceptional divisor $C_1=e_9$. As $k=9$, $-K$ is no longer positive. Indeed, it can be identified with the elliptic fiber $E$ of this surface. Setting $\alpha_0=e_9-e_8$, the divisors can be mapped to simple roots of the affine $E_8$ Lie algebra, with $\alpha_0$ constituting the affine 
root. The additional intersection numbers are $C_1^2 =-1$, $C_1 \cdot E=1$, and $E^2=0$.

The explicit geometric interpretation of the mass parameter in the case of the M-string, and more generally in the case of elliptic fibrations over resolved ADE singularities, is not equally transparent to us. Following \cite{Witten:1997sc}, one could attempt to interpret the mass parameter as deforming the geometry $\check M = \check B \times T^2$ away from the trivial product by fibering the non-compact surface $\check B$ over $T^2$. We leave the exploration of this possibility (and in particular the analysis of the $D$ and $E$ case, the mass deformation of which is excluded in \cite{Gadde:2015tra} based on physical considerations) to future work. In section \ref{s:BPSvanishing}, we will discuss a related proposal for the mass deformation.

\section{Holomorphic anomaly and Jacobi forms}
\label{Anomaly}  
The holomorphic anomaly equations for elliptic fibrations can be written in a form which implies that $Z_\beta$ is a Jacobi form of weight 0 
and index a fixed function of $\beta$. For the relevant properties of Jacobi forms, see appendix \ref{Appendixmod}. In the following, we will use the notation 
\be
\cA(\tau,\zeta)=A\left(\tau,z\right) = \phi_{-2,1}(\tau,z) \,, \quad \cB(\tau,\zeta) = B\left(\tau,z\right) = \phi_{0,1} (\tau,z)  \ ,
\ee
with $\zeta = 2\pi z $, for the two generators of the ring of holomorphic weak Jacobi forms interchangeably. When only one argument is given, the elliptic parameter is meant.

\subsection{A differential equation for Jacobi forms} 
\label{s:diff_equ_J}
  
A weak Jacobi form of weight $k$ has a Taylor expansion 
\be  \label{j_taylor}
\phi(\tau,z) = \sum_{j=0}^\infty \eta_j(\tau) z^j  \,,
\ee
with the $\eta_j$ being quasi-modular forms of weight $j+k$ (in fact, one can be more precise about the form of these coefficients, expressing them as functions of modular forms, their derivatives, and the weight and index of $\phi$ \cite{EZ}).
Considering (\ref{j_taylor}) term by term, we deduce that the phase factor arising upon modular transformation of $\phi$ is due to the occurrence of factors of $E_2$ in the coefficients $\eta_k$. Since $E_2$ transforms under modular transformations as
\be
E_2\(\frac{a \tau + b}{c \tau + d}\) = (c \tau + d)^2 E_2(\tau) + \frac{6}{\pi \ri} c (c\tau +d) \,,
\ee
the exponential 
\be
\exp \( \frac{ \pi^2}{3} m z^2 E_2\)
\ee
generates the inverse of the phase factor occurring in the transformation of the Jacobi form. The product $\exp ( \frac{ \pi^2}{3} m z^2 E_2) \phi$ thus transforms without phase factor. Arguing again termwise, we conclude that the $E_2$ dependence in the Taylor coefficients is cancelled by this prefactor. Hence a Jacobi form of index $m$ satisfies the equation
\be
\frac{\partial}{\partial E_2} \exp \( \frac{ \pi^2}{3} m z^2 E_2\) \phi = \exp \( \frac{ \pi^2}{3} m z^2 E_2\) \left( \frac{\partial}{\partial E_2} + \frac{ \pi^2}{3} m z^2 \right) \phi = 0 \,.
\ee
Conversely, a power series in $z$ with quasi-modular coefficients that satisfies the equation
\be
\left( \frac{\partial}{\partial E_2} + \frac{ \pi^2}{3} m z^2 \right) \phi = 0
\ee
transforms under modular transformations as a Jacobi form of index $m$. If it is also periodic under $z \rightarrow z +1$, its transformation behavior under $z \rightarrow z+ \tau$ as a Jacobi form follows by combining periodicity and modular transformations. In the following, we will use elliptic parameters $\zeta$ with period $2 \pi + 2 \pi \tau$; they are related to $z$ by $z\mapsto \zeta/(2\pi)$. For these,
\be
\left( \frac{\partial}{\partial E_2} + \frac{1}{12} m \zeta^2 \right) \phi = 0 \,. \label{JacE2}
\ee

\subsection{The index from the holomorphic anomaly equations}
\label{AnomalyEM}

For the geometries described in the previous section, the holomorphic anomaly equations have been motivated in the form \cite{Huang:2015sta}
\be
\left( \partial_{E_2} + \frac{1}{12} \frac{\beta \cdot (\beta + K_{\ckB})}{2} g_s^2 \right) Z_\beta = 0  \,.
\ee
Here, the modular parameter of $E_2$ is the K\"ahler parameter $\tau$ of the elliptic fiber, and $\beta = \sum_{i=1}^r n_{b_i} D_{b_i} = (n_{b_i})$, where $D_{b_i}$ denote compact divisors in the non-compact base surface $\ckB$ of the elliptic fibration, whose canonical divisor is denoted as $K_\ckB$. Comparing to equation (\ref{JacE2}), we conclude that the $Z_\beta$ are Jacobi forms with modular parameter $\tau$, elliptic parameter the string coupling $g_s$, and of index 
\be
m = \frac{\beta \cdot (\beta + K_{\ckB})}{2} \,.
\ee
This expression for the index can easily be evaluated using the intersection matrix of the $D_{b_i}$ and the adjunction formula in the form (\ref{KC_from_adjunction}). For the E- and the M-string, with $\beta = (n_b)$, we obtain
\be
\beta \cdot (\beta + K_\ckB) = \begin{cases}
	-n_b (n_b+1) & \text{for the E-string,} \\
	-2 n_b^2 & \text{for the M-string.} \\
\end{cases}
\ee
Already in \cite{Minahan:1998vr}, it was argued that the masses of the E-string should enter the partition function as the elliptic parameters of Jacobi forms, with index proportional to the number of strings. The same holds true for the M-string. This gives rise to the equation 
\be
\left( \partial_{E_2} + \frac{1}{12} \left[\frac{\beta \cdot (\beta + K_\ckB)}{2} g_s^2 + c_m n_b \, Q(\md m) \right] \right) Z_\beta = 0  \,,
\ee
where the quadratic form $Q(\md m)$ is
\begin{equation}
Q(\md m) = \begin{cases}
(\md m, \md m)_{\mathfrak{e}_8}  & \textrm{for the E-string \ ,} \\
m^2 & \textrm{for the M-string \ .}
\end{cases}
\end{equation}
$(\cdot,\cdot)_{\mathfrak{e}_8}$ is the $E_8$ Weyl invariant inner product on the root lattice $\Lambda_{\mathfrak{e}_8}$, normalized so that the norm square of the highest (and hence every) root of $\mathfrak{e}_8$ is 2 (the Killing form will also make an appearance below, for which we will use the notation $(\cdot,\cdot)$, dropping the index). $\md m$ is the 8 dimensional mass parameter of the E-string and takes values in the complexified root lattice $\Lambda_{\mathfrak{e}_8} \otimes \IC$ of $\mathfrak{e}_8$. $m$ is the mass associated to the M-string. The value of $c_m$ is determined experimentally to be
\be
c_m = \begin{cases}
	\frac{1}{2} & \text{for the E-string,} \\
	1 & \text{for the M-string.} \\
\end{cases}
\ee
The coupling constant $s$ of the refined topological string in many respects behaves as a mass parameter \cite{Huang:2011qx}. It is hence to be expected that it also enters the holomorphic anomaly equations with a coefficient proportional to $n_b$.
\be\label{eq:holomorphic-anomaly}
\left( \partial_{E_2} + \frac{1}{12} \left[\frac{\beta \cdot (\beta + K_\ckB)}{2} g_s^2 + c_m n_b \,Q(\md m) + c_s n_b s^2\right] \right) Z_\beta = 0  \,.
\ee
We determine the values of $c_s$ experimentally to be 
\be
c_s = \begin{cases}
	\frac{1}{2} & \text{for the E-string,} \\
	\frac{1}{4} & \text{for the M-string.} \\
\end{cases}
\ee
The value of $c_s$ for the E-string was already reported in \cite{Huang:2013}. We will call
\begin{equation}
M_{\beta} = \frac{\beta \cdot (\beta + K_\ckB)}{2} g_s^2 + c_m n_b \,Q(\md m) + c_s n_b s^2
\end{equation}
the index polynomial of $Z_\beta$.

Assuming that the contributions to the index polynomial which depend on $s$ and the masses remain linear in the base class, $M_\beta$ has a unique generalization to geometries with base $\check B$ a resolved ADE quotient of $\IC^2$, possibly with a $(-1)$ curve added at the end of the chain of $(-2)$ curves, given by
\begin{equation}
M_{\beta} = \frac{\beta \cdot (\beta + K_\ckB)}{2} g_s^2 +  \frac{1}{4}\(2n_0+\sum_{i=1}^{n} n_i \)s^2 +\frac{n_0}{2}(\md m, \md m)_{\mathfrak{e}_8} + m^2 \sum_{i=1}^{n} n_i  \ .
\end{equation}
We have here set $\beta = (n_0, \mathbf{n})$, with $n_0$ specifying the wrapping number of the $(-1)$ curve, and called the rank of the ADE quotient $n$. With recourse to \eqref{KC_from_adjunction}, we can easily evaluate the intersection products. This yields
\begin{equation} \label{general_index_poly}
M_{\beta} = -\frac{1}{2}\( \sum_{i,j=0}^{n} C'_{ij} n_i n_j + n_0\) g_s^2 +  \frac{1}{4}\(2n_0+\sum_{i=1}^{n} n_i \)s^2 +\frac{n_0}{2}(\md m, \md m)_{\mathfrak{e}_8} + m^2 \sum_{i=1}^{n} n_i  \ ,
\end{equation}
where we have introduced the intersection matrix
\begin{equation}
C'=\left(
\begin{array}{c c}
1 &\!\! -1 \,\,\,\, 0 \,\,\,\cdots  \\ 
-1 & \raisebox{-15pt}{{\huge\mbox{{$C_\Gamma$}}}} \\[-3ex]
0 & \\
\vdots & \\[-0.5ex]
\end{array}
\right)\,.
\end{equation}

This index polynomial can in principle be applied to any of the ADE quotients. For the $D$ and the $E$ series, it is not immediately obvious that twisting by the mass parameter is well-defined geometrically. In appendix \ref{Modularandenumerative}, we give some coefficients $N_{j_{-}j_{+}}^\kappa$ extracted from our preliminary results for the $D_4$ quotient which pass basic consistency checks: they are positive integers, and follow a checkerboard pattern \cite{CKK}.

\subsection{The index from anomalies}
\label{sc:anomaly}

The contribution $Z_\beta (\tau, \md m, \epsilon_1, \epsilon_2) \boldsymbol{Q}^\beta$ in the expansion \eqref{betaexpansion} of the refined topological string partition function can be identified with the elliptic genus of the non-critical string coming from $D3$ branes wrapping the curve class $\beta$ in the Calabi-Yau. From the elliptic genus point of view, $\epsilon_1,\epsilon_2$ and $\md m$ are interpreted as fugacities $u_a$ associated to global symmetries in the worldsheet theory of the non-critical string. This connection gives us another way to derive the index polynomial \eqref{general_index_poly}.

The elliptic genus transforms under the modular group as \cite{Benini:2013xpa}
\begin{equation}
Z\(\frac{a \tau+b}{c\tau+d}, \frac{u_a}{c\tau+d}\) = \eta \exp\left[ -\frac{\pi\ri c}{c \tau +d} \sum_{a} \mathcal{A}^{a} u_a^2 \right] Z(\tau, u_a) \ , \quad \begin{pmatrix}
a & b \\ c & d
\end{pmatrix} \in \SLtz \ .
\end{equation}
Here $\eta$ is a phase factor which will be reproduced in our ansatz by inclusion of factors of the Dedekind $\eta$ function; $\mathcal{A}^{a}$ are the 't Hooft anomaly coefficients of the global symmetry currents
\begin{equation}
\mathcal{A}^{a} = \Tr_{\textrm{Weyl}} \gamma_3 Q^a Q^a \ ,
\end{equation}
$Q^a$ being the charge operators of the global symmetries with fugacities $u_a$. The index polynomial of the elliptic genus is thus given by
\begin{equation}
M= \frac{1}{2}\sum_{a} \mathcal{A}^a u_a^2 \ .
\end{equation}
On the other hand, the 't Hooft anomaly coefficients are naturally contained in the anomaly polynomial $I_4$ of the 2d worldsheet theory, since the latter contains terms like \cite{Bobev:2015kza}
\begin{equation}
I_4  \ni \sum_a \mathcal{A}^a \textrm{ch}_2(\mathcal{F}_a) \ ,
\end{equation}
where $\mathcal{F}_a$ are the field strength of the global symmetries which are weakly gauged. In fact, as pointed out in \cite{DelZotto:2016pvm}, the index polynomial $M$ of the elliptic genus can be obtained from the anomaly polynomial $I_4$ by a simple substitution. This procedure reproduces the index polynomial \eqref{general_index_poly} argued for above, as we now demonstrate.

The anomaly polynomial of the 2d worldsheet theory of the non-critical string reads \cite{Shimizu:2016lbw}
\begin{equation}\label{eq:anomaly}
\begin{aligned}
I_4 = & \sum_{i,j}\frac{\eta^{ij} n_i n_j}{2}(c_2(L) - c_2(R) ) \\
& + \sum_{i} n_i \(\frac{1}{4}\sum_{a}\eta^{ia} \Tr F_a^2 -\frac{2-\eta^{ii}}{4}(p_1(T) - 2c_2(L) - 2 c_2(R))  +h_{\mathfrak g_i}^\vee c_2(I) \) \ .
\end{aligned}
\end{equation}
Here $c_2(L)$, $c_2(R)$ are the second Chern classes of the $SU(2)_-$, $SU(2)_+$ bundles associated to the Omega background, while $c_2(I)$ is the second Chern class of the $SU(2)_I$ bundle associated to the $SU(2)$ R-symmetry of the 6d $(1,0)$ theory inherited by the worldsheet of the non-critical string. $p_1(T)$ is the first Pontryagin class of the tangent bundle of the worldsheet, and it vanishes in the elliptic genus calculation as the worldsheet is a torus. $F_a$ are the field strength of weakly gauged flavor symmetries. They include those $G_i$ induced by singular elliptic fibers over the compact divisors $D_{b_i}$ in the base, which manifest themselves as gauge symmetries in the 6d theory, as well as the flavor symmetries inherited from the 6d theory, for instance the $E_8$ symmetry associated to the E-string and the $U(1)_m$ symmetry associated to the M-string. If $G_i=\emptyset$, one sets $h^\vee_{\mathfrak g_i}=1$. $n_i$ are the wrapping numbers of the non-critical string on $D_{b_i}$. $\eta^{ij}$ can be identified with the intersection numbers $C'_{ij}$ of $D_{b_i}$, and $\eta^{ia}$ are the charges of the flavor symmetries carried by the non-critical string. For the flavor symmetries of the E-string and the M-string, $\eta^{ia} = -1$ \cite{Shimizu:2016lbw}.

It is claimed in \cite{DelZotto:2016pvm} that the index polynomial (called modular anomaly in that paper) of the elliptic genus can be obtained from the anomaly polynomial by the substitution
\begin{equation}\label{eq:relacement}
\begin{gathered}
c_2(L) \mapsto -\epsilon_-^2 \ ,  \quad c_2(R) \mapsto -\epsilon_+^2 \ , \quad c_2(I) \mapsto - \epsilon_+^2 \ ,\\
\frac{1}{2}\Tr F_G^2 \mapsto -\frac{1}{2 h_{\mathfrak g}^\vee}\sum_{\alpha\in \Phi_\G} (\alpha, \md m)_{\mathfrak{g}} (\alpha, \md m)_{\mathfrak{g}} \ ,\quad p_1(T)\mapsto 0 \ .\\
\end{gathered}
\end{equation}
Here, $\Phi_\G$ indicates the set of roots. The inner product $(\cdot,\cdot)_{\G}$ was introduced above as the scalar product normalized such that long roots have length squared 2. To relate to our discussion in section \ref{AnomalyEM}, note that the only flavor symmetries are the $E_8$ symmetry of the E-string and the $U(1)_m$ symmetry associated to the chain of $(-2)$ curves. 
In the latter case, one simply needs 
\begin{equation}\label{eq:M-m-rep}
\Tr F_{U(1)_m}^2 \mapsto -  4m^2 \ .
\end{equation}
In the case of the E-string, via the equalities
\be
(\md m,\md m) = \sum_{\alpha\in \Phi_\G} (\alpha, \md m)(\alpha,\md m)  \quad \mbox{and} \quad (\theta,\theta) = \frac{1}{h_\G^\vee} \,,
\ee 
where $\theta$ indicates a highest root and the inner product $(\cdot,\cdot)$ on the complexified root lattice is the one induced by the Killing form, we can simplify the replacement rule and obtain
\begin{equation}\label{eq:E-m-rep}
\frac{1}{2}\Tr F_{E_8}^2 \mapsto - (\md m, \md m)_{\mathbf{e}_8} \ .
\end{equation}
The replacement rules \eqref{eq:relacement} together with \eqref{eq:M-m-rep} and \eqref{eq:E-m-rep} reproduce the index polynomial \eqref{general_index_poly}.

Note that given the information that $Z_\beta$ is a Jacobi form, the holomorphic anomaly equations are completely determined by the index polynomial of the Jacobi form, as we argue in section \ref{s:diff_equ_J} and \ref{AnomalyEM}. We have here seen that the index polynomial is also fully determined by the chiral anomaly of the 2d worldsheet theory of the 6d BPS string. The holomorphic anomaly in this setting thus follows from the chiral anomaly.

\subsection{The ansatz for $Z_\beta $} 

\label{s:ansatz_z_base}

The space of holomorphic weak Jacobi forms of fixed weight and index is finite dimensional. We will make the ansatz, following \cite{Huang:2015sta} for the unrefined case, that the meromorphic Jacobi form $Z_\beta$ is a rational function of Jacobi forms, with a universal denominator. Once the form of the denominator is fixed, we have the following situation:
\begin{itemize}
\item In the massless unrefined case, the structure of the numerator of $Z_\beta$ is completely fixed, and the computation of $Z_\beta$ reduces to fixing a finite number of coefficients. 

\item Considering masses increases the number of elliptic parameters. The simplest generalisation of the massless ansatz is to consider the numerator to be a polynomial of Jacobi forms of elliptic parameter {\it either} $g_s$ {\it or} the masses. This proves to be sufficient for the smooth elliptic fibrations that we are considering. The M-string exhibits a single mass parameter, Jacobi forms of index $n_b$ with regard to this mass are hence polynomials in $\cA(\tau, m)$ and $\cB(\tau, m)$. The 8 mass parameters of the E-string enjoy the action of the $E_8$ Weyl group, under which the theory is symmetric. A set of 9 $E_8$ Weyl invariant Jacobi forms here generates the ring of Jacobi forms required. The main properties of these forms are reviewed in appendix~\ref{sc:jacobi_weyl}. In the case of the E-M$^n$ string chain, one includes both the 8 masses of the E-string and the extra mass associated to the M-string chain.

\item Refinement increases yet again the number of elliptic parameter. Now, Jacobi forms of different combinations of the elliptic parameters are required. 
\end{itemize}

To arrive at the correct ansatz in the massive refined case, we take the form of the refined topological string free energy \cite{Hollowood:2003cv} in terms of 5d BPS invariants $n^{g_-,g_+}_{\boldsymbol{\beta}}$ as inspiration.\footnote{These can easily be related to the BPS numbers $N_{j_- j_+}^\beta$ introduced above; see e.g. \cite{Huang:2011qx}.} Recall that this takes the form
\be \label{refined_GV}
F = \sum \frac{n^{g_-,g_+}_{\boldsymbol{\beta}}}{w}\frac{(2\sin \frac{w \epsilon_+}{2})^{2g_+} (2\sin \frac{w \epsilon_-}{2})^{2g_-}}{2 \sin \frac{w \epsilon_1}{2}2 \sin \frac{w \epsilon_2}{2}} \boldsymbol{Q}^{w\boldsymbol{\beta}} \,,
\ee
where $\epsilon_{\pm}=\frac{\epsilon_1 \pm \epsilon_2}{2}$. Note that the argument of the sine functions are proportional to $\epsilon_{1,2}$ in the denominator, and proportional to $\epsilon_\pm$ in the numerator. Given that the generators of integer index Jacobi forms, $A(\tau,z)$ and $B(\tau,z)$, have an expansion in 
\be  \label{def_x}
x = (2 \sin \pi z)^2 \,,
\ee
with $A \sim x$ to leading order, we make the ansatz 
\be \label{ansatz_z_nb}
Z_{\beta} = \left(\frac{\sqrt{q}}{\eta(\tau)^{12}}\right)^{-\beta \cdot K_\ckB}\frac{\num_{k,\np,\nm,\beta}(\tau,\boldsymbol{m},\epsilon_+,\epsilon_-)}{ \prod_{i=1}^r \prod_{s=1}^{\beta_i} \left[ \phi_{-1,\frac{1}{2}}(\tau, \tfrac{s \epsilon_1}{2\pi}) \phi_{-1,\frac{1}{2}}(\tau, \tfrac{s\epsilon_2}{2\pi}) \right]} \,,
\ee
where $\phi_{-1,\frac{1}{2}}^2 = \phi_{-2,1}$ and
\ba \label{ansatz_numerator}
\num_{k,\np,\nm,\beta}(\tau,\boldsymbol{m},\epsilon_+,\epsilon_-) &=&
\sum c_{k_4,k_6,k_{\cA,+},k_{\cB,+},k_{\cA,-},k_{\cB,-},\boldsymbol{k_m}} \\&&  E_4^{k_4}E_6^{k_6} \num_{\boldsymbol{k_m}}(\boldsymbol{m})  (\cA^{k_{\cA,+}}\cB^{k_{\cB,+}})(\epsilon_+)( \cA^{k_{\cA,-}}\cB^{k_{\cB,-}})(\epsilon_-) \,, \nn
\ea
with $\num_{\boldsymbol{k_m}}(\boldsymbol{m})$ a polynomial of appropriate weight and index in the generators $\cA(\tau,m)$, $\cB(\tau,m)$, and the $E_8$ Weyl invariant Jacobi forms. The $\eta(\tau)$ dependent prefactor is argued for in \cite{HKK2}, and entails that $Z_\beta$ for $\beta$ including a (-1) curve class transforms as a Jacobi form only up to a non-trivial multiplier system. The factor $\sqrt{q}$ comes from a redefinition of the curve classes $\boldsymbol{Q}$ in the base such that the latter behave properly under the modular transformation of the elliptic fiber \cite{Huang:2015sta}. We include it in $Z_\beta$ so that the combination $\sqrt{q}/\eta(\tau)^{12}$ is invariant under the shift $\tau \mapsto \tau+1$.
For $Z_{\beta}$ to have vanishing weight, we restrict the sum such that the total weight $k$ of each summand cancels that of the denominator in (\ref{ansatz_z_nb}). For the E- and the M-string, with $\beta = (n_b)$, this yields
\be
k = \begin{cases}
	(6-2) n_b = 4 n_b & \text{for the E-string,} \\
	-2 n_b & \text{for the M-string.} \\
\end{cases}
\ee
The index of $Z_{\beta}$ determined in section~\ref{AnomalyEM} then constrains the indices $(n_+,n_-)$ of the Jacobi forms in each summand as follows:
\ba
\np \left(\frac{\epsilon_1+\epsilon_2}{2} \right)^2 &+& \nm \left(\frac{\epsilon_1-\epsilon_2}{2} \right)^2 - \frac{n_b(n_b+1)(2n_b+1)}{12}( \epsilon_1^2 + \epsilon_2^2) = \\
 &=& \begin{cases}
	\frac{n_b(n_b+1) \epsilon_1 \epsilon_2 - n_b (\epsilon_1 + \epsilon_2)^2}{2} & \text{for the E-string,} \\
	\frac{4n_b^2 \epsilon_1 \epsilon_2 - n_b (\epsilon_1 + \epsilon_2)^2}{4} & \text{for the M-string.} \\
\end{cases} \nn
\ea
yielding
\be \label{nump_indices}
n_+ = \begin{cases}
	\frac{n_b}{3} (n_b^2 + 3n_b-4)  & \text{for the E-string,} \\
	\frac{n_b}{6}(2n_b^2+9n_b-5) & \text{for the M-string,} \\
\end{cases}
\ee
and
\be \label{numm_indices}
n_- = \begin{cases}
	\frac{n_b}{3}(n_b^2-1)    & \text{for the E-string,} \\
	\frac{n_b}{6}(2n_b^2 -3n_b +1)& \text{for the M-string.} \\
\end{cases}
\ee
Similar calculations can be performed for the geometries based on chains of $(-1)$ and $(-2)$ curves.

\section{Computing $Z_{\beta}$ in terms of meromorphic Jacobi forms}
\label{Determining} 

To determine the coefficients of the numerator (\ref{ansatz_numerator}) of $Z_\beta$, we proceed as follows:\footnote{For ease of exposition, we set the E-string masses to zero here. Adapting the algorithm to the massive E-string case is immediate. In our computations, we have considered the fully massive case.} Assume all $Z_{\beta'}$ below a certain base degree $\beta$ have been determined.
\begin{itemize}
	\item Express $F_\beta$ as defined in \eqref{betaexpansion} in terms of $Z_{\beta'}$ for $\beta' \le \beta$,
	\be
	F_\beta = Z_\beta + \mathrm{polynomial\,\, in}\,\, Z_{\beta'} \,, \quad \beta' < \beta \ .
	\ee
	\item Express $Z_{\beta'}$ for $\beta' \le \beta$ in the form \eqref{ansatz_z_nb}, with the coefficients ${\mathbf c}$ in \eqref{ansatz_numerator} known for $\beta' < \beta$.
	\item Expand $F_\beta$ in $q$, with its coefficients expressed as polynomials in 
	\be 
	x_\pm = x\left(\frac{\epsilon_\pm}{2\pi}\right) \,, \quad x_m = x\left(\frac{m}{2\pi}\right)\,,
	\ee 
	with $x$ as defined in (\ref{def_x}).
	\item Obtain the same expansion of $F_\beta$ expressed in terms of BPS invariants by expressing (\ref{refined_GV}) in terms of $x_\pm$ and $x_m$ and expanding appropriately.
	\item Imposing the boundary conditions on the BPS invariants that we discuss in section \ref{s:sufficiency} and comparing coefficients suffices to fix all unknown coefficients.  
\end{itemize}
Below, we will give a summary of the partition functions we have computed, including the weight, the indices, and the number of summands in the numerator $\num_{k,n_+,n_-,\beta}$ for the E-string, the M-string, and the E-M$^{n}$ string chain. In each case, we also give explicit expressions for the numerator $\num_{k,n_+,n_-,\beta}$ for relatively small base degrees in terms of weak Jacobi forms, such that $Z_\beta$ can easily be reconstructed in accordance with \eqref{ansatz_z_nb}. The $\num_{k,n_+,n_-,\beta}$ for higher base degrees are available upon request. In the following, we will use the following notation: 
\be
\cA_\pm = \cA(\tau, \epsilon_\pm) \,, \quad \cA_m = \cA(\tau, m) \,,\quad \cB_\pm = \cB(\tau, \epsilon_\pm) \,, \quad \cB_m = \cB(\tau, m) \,.
\label{JacobiringI}
\ee
$A_i$ and $B_i$ denote the $W(E_8)$-invariant Jacobi forms reviewed in appendix \ref{sc:jacobi_weyl}.

\subsection{Sufficiency of vanishing conditions to fix ansatz} \label{s:sufficiency}
In this section, we study to what extent the vanishing conditions on the BPS numbers $N_{j_- j_+}^\kappa$ suffice to fix the coefficients in the ansatz \eqref{ansatz_numerator}.

\subsubsection{The unrefined case}
For ease of exposition, we begin with the unrefined case. The ansatz \eqref{ansatz_z_nb} simplifies in this case to
\be \label{ansatz_z_nb_unref}
Z_{\beta} \bigg/ \left(\frac{\sqrt{q}}{\eta(\tau)^{12}}\right)^{-\beta \cdot K_\ckB} = \frac{\num_{k,n,\beta}(\tau,\boldsymbol{m},\epsilon)}{ \prod_{i=1}^r \prod_{s=1}^{\beta_i} \left[ \phi_{-2,1}(\tau, \tfrac{s \epsilon}{2\pi}) \right]} \,,
\ee
with
\be \label{ansatz_numerator_unref}
\num_{k,n,\beta}(\tau,\boldsymbol{m},\epsilon)=
\sum c_{k_4,k_6,k_{\cA},k_{\cB},\boldsymbol{k_m}}  E_4^{k_4}E_6^{k_6} \num_{\boldsymbol{k_m}}(\boldsymbol{m})  (\cA^{k_{\cA}}\cB^{k_{\cB}})(\epsilon) \,. 
\ee
We will think of $Z_\beta$ as a power series in $q$ with coefficients that are Laurent series in $x=(2 \sin \pi z)^2$, see appendix \ref{Appendixmod}. For the ensuing discussion, it will be convenient to introduce the following terminology: we will call a series
\be
\phi(q,x) = \sum_{n=0}^{\infty} q^n f_n(x) 
\ee 
{\it geometrically bounded} if for all $n$, the series $f_n(x)$ are finite polynomials. For example, $1+q x$ is geometrically bounded, while $1+\frac{q}{1-x}$ is not. Clearly, the product of two geometrically bounded series is also geometrically bounded. It is easy to show that $A(\tau, z)$ and $B(\tau, z)$ are geometrically bounded as power series in $q$ and $x$; hence, all weak Jacobi forms are, as they can be written as polynomials in $A(\tau, z)$, $B(\tau, z)$, $E_4(\tau)$ and  $E_6(\tau)$. The inverse  $\phi(q,x)^{-1}$ of a geometrically bounded series $\phi(q,x)$ is geometrically bounded if and only if the leading term $f_0(x)$ is a non-vanishing constant.

As we discussed in section \ref{IntegerGeometric}, Castelnuovo theory asserts the vanishing of Gopakumar-Vafa invariants of a given curve class at sufficiently high genus. Defining the single wrapping contribution to $Z_\beta$ as
\ba \label{zsing}
\zbetasing&=& Z_\beta - \sum \mathrm{contributions\,\, from\, }I^{\beta'}_g \mathrm{\,\,for\,\,}\beta'<\beta \nn\\
&=& 
\sum_{g=0}^\infty \sum_{\kappa}
I^{\beta,\kappa}_g \left(2 \sin \left( \frac{g_s}{2}\right)\right)^{2 g -2} \boldsymbol{Q}^{\kappa}  \,, 
\ea
where $g_s = 2\pi z$ and $\boldsymbol{Q}=(q,\boldsymbol{Q}_{\boldsymbol{m}})$, we conclude that $x \zbetasing$ must be geometrically bounded. We will call a series that becomes geometrically bounded upon multiplication with $x$ a {\it Gopakumar-Vafa series}. 

$\zbetasing$ will generically not be of the form \eqref{ansatz_z_nb_unref}. Assume however that we have fixed all Gopakumar-Vafa invariants at base degrees $\beta'<\beta$. Assume further that at base degree $\beta$, we find two different sets of constants ${\bf c_1}$ and ${\bf c_2}$ in \eqref{ansatz_numerator_unref} that are compatible with the vanishing conditions on Gopakumar-Vafa invariants. Multi-wrapping contributions cancel out in the difference of $Z_\beta({\bf c_1})$ and $Z_\beta({\bf c_2})$, as these depend on $\beta'<\beta$. The ambiguity $\zbetaambig$ in our procedure,
\be \label{def_amb}
\zbetaambig = Z_\beta({\bf c_1}) - Z_\beta({\bf c_2}) = \zbetasing({\bf c_1}) - \zbetasing({\bf c_2}) \,,
\ee
is therefore a Gopakumar-Vafa series, and it {\it can} be written in the form \eqref{ansatz_z_nb_unref}, as evident from the first equality in \eqref{def_amb}. Our argument will now rely on the lemma that the only rational functions of weak Jacobi forms of the form \eqref{ansatz_z_nb_unref} which are Gopakumar-Vafa series have at most a single power of $A(\tau,z)$ in the denominator. Hence,
\be \label{ambiguity}
\zbetaambig \bigg/ \left(\frac{\sqrt{q}}{\eta(\tau)^{12}}\right)^{-\beta \cdot K_\ckB} = \frac{\phibetaambig(\tau,\epsilon,\boldsymbol{m})}{  \phi_{-2,1}(\tau, \tfrac{\epsilon}{2\pi} ) } \,.
\ee
Note that $\zbetaambig$ has the same index as $Z_\beta$. $\zbetaambig$ vanishes, and $Z_\beta$ is therefore completely determined by the vanishing conditions, when this index would require the index of $\phibetaambig$ to be negative.

To prove the lemma, assume that $\zbetaambig$ is a Gopakumar-Vafa series. Then 
\be \label{arguing_div}
A(\tau,z) \zbetaambig \bigg/ \left(\frac{\sqrt{q}}{\eta(\tau)^{12}}\right)^{-\beta \cdot K_\ckB}= \frac{\phi(\tau,z)}{D(\tau,z)}
\ee
is geometrically bounded, with $D(\tau,z)$ in the product form implied by \eqref{ansatz_z_nb_unref}. By the factorized form \eqref{Aproduct} of $A(\tau,z)$, $D(\tau,z)$ can be decomposed as $D(\tau,z) = d(z) \tilde{D}(\tau,z)$, such that $\tilde{D}(\tau,z) = 1 + \cO(q,z)$. Geometric boundedness of \eqref{arguing_div} thus implies that each coefficient of $\phi(\tau,z)$ in a $q$ expansion must be divisible by $d(z)$. We can conclude that $\phi(\tau,z)$ vanishes at the zeros of $d(z)$ with at least equal multiplicity. Now $d(z)$ has only real zeros, and these coincide (counted with multiplicities) with the real zeros in $z$ of $D(\tau,z)$. Invoking the $\SLtz$ transformation properties of both $\phi(\tau,z)$ and $D(\tau,z)$ to obtain the $\SLtz$ images of these zeros in the complex plane, we can conclude that $\phi(\tau,z)$ vanishes at all the zeros of not only $d(z)$ but $D(\tau,z)$, with at least equal multiplicity. Hence, $A(\tau,z) \zbetaambig \big/ \left(\frac{\sqrt{q}}{\eta(\tau)^{12}}\right)^{-\beta \cdot K_\ckB}$ only exhibits removable singularities, is thus a weak Jacobi form, and $\zbetaambig$ is of the form \eqref{ambiguity}. This concludes the argument for the lemma.

\subsubsection{The refined case}
To adapt the terminology introduced above to the refined case, we introduce the notation $x_{\pm} = (2\sin(\frac{\epsilon_{\pm}}{2}))^2$, and call a power series in $q, x_{\pm}$ organized in the form
\be
\phi(q,x_+,x_-) = \sum_{n=0}^{\infty} q^n f_n(x_+,x_-)  
\ee 
{\it geometrically bounded} if the $f_n$ are polynomials in both $x_+$ and $x_-$. The numerator \eqref{ansatz_numerator} is evidently such a geometrically bounded power series. Likewise, we introduce the terminology {\it refined Gopakumar-Vafa series} for a power series in $(q,x_+,x_-)$, which, like the single wrapping contribution to $Z_\beta$ in the refined case, becomes geometrically bounded when multiplied by $x_{+} -x_{-} = 4\sin(\frac{\epsilon_{1}}{2}) \sin(\frac{\epsilon_{2}}{2})$. 

To adapt our argument regarding $\zbetaambig$ to the refined case, we require a lemma stating that the only rational functions of weak Jacobi forms of the form \eqref{ansatz_z_nb} which are refined Gopakumar-Vafa series have at most a single power of
\be
\phi_{-1,\frac{1}{2}}(\tau, \tfrac{\epsilon_{1}}{2\pi}) \phi_{-1,\frac{1}{2}}(\tau, \tfrac{\epsilon_{2}}{2\pi}) = \frac{1}{12}(\cA(\tau, \epsilon_{-}) \cB(\tau, \epsilon_{+}) - \cA(\tau, \epsilon_{+}) \cB(\tau, \epsilon_{-}))
\ee
in the denominator. This will allow us to conclude that the ambiguity in determining $Z_\beta$ upon imposing vanishing conditions takes the form
\be \label{ambiguity_ref}
\zbetaambig \bigg /\left(\frac{\sqrt{q}}{\eta(\tau)^{12}}\right)^{-\beta \cdot K_\ckB} = \frac{\phibetaambig(\tau,\epsilon_+,\epsilon_-,\boldsymbol{m})}{  \phi_{-1,\frac{1}{2}}(\tau, \tfrac{\epsilon_{1}}{2\pi}) \phi_{-1,\frac{1}{2}}(\tau, \tfrac{\epsilon_{2}}{2\pi})  } \,.
\ee
The negativity of either the $\epsilon_+$ or the $\epsilon_-$ index of $\phibetaambig$ will then suffice to conclude that the ambiguity must vanish.

To establish the lemma, we argue as above: if $\zbetaambig$ is to be a Gopakumar-Vafa series, 
\be \label{lemma_ref}
(\cA(\tau, \epsilon_{-}) \cB(\tau, \epsilon_{+}) - \cA(\tau, \epsilon_{+}) \cB(\tau, \epsilon_{-})) \zbetaambig = \frac{\phi(\tau,\epsilon_+,\epsilon_-)}{D(\tau,\epsilon_+,\epsilon_-)}
\ee
must be geometrically bounded. Replacing $D(\tau,\epsilon_+,\epsilon_-)$ by $d(\epsilon_+,\epsilon_-)$ and arguing term-wise, we again conclude that \eqref{lemma_ref} exhibits no poles. At this point, we would require a structure theorem stating that a weak Jacobi form in two integer indices $\epsilon_+$ and $\epsilon_-$ is generated by $A_\pm$, $B_\pm$, $E_4$ and $E_6$. Short of this, we invoke the conjecture put forth in section \ref{strategy}, stating that $Z_\beta$ can be expressed in the form (\ref{ansatz_z_nb}), to conclude the argument.

Let us now apply these considerations to the massless E-string and M-string. For the leading base degree $\beta = C_{1,2}$, the general ansatz (\ref{ansatz_z_nb}) is a refined Gopakumar-Vafa series, so imposing the vanishing of BPS invariants is not sufficient to determine the coefficients ${\mathbf c}$. It turns out that for both models, providing one non-vanishing BPS invariant is sufficient to completely fix the ansatz. Beyond this base degree, the indices for $\epsilon_{-}$ in both models are negative and smaller than $-1$. While for the ansatz (\ref{ansatz_z_nb}), the numerator still has non-negative index, the numerator of $\zbetaambig$ would now require negative index for $\epsilon_{-}$. As this is not possible, imposing the vanishing conditions suffices to fix the ansatz completely for all base degrees beyond 
$\beta = C_{1,2}$. The analysis for the more general geometries we consider proceeds analogously.

\subsection{The M-string}
\label{Mstring} 

For the M-string with its associated mass $m$ turned on, we have computed the partition function up to base degree three. The results are summarized in Tab.~\ref{tb:M-summary}. 
The explicit forms of the numerator with base degree one and two are given in the following.

\begin{table}
	\centering
	\begin{tabular}{*{6}{>{$}c<{$}}}\toprule
		(n_b) & k & n_+ & n_-& \# \\\midrule
		(1) & -2 & 1 & 0 & 2/1 \\
		(2) & -4 & 7 & 1 & 28/7 \\
		(3) & -6 & 20 & 5 & 982/215 \\
		(4) & -8 & 42 & 14 & -/2694 \\\bottomrule
	\end{tabular}
	\caption{The weight $k$, the indices $n_{\pm}$, and the number of terms $\#$ in the numerator $\num_{k,n_+,n_-,(n_b)}$ of the partition function of the massive/massless M-string.}\label{tb:M-summary}
\end{table}

\noindent $\mathbf{n_b=1}$
\be \label{Mnb1}
\num_{-2,1,0,(1)}^{\scriptscriptstyle \Ms} = \frac{-\cA_m \cB_+ + \cA_+ \cB_m}{12} \,.
\ee

\noindent $\mathbf{n_b=2}$
\be \label{Mnb2}
\begin{array}{rl}
	\num_{-4,7,1,(2)}^{\scriptscriptstyle \Ms} &= \frac{E_4^3 \cA_- \cA_+^7 \cB_m^2}{7962624}-\frac{E_4^3 \cB_- \cA_+^7 \cA_m \cB_m}{7962624}+\frac{E_4^3 \cB_- \cA_+^6 \cB_+ \cA_m^2}{7962624}-\frac{E_4^3 \cA_- \cA_+^6 \cB_+ \cA_m \cB_m}{7962624}\\
	&+\frac{5 E_4^2 \cB_- \cA_+^5 \cB_+^2 \cA_m \cB_m}{23887872}-\frac{5 E_4^2 \cA_- \cA_+^5 \cB_+^2 \cB_m^2}{23887872}+\frac{5 E_4^2 \cA_- \cA_+^4 \cB_+^3 \cA_m \cB_m}{23887872}-\frac{5 E_4^2 \cB_- \cA_+^4 \cB_+^3 \cA_m^2}{23887872}\\
	&+\frac{E_4 E_6 \cA_- \cA_+^6 \cB_+ \cB_m^2}{8957952}-\frac{E_4 E_6 \cB_- \cA_+^6 \cB_+ \cA_m \cB_m}{8957952}+\frac{E_4 E_6 \cB_- \cA_+^5 \cB_+^2 \cA_m^2}{8957952}-\frac{E_4 E_6 \cA_- \cA_+^5 \cB_+^2 \cA_m \cB_m}{8957952}\\
	&+\frac{5 E_4 \cB_- \cA_+^3 \cB_+^4 \cA_m \cB_m}{71663616}-\frac{5 E_4 \cA_- \cA_+^3 \cB_+^4 \cB_m^2}{71663616}+\frac{5 E_4 \cA_- \cA_+^2 \cB_+^5 \cA_m \cB_m}{71663616}-\frac{5 E_4 \cB_- \cA_+^2 \cB_+^5 \cA_m^2}{71663616}\\
	&+\frac{E_6^2 \cB_- \cA_+^7 \cA_m \cB_m}{6718464}-\frac{E_6^2 \cA_- \cA_+^7 \cB_m^2}{6718464}+\frac{E_6^2 \cA_- \cA_+^6 \cB_+ \cA_m \cB_m}{6718464}-\frac{E_6^2 \cB_- \cA_+^6 \cB_+ \cA_m^2}{6718464}\\
	&+\frac{5 E_6 \cA_- \cA_+^4 \cB_+^3 \cB_m^2}{26873856}-\frac{5 E_6 \cB_- \cA_+^4 \cB_+^3 \cA_m \cB_m}{26873856}+\frac{5 E_6 \cB_- \cA_+^3 \cB_+^4 \cA_m^2}{26873856}-\frac{5 E_6 \cA_- \cA_+^3 \cB_+^4 \cA_m \cB_m}{26873856}\\
	&+\frac{\cA_- \cA_+ \cB_+^6 \cB_m^2}{214990848}-\frac{\cB_- \cA_+ \cB_+^6 \cA_m \cB_m}{214990848}+\frac{\cB_- \cB_+^7 \cA_m^2}{214990848}-\frac{\cA_- \cB_+^7 \cA_m \cB_m}{214990848} \,.
\end{array}
\ee
The massless limit is smooth and is obtained by replacing $\cA_m$ by $0$ and $\cB_m$ by $12$. Alternatively, the partition function of the massless M-string can be computed directly from the ansatz \eqref{ansatz_z_nb} with mass set to zero. We pushed this calculation up to base degree 4 as summarized in Tab.~\ref{tb:M-summary}.
%

\subsection{The E-string}
\label{Estring}

We have computed the partition functions for the E-string with its eight mass parameters turned on up to base degree three. The results are summarized in Tab.~\ref{tb:E-summary}. 
The explicit form of the numerator with base degrees one and two are given in the following. 

\begin{table}
	\centering
	\begin{tabular}{*{6}{>{$}c<{$}}}\toprule
		(n_b) & k & n_+ & n_-& \# \\\midrule
		(1) & 4 & 0 & 0 & 1/1 \\
		(2) & 8 & 4 & 2 & 43/24 \\
		(3) & 12 & 14 & 8 & 1663/431 \\
		(4) & 16 & 32 & 20 & -/4207 \\ \bottomrule
	\end{tabular}
	\caption{The weight $k$, the indices $n_{\pm}$, and the number of terms $\#$ in the numerator $\num_{k,n_+,n_-,(n_b)}$ of the partition function of the massive/massless $E$-string.}\label{tb:E-summary}
\end{table}

\noindent $\mathbf{n_b=1}$
\begin{equation} \label{Enb1}
	\num_{4,0,0,(1)}^{\scriptscriptstyle \Es} = - A_1 \,.
\end{equation}

\noindent $\mathbf{n_b=2}$
\begin{equation} \label{Enb2}
	\begin{array}{rl}
		\num_{8,4,2,(2)}^{\scriptscriptstyle \Es} &=-\frac{5 E_4^3 A_1^2 \cA_-^2
			\cA_+^4}{1492992}+\frac{\left(E_4^3-E_6^2\right) A_1^2 \cA_-^2
			\cA_+^4}{46656}-\frac{E_4^2 A_1^2 \cB_-^2 \cA_+^4}{31104}-\frac{E_4^3 A_2 \cB_-^2
			\cA_+^4}{124416}+\frac{\left(E_4^3-E_6^2\right) A_2 \cB_-^2
			\cA_+^4}{31104} \\ &-\frac{5 E_4 E_6 B_2 \cB_-^2 \cA_+^4}{373248}+\frac{E_4
			E_6 A_1^2 \cA_- \cB_- \cA_+^4}{46656}+\frac{E_4^2 E_6 A_2 \cA_- \cB_-
			\cA_+^4}{497664}+\frac{5 E_4^3 \cA_- B_2 \cB_- \cA_+^4}{1492992} \\& +\frac{5
			\left(E_4^3-E_6^2\right) \cA_- B_2 \cB_- \cA_+^4}{186624}  +\frac{E_4
			E_6 A_1^2 \cA_-^2 \cB_+ \cA_+^3}{23328}-\frac{E_4^2 E_6 A_2 \cA_-^2 \cB_+
			\cA_+^3}{497664}+\frac{E_6 A_1^2 \cB_-^2 \cB_+ \cA_+^3}{2916} \\&+\frac{E_4 E_6
			A_2 \cB_-^2 \cB_+ \cA_+^3}{10368}+\frac{5 E_4^2 B_2 \cB_-^2 \cB_+ \cA_+^3}{31104}-\frac{5
			E_4^3 \cA_-^2 B_2 \cB_+ \cA_+^3}{1492992} -\frac{5 \left(E_4^3-E_6^2\right)
			\cA_-^2 B_2 \cB_+ \cA_+^3}{186624} \\& -\frac{E_4^2 A_1^2 \cA_- \cB_- \cB_+
			\cA_+^3}{3888}  -\frac{E_4^3 A_2 \cA_- \cB_- \cB_+
			\cA_+^3}{62208}+\frac{\left(E_4^3-E_6^2\right) A_2 \cA_- \cB_- \cB_+
			\cA_+^3}{15552}-\frac{5 E_4 E_6 \cA_- B_2 \cB_- \cB_+
			\cA_+^3}{186624} \\& -\frac{E_4^2 A_1^2 \cA_-^2 \cB_+^2 \cA_+^2}{5184}+\frac{E_4^3 A_2
			\cA_-^2 \cB_+^2 \cA_+^2}{41472}-\frac{\left(E_4^3-E_6^2\right) A_2 \cA_-^2 \cB_+^2
			\cA_+^2}{10368}-\frac{ E_4 A_1^2 \cB_-^2 \cB_+^2 \cA_+^2}{972} \\& -\frac{E_4^2 A_2
			\cB_-^2 \cB_+^2 \cA_+^2}{2592}-\frac{5 E_6 B_2 \cB_-^2 \cB_+^2 \cA_+^2}{7776}+\frac{5
			E_4 E_6 \cA_-^2 B_2 \cB_+^2 \cA_+^2}{124416}+\frac{E_6 A_1^2 \cA_-
			\cB_- \cB_+^2 \cA_+^2 }{972} \\&+\frac{E_6 A_1^2 \cA_-^2 \cB_+^3 \cA_+}{2916}-\frac{E_4 E_6
			A_2 \cA_-^2 \cB_+^3 \cA_+}{10368}+\frac{E_6 A_2 \cB_-^2 \cB_+^3 \cA_+}{1944}+\frac{5
			E_4 B_2 \cB_-^2 \cB_+^3 \cA_+}{5832}-\frac{5 E_4^2 \cA_-^2 B_2 \cB_+^3
			\cA_+}{31104} \\&-\frac{E_4 A_1^2 \cA_- \cB_- \cB_+^3 \cA_+}{729} +\frac{E_4^2 A_2 \cA_-
			\cB_- \cB_+^3 \cA_+}{3888}+\frac{5 E_6 \cA_- B_2 \cB_- \cB_+^3 \cA_+}{11664}-\frac{E_4
			A_1^2 \cA_-^2 \cB_+^4}{5832} \\&+\frac{E_4^2 A_2 \cA_-^2 \cB_+^4}{7776}+\frac{2}{729} A_1^2
		\cB_-^2 \cB_+^4+\frac{5 E_6 \cA_-^2 B_2 \cB_+^4}{23328}-\frac{E_6 A_2 \cA_- \cB_-
			\cB_+^4}{1944}-\frac{5 E_4 \cA_- B_2 \cB_- \cB_+^4}{5832} \,.
	\end{array}
\end{equation}
Just as for the M-string, the massless limit is smooth, and yields a partition function summed over contributions from curve classes corresponding to the masses. For $n_b=1,2$, the expressions above reduce to
\be
\num_{4,0,0,(1)}^{\scriptscriptstyle \Es,m=0} = - E_4
\ee
and
\begin{equation} \label{3.16nb2}
	\begin{array}{rl} \num_{8,4,2,(2)}^{\scriptscriptstyle \Es,m=0}&=\frac{\cA_-^2 \cA_+^4 E_4^5}{55296}-\frac{\cA_-^2 \cA_+^4 E_4^2 E_6^2}{46656}+\frac{\cA_-^2 \cA_+^3 \cB_+ E_4^3 E_6}{373248}+\frac{5 \cA_-^2 \cA_+^3 \cB_+ E_6^3}{746496}-\frac{11 \cA_-^2 \cA_+^2 \cB_+^2 E_4^4}{663552} \\ 
		&+\frac{17 \cA_-^2 \cA_+^2 \cB_+^2 E_4 E_6^2}{1990656}+\frac{\cA_-^2 \cA_+ \cB_+^3 E_4^2 E_6}{746496}-\frac{\cA_-^2 \cB_+^4 E_4^3}{5971968}+\frac{5 \cA_-^2 \cB_+^4 E_6^2}{5971968}+\frac{5 \cA_- \cA_+^4 \cB_- E_4^3 E_6}{373248}\\ 
		&-\frac{5 \cA_- \cA_+^4\cB_- E_6^3}{746496}-\frac{13 \cA_- \cA_+^3 \cB_- \cB_+ E_4^4}{995328}-\frac{17 \cA_- \cA_+^3 \cB_- \cB_+ E_4 E_6^2}{2985984}+\frac{\cA_- \cA_+^2 \cB_-\cB_+^2 E_4^2 E_6}{62208}\\ 
		&-\frac{13 \cA_- \cA_+ \cB_- \cB_+^3 E_4^3}{2985984}+\frac{5 \cA_- \cA_+ \cB_- \cB_+^3 E_6^2}{2985984}-\frac{\cA_- \cB_- \cB_+^4 E_4 E_6}{746496}-\frac{\cA_+^4 \cB_-^2 E_4^4}{1990656}-\frac{17 \cA_+^4 \cB_-^2 E_4 E_6^2}{5971968}\\ 
		&+\frac{7 \cA_+^3 \cB_-^2 \cB_+ E_4^2 E_6}{746496}-\frac{11 \cA_+^2\cB_-^2 \cB_+^2 E_4^3}{1990656}-\frac{5 \cA_+^2 \cB_-^2 \cB_+^2 E_6^2}{1990656}+\frac{\cA_+ \cB_-^2 \cB_+^3 E_4 E_6}{746496}+\frac{\cB_-^2 \cB_+^4 E_4^2}{1492992} \,.
	\end{array}
\end{equation}
Similar to the M-string, the partition function of the massless E-string can alternatively be computed from the ansatz \eqref{ansatz_z_nb} with all masses set to zero. We have pushed this calculation up to $n_b=4$ as summarized in Tab.~\ref{tb:E-summary}.

\subsection{The E-M$^{n}$ string chain}
\label{EMchain}

Here we consider the E-M$^{n}$ string chain, which is engineered by the geometric configuration of one $(-1)$ curve connected to one end of a linear chain of $n$ $(-2)$ curves. To simplify the computation, we turn off the eight mass parameters associated to the E-string, but we keep the mass $m$ associated to the M-string chain finite. The weight and the indices of the numerator $\num_{k,n_+,n_-,(n_b)}$ read
\begin{equation}
	\begin{aligned}
		& k = 4n_0 - 2\sum_{i=1}^n n_i \ ,\\
		& n_+ = \frac{1}{2}\sum_{i,j=0}^n C'_{ij}n_i n_j - \(\frac{3}{2}n_0 + \sum_{i=1}^n n_i\) + \sum_{i=0}^n \frac{n_i(n_i+1)(2n_i + 1)}{6} \ ,\\
		& n_- = -\frac{1}{2}\sum_{i,j=0}^n C'_{ij}n_i n_j - \frac{1}{2}n_0 + \sum_{i=0}^n \frac{n_i(n_i+1)(2n_i + 1)}{6} \ . 
	\end{aligned}
\end{equation}

We have computed the partition functions up to base degree $(2,1), (1,2)$ for $n=1$ ($(-1)$ curve connected to one $(-2)$ curve), base degree $(2,1,1), (1,2,1), (1,1,2)$ for $n=2$ ($(-1)$ curve connected to two $(-2)$ curves), and base degree $(2,1,1,1)$ for $n=3$. The numbers of terms in the numerators are respectively 56, 53; 89, 80, 80; 126. We give in the following the explicit expression for base degrees (1,1), (1,2), (2,1) for $n=1$.

\noindent  $\mathbf{n_b=(1,1)}$
\begin{equation}
	\num_{2,0,1,(1,1)} = \frac{E_4}{12}(\cA_m \cB_- -\cA_- \cB_m)
\end{equation}

\noindent $\mathbf{n_b=(1,2)}$
\begin{equation}
	\begin{array}{rl}
		\num_{0,5,3,(1,2)} &= 
		\frac{E_4 E_6^2 \cA_-^3 \cB_m^2 \cA_+^5}{6718464}+\frac{E_4 E_6 \cB_-^3 \cB_m^2 \cA_+^5}{53747712}+\frac{E_4^2 E_6 \cA_- \cA_m \cB_-^2 \cB_m \cA_+^5}{8957952}+\frac{E_4^4 \cA_-^2 \cA_m \cB_- \cB_m \cA_+^5}{2654208}\\
		&-\frac{E_4 E_6^2 \cA_-^2 \cA_m \cB_- \cB_m \cA_+^5}{2239488}-\frac{E_4^4 \cA_-^3 \cB_m^2 \cA_+^5}{7962624}-\frac{E_4^3 \cA_- \cB_-^2 \cB_m^2 \cA_+^5}{23887872}-\frac{E_4^3 \cA_m \cB_-^3 \cB_m \cA_+^5}{23887872}\\
		&+\frac{E_4 E_6 \cA_- \cB_-^2 \cB_+ \cB_m^2 \cA_+^4}{17915904}+\frac{E_4^3 \cA_-^2 \cB_- \cB_+ \cB_m^2 \cA_+^4}{11943936}+\frac{E_4^3 \cA_m^2 \cB_-^3 \cB_+ \cA_+^4}{23887872}+\frac{E_4 E_6^2 \cA_-^2 \cA_m^2 \cB_- \cB_+ \cA_+^4}{2239488}\\
		&+\frac{E_4 E_6 \cA_m \cB_-^3 \cB_+ \cB_m \cA_+^4}{13436928}+\frac{E_4 E_6^2 \cA_-^3 \cA_m \cB_+ \cB_m \cA_+^4}{6718464}+\frac{E_4^2 E_6 \cA_-^2 \cA_m \cB_- \cB_+ \cB_m \cA_+^4}{8957952}\\
		&-\frac{E_4^4 \cA_-^2 \cA_m^2 \cB_- \cB_+ \cA_+^4}{2654208}-\frac{E_4^4 \cA_-^3 \cA_m \cB_+ \cB_m \cA_+^4}{7962624}-\frac{E_4^2 E_6 \cA_-^3 \cB_+ \cB_m^2 \cA_+^4}{8957952}-\frac{E_4^2 E_6 \cA_- \cA_m^2 \cB_-^2 \cB_+ \cA_+^4}{8957952}\\
		&-\frac{5 E_4^3 \cA_- \cA_m \cB_-^2 \cB_+ \cB_m \cA_+^4}{23887872}-\frac{E_4^2 \cB_-^3 \cB_+ \cB_m^2 \cA_+^4}{35831808}+\frac{E_4^4 \cA_-^3 \cA_m^2 \cB_+^2 \cA_+^3}{3981312}+\frac{E_4^3 \cA_- \cA_m^2 \cB_-^2 \cB_+^2 \cA_+^3}{3981312}\\
		&+\frac{E_4^3 \cA_-^3 \cB_+^2 \cB_m^2 \cA_+^3}{5971968}+\frac{E_4 E_6 \cA_- \cA_m \cB_-^2 \cB_+^2 \cB_m \cA_+^3}{8957952}-\frac{E_4 E_6^2 \cA_-^3 \cA_m^2 \cB_+^2 \cA_+^3}{3359232}-\frac{E_4 E_6 \cA_-^2 \cB_- \cB_+^2 \cB_m^2 \cA_+^3}{5971968}\\
		&-\frac{E_4^2 E_6 \cA_-^2 \cA_m^2 \cB_- \cB_+^2 \cA_+^3}{8957952}-\frac{E_4^3 \cA_-^2 \cA_m \cB_- \cB_+^2 \cB_m \cA_+^3}{11943936}-\frac{E_4^2 \cA_m \cB_-^3 \cB_+^2 \cB_m \cA_+^3}{35831808}-\frac{5 E_4 E_6 \cA_m^2 \cB_-^3 \cB_+^2 \cA_+^3}{53747712}\\
		&+\frac{E_4^2 \cA_m^2 \cB_-^3 \cB_+^3 \cA_+^2}{17915904}+\frac{E_4^2 E_6 \cA_-^3 \cA_m^2 \cB_+^3 \cA_+^2}{8957952}+\frac{E_4 \cB_-^3 \cB_+^3 \cB_m^2 \cA_+^2}{107495424}+\frac{E_4^2 \cA_-^2 \cB_- \cB_+^3 \cB_m^2 \cA_+^2}{11943936}\\
		&+\frac{E_4 E_6 \cA_-^2 \cA_m \cB_- \cB_+^3 \cB_m \cA_+^2}{8957952}-\frac{E_4 E_6 \cA_- \cA_m^2 \cB_-^2 \cB_+^3 \cA_+^2}{5971968}-\frac{E_4^3 \cA_-^3 \cA_m \cB_+^3 \cB_m \cA_+^2}{11943936}\\
		&-\frac{E_4^2 \cA_- \cA_m \cB_-^2 \cB_+^3 \cB_m \cA_+^2}{35831808}-\frac{5 E_4 E_6 \cA_-^3 \cB_+^3 \cB_m^2 \cA_+^2}{53747712}+\frac{E_4^2 \cA_- \cA_m^2 \cB_-^2 \cB_+^4 \cA_+}{35831808}+\frac{E_4 E_6 \cA_-^2 \cA_m^2 \cB_- \cB_+^4 \cA_+}{17915904}\\
		&+\frac{E_4^2 \cA_-^3 \cB_+^4 \cB_m^2 \cA_+}{71663616}+\frac{E_4 E_6 \cA_-^3 \cA_m \cB_+^4 \cB_m \cA_+}{13436928}-\frac{E_4^3 \cA_-^3 \cA_m^2 \cB_+^4 \cA_+}{11943936}-\frac{E_4 \cA_- \cB_-^2 \cB_+^4 \cB_m^2 \cA_+}{71663616}\\
		&-\frac{5 E_4^2 \cA_-^2 \cA_m \cB_- \cB_+^4 \cB_m \cA_+}{71663616}-\frac{E_4 \cA_m \cB_-^3 \cB_+^4 \cB_m \cA_+}{214990848}+\frac{E_4 E_6 \cA_-^3 \cA_m^2 \cB_+^5}{53747712}+\frac{E_4 \cA_- \cA_m \cB_-^2 \cB_+^5 \cB_m}{71663616}\\
		&-\frac{E_4^2 \cA_-^2 \cA_m^2 \cB_- \cB_+^5}{71663616}-\frac{E_4^2 \cA_-^3 \cA_m \cB_+^5 \cB_m}{71663616}-\frac{E_4 \cA_m^2 \cB_-^3 \cB_+^5}{214990848} \ .	
	\end{array}
\end{equation}

\noindent $\mathbf{n_b=(2,1)}$
\begin{equation}
	\begin{array}{rl}
		\num_{6,3,4,(2,1)} &= \frac{\cA_-^4 \cA_+^2 \cA_m \cB_+ E_4^5}{663552}+\frac{\cA_-^4 \cA_+^3 \cB_m E_4^5}{663552}-\frac{\cA_-^3 \cA_+^3 \cA_m \cB_- E_4^5}{331776}+\frac{\cA_- \cA_+^3 \cA_m \cB_-^3 E_4^4}{995328}\\
		&+\frac{13 \cA_-^2 \cA_+^2 \cA_m \cB_-^2 \cB_+ E_4^4}{7962624}+\frac{\cA_-^2 \cA_+^3 \cB_-^2 \cB_m E_4^4}{2654208}-\frac{\cA_-^3 \cA_+ \cA_m \cB_- \cB_+^2 E_4^4}{11943936}-\frac{23 \cA_-^3 \cA_+^2 \cB_- \cB_+ \cB_m E_4^4}{11943936}\\
		&-\frac{\cA_-^4 \cA_m \cB_+^3 E_4^4}{23887872}-\frac{23 \cA_-^4 \cA_+ \cB_+^2 \cB_m E_4^4}{23887872}+\frac{23 \cA_- \cA_+ \cA_m \cB_-^3 \cB_+^2 E_4^3}{35831808}+\frac{5 E_6 \cA_-^4 \cA_+ \cA_m \cB_+^2 E_4^3}{4478976}\\
		&+\frac{23 \cA_+^2 \cA_m \cB_-^4 \cB_+ E_4^3}{71663616}+\frac{\cA_+^3 \cB_-^4 \cB_m E_4^3}{71663616}+\frac{5 E_6 \cA_-^3 \cA_+^3 \cB_- \cB_m E_4^3}{4478976}+\frac{\cA_- \cA_+^2 \cB_-^3 \cB_+ \cB_m E_4^3}{35831808}\\
		&+\frac{E_6 \cA_-^4 \cA_+^2 \cB_+ \cB_m E_4^3}{4478976}-\frac{E_6 \cA_-^2 \cA_+^3 \cA_m \cB_-^2 E_4^3}{746496}-\frac{\cA_-^3 \cB_- \cB_+^3 \cB_m E_4^3}{2985984}-\frac{5 E_6 \cA_-^3 \cA_+^2 \cA_m \cB_- \cB_+ E_4^3}{4478976}\\
		&-\frac{\cA_-^2 \cA_m \cB_-^2 \cB_+^3 E_4^3}{7962624}-\frac{13 \cA_-^2 \cA_+ \cB_-^2 \cB_+^2 \cB_m E_4^3}{23887872}+\frac{5 E_6 \cA_-^3 \cA_m \cB_- \cB_+^3 E_4^2}{8957952}+\frac{E_6^2 \cA_-^3 \cA_+^3 \cA_m \cB_- E_4^2}{279936}\\
		&+\frac{E_6 \cA_- \cA_+^3 \cB_-^3 \cB_m E_4^2}{8957952}+\frac{E_6 \cA_-^4 \cB_+^3 \cB_m E_4^2}{4478976}+\frac{\cA_- \cB_-^3 \cB_+^3 \cB_m E_4^2}{8957952}+\frac{11 E_6 \cA_-^3 \cA_+ \cB_- \cB_+^2 \cB_m E_4^2}{8957952}\\
		&+\frac{E_6 \cA_-^2 \cA_+^2 \cB_-^2 \cB_+ \cB_m E_4^2}{1492992}-\frac{E_6^2 \cA_-^4 \cA_+^2 \cA_m \cB_+ E_4^2}{559872}-\frac{E_6^2 \cA_-^4 \cA_+^3 \cB_m E_4^2}{559872}-\frac{E_6 \cA_-^2 \cA_+ \cA_m \cB_-^2 \cB_+^2 E_4^2}{1492992}\\
		&-\frac{E_6 \cA_+^3 \cA_m \cB_-^4 E_4^2}{4478976}-\frac{17 E_6 \cA_- \cA_+^2 \cA_m \cB_-^3 \cB_+ E_4^2}{8957952}-\frac{\cA_m \cB_-^4 \cB_+^3 E_4^2}{17915904}-\frac{\cA_+ \cB_-^4 \cB_+^2 \cB_m E_4^2}{17915904}\\
		&+\frac{E_6 \cA_- \cA_m \cB_-^3 \cB_+^3 E_4}{8957952}+\frac{17 E_6^2 \cA_-^2 \cA_+^2 \cA_m \cB_-^2 \cB_+ E_4}{23887872}+\frac{17 E_6^2 \cA_-^4 \cA_+ \cB_+^2 \cB_m E_4}{71663616}+\frac{E_6 \cA_+^2 \cB_-^4 \cB_+ \cB_m E_4}{8957952}\\
		&+\frac{17 E_6^2 \cA_-^3 \cA_+^2 \cB_- \cB_+ \cB_m E_4}{35831808}-\frac{E_6 \cA_+ \cA_m \cB_-^4 \cB_+^2 E_4}{8957952}-\frac{E_6 \cA_- \cA_+ \cB_-^3 \cB_+^2 \cB_m E_4}{8957952}-\frac{17 E_6^2 \cA_-^2 \cA_+^3 \cB_-^2 \cB_m E_4}{23887872}\\
		&-\frac{17 E_6^2 \cA_-^3 \cA_+ \cA_m \cB_- \cB_+^2 E_4}{35831808}-\frac{17 E_6^2 \cA_-^4 \cA_m \cB_+^3 E_4}{71663616}+\frac{5 E_6^2 \cA_- \cA_+ \cA_m \cB_-^3 \cB_+^2}{35831808}+\frac{5 E_6^2 \cA_+^2 \cA_m \cB_-^4 \cB_+}{71663616}\\
		&+\frac{5 E_6^3 \cA_-^3 \cA_+^2 \cA_m \cB_- \cB_+}{8957952}+\frac{5 E_6^2 \cA_-^2 \cA_+ \cB_-^2 \cB_+^2 \cB_m}{23887872}+\frac{5 E_6^3 \cA_-^4 \cA_+^2 \cB_+ \cB_m}{8957952}-\frac{5 E_6^3 \cA_-^4 \cA_+ \cA_m \cB_+^2}{8957952}\\
		&-\frac{5 E_6^3 \cA_-^3 \cA_+^3 \cB_- \cB_m}{8957952}-\frac{5 E_6^2 \cA_-^2 \cA_m \cB_-^2 \cB_+^3}{23887872}-\frac{5 E_6^2 \cA_- \cA_+^2 \cB_-^3 \cB_+ \cB_m}{35831808}-\frac{5 E_6^2 \cA_+^3 \cB_-^4 \cB_m}{71663616}
	\end{array}
\end{equation}


The partition functions with $(n_b)= (1,1), (1,2), (2,1)$ are also given in the appendix of \cite{Gadde:2015tra} in terms of Jacobi theta functions, and we have checked that our results coincide with theirs in the small $q = \exp(2\pi\ri \tau)$ expansion. Furthermore, the partition functions with base degrees $(n_b) = (0,n_1,n_2)$ can also be computed with the multi-M-string setup where $M2$ branes are suspended between three parallel $M5$ branes, as discussed in \cite{Haghighat:2013}. We have checked that our results of $(n_b)= (0,1,1), (0,2,1)$ are identical with those in \cite{Haghighat:2013}.

\subsubsection{Factorization of the partition function}

\begin{figure}
	\centering
	\includegraphics[width=0.5\linewidth]{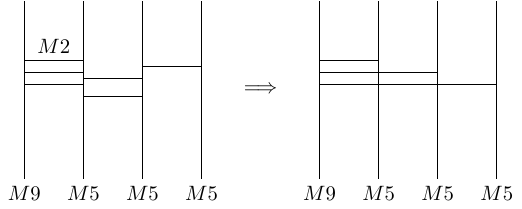}
	\caption{All the $M2$ branes must form E-string-like bound states in the limit $m=\epsilon_+$.}\label{fg:EM-bound}
\end{figure}

It is claimed in \cite{Haghighat:2013} that when the mass $m$ associated to $(-2)$ curves is set to $\epsilon_+$, the partition function for the E-M string chain factorizes
\begin{equation}\label{eq:E-M}
	Z_{(n_1, n_2)}  =
	\begin{cases} 
		Z^{E}_{(n_2)} Z^E_{(n_1- n_2)}  \quad & n_1\geq n_2  \\
		0 \quad  & n_1 < n_2 
	\end{cases} \ ,
\end{equation}
where $Z^E_{(n_b)}$ is the partition function of the E-string with base degree $n_b$. We find that factorization of this type is universal for E-M$^n$ string chains. For instance, we found that with $m= \epsilon_+$
\begin{equation}
	\begin{gathered}
		Z_{(1,0,1)} = Z_{(1,2,1)} = Z_{(1,1,2)} = 0 \ ,\\
		Z_{(1,1,1)} = Z_{(1,1,1,1)} = Z^E_{(1)}\ , \; Z_{(2,1,1)} = Z_{(2,1,1,1)} = (Z^E_{(1)})^2 \ ,\; \ldots
	\end{gathered}
\end{equation}
All these factorization properties can be summarized in the following formula
\begin{equation}
	Z_{(k_1,k_2,\dots,k_n)} =\begin{cases}
		\prod_{i=1}^{n} Z^E_{(k_i-k_{i+1})} & k_1\geq k_2 \geq \ldots \geq k_n\geq k_{n+1} = 0 \ ,\\
		0 & \textrm{otherwise} \ .
	\end{cases}
\end{equation}
It can be explained by a mechanism similar to that behind \eqref{eq:E-M}. When $m=\epsilon_+$, as argued in \cite{Gadde:2015tra}, all $M2$ branes must form bound states like $(\underbrace{k,k,\ldots,k}_j)$, in other words, an $M2$ brane suspended between the $M9$ brane and the $j$-th $M5$ brane, and the bound state behaves like an E-string state; see Fig.~\ref{fg:EM-bound}. Any remaining M-string state that is not part of a bound state leads to the vanishing of the partition function.

\section{BPS invariants and vanishing conditions} \label{s:BPSvanishing}
In this section, we will discuss the BPS invariants that we can extract from our modular expressions for $Z_\beta$. We will supplement the vanishing conditions \eqref{leftbound} and \eqref{rightbound} derived in section \ref{IntegerGeometric} with more stringent conjectured bounds based on these results, and get additional hints with regard to the geometric origin of the mass parameter in geometries involving $(-2)$ curves.

\subsection{The E-string}
The topological string on the compact elliptic fibration over the Hirzebruch surface $\IF_1$ can be solved to low genus in the B-model using mirror symmetry. This allows the computation of the BPS invariants of local $\halfKthree$ which are accessible in the compact model, corresponding to curve classes $C= b C_1 + e E$. In this manner, \cite{Klemm:1996} obtained the result
\be 
\sum_{n_e=0}^\infty  I^{1,n_e}_0 q^{n_e}=\frac{\sqrt{q} E_4}{\eta^{12}}=1 + 252 q + 5130 q^2 + 54760 q^3 +\ldots \ ,      
\ee
The base degree $b=0$ invariants are given by \cite{HKK2}
\be 
I_0^{0,n_e}= 480, \quad  I_1^{0,n_e}= 4, \quad \forall \ n_e\in \mathbb{N}_0, \quad {\rm while}  \quad I_g^{0,n_e}=0  \quad  \forall g>1, \ \ \  n_e\in \mathbb{N}_0 .  
\label{deg0inbase}
\ee
In fact, these numbers are also related to modular generating functions, see~\cite{HKK2} for 
details. 

As discussed in section \ref{sc:mass}, eight additional homology classes arise in the decompactification limit of the geometry, given by the total space of the canonical bundle over $\halfKthree$. A general compact curve class in this geometry is of the form
\be
C= b C_1 + e E + \sum_{i=1}^8 \mu^i \alpha_i\ .
\ee
With the intersection numbers discussed in section \ref{sc:mass}, we obtain
\be 
C^2=-b^2+ 2 be - (\bmu, \bmu)_{\mf e_8}, \qquad  K\cdot C=-b 
\ee
and can evaluate the Castelnuovo bound \eqref{gmax} for this geometry. For the case $e<b$, our computations suggest a stricter bound than the one that follows from \eqref{gmax}. Overall, the vanishing conditions for the unrefined invariants $I_g^{b,e}=0$ that we obtain are
\be 
I_g^{b,e,\bmu}=0 \quad  \left\{\begin{array}{rl}  & {\rm for }\  g>0 \ {\rm or}\ \md \bmu >0\ {\rm if}\quad   e=0 \ {\rm and}\   b=1\\
	&\forall\  g \ {\rm if}\quad   e<b \ {\rm and}\   b>1\\
	&{\rm for } \ g>g^{max}(b,e,\bmu)=\frac{1}{2}(2 e b-b(b+1)- (\bmu, \bmu)_{\mf e_8})+1  \quad {\rm if}\  e\ge b  \ .
\end{array} \right.
\label{boundsmasslessestring}  
\ee 
Note that due to the positivity of the Cartan matrix of $E_8$, the bound is weakest at ${\bmu = 0}$. We further conclude that only a finite number of $E_8$ Weyl orbits of the mass parameters, as specified by the Weyl invariant inner product $M= (\bmu, \bmu)_{\mf e_8} \in 2 \IN$ for $\mu \in \Lambda_{\mathfrak{e}_8} \otimes \IZ$, can contribute to BPS invariants at fixed $(b,e)$:
\be 
M \le M^{max}(e,b)= 2 e b-b(b+1)+2 \,.
\ee

For the massless case, we can revert to (\ref{leadingunrefined}) to predict the leading unrefined BPS invariants at $g^{max}(b,e)$ to be
\be 
I^{b,e}_{g^{max}(b,e)}= (-1)^{g_{max}(b,e)+b} (g_{max}(b,e)+b) \,.
\ee
This is confirmed by our explicit determination of $Z_\beta$ in section \ref{Estring}. We list some of the invariants $I_g^{b,e}$ at base degree $b=1,2,3$ in table \ref{estringb123}. Some invariants at base degrees $b=4,5$ can be found in appendix \ref{Modularandenumerative}.  
\begin{table}[h!]
	\centering
	{\tiny
		\begin{tabular}{|r|ccccc|}\hline
			$I^{1,e}_g$& $e=$0&1&2&3&4\\ \hline
			$g=$0&1&252&5130&54760&419895\\ 
			1 &&-2&-510&-11780&-142330\\ 
			2 &&&3&772&19467\\ 
			3 &&&&-4&-1038\\ 
			4 &&&&&5 \\
			\hline \end{tabular}
		\begin{tabular}{|r|ccc|}\hline
			$I^{2,e}_g$& $e=$2&3&4\\ \hline
			$g=$0&-9252&-673760&-20534040\\ 
			1 &760&205320&11361360\\ 
			2 &-4&-25604&-3075138\\ 
			3 &&1296&494144\\ 
			4 &&-6&-45172\\ 
			5 &&&1844\\ 
			6 &&&-8\\  
			\hline \end{tabular}
		\begin{tabular}{|r|cc|}\hline
			$I^{3,e}_g$& $e=$3&4\\ \hline
			$g=$0&848628&115243155\\ 
			1 &-246790&-76854240\\ 
			2 &30464&26356767\\ 
			3 &-1548&-5707354\\ 
			4 &7&790293\\ 
			5 &&-64252\\ 
			6 &&2388\\ 
			7 &&-10\\ 
			\hline \end{tabular}
	}
\caption{Unrefined BPS invariants for the massless E-string at base degrees $b=1,2,3$.}\label{estringb123}
\end{table}

We can similarly evaluate the Castelnuovo bounds \eqref{leftbound} and \eqref{rightbound} for the refined invariants $N_{j_- j_+}^{b,e}$ to obtain
\be 
N_{j_-j_+}^{b,e,\bmu}=0 \quad  \left\{\begin{array}{rl} 
	&{\rm for }\  j_-> 0\ {\rm or }\ j_+>0 \ {\rm or }\ \bmu>0 \ {\rm if} \quad   e=0 \ {\rm and}\   b=1\\                                                             
	&\forall\  j_-,j_+ \ {\rm if} \quad   e<b \ {\rm and}\   b>1\\
	&{\rm for } \ \left\{\begin{array}{rl} 2 j_- &> 2 j^{max}_-(b,e)= \frac{1}{2}(-b^2+2b e - (\bmu, \bmu)_{\mf e_8}-b)+1 \\ & 
		{\rm or}  \\   
		2 j_+&> 2 j^{max}_+(b,e)=\frac{1}{2}(-b^2+2 be - (\bmu, \bmu)_{\mf e_8}+b)
	\end{array} \right. \quad {\rm if}\  e\ge b
	
\end{array} \right.
\label{Ecastelnouvorefined} 
\ee
We have again supplemented the bound with a stricter observational bound for $e<b$.

From the Lefschetz decomposition of the moduli space in the massless case, we conclude 
\be 
N^{(b,e)}_{j_-^{max}(b,e) j_+^{max}(b,e)}=1 \,,
\ee
as well as the checkerboard pattern for vanishing and non-vanishing BPS invariants inside the bound 
$(j_-,j_+)\le $ $(j_-^{max}(b,e),$ $ j_+^{max}(b,e))$,      
\be
N^{(b,e)}_{j_-j_+}=0 \ \ {\rm if }\ \  2( j_-^{max}(b,e) +j_+^{max}(b,e)- j_- -j_+) \neq 0\ {\rm mod} \ 2 . 
\label{checkerboard}
\ee

Note that the Weyl symmetry that acts on the curves in $\halfKthree$ guaranties that the BPS states are organized 
in terms of representations of the Weyl group. Setting $\md m=0$ yields BPS invariants that are sums over the 
contributions of different Weyl orbits. 
This is visible for the refined BPS invariants in table \ref{refinedEstringb12}, as all orbits contribute with 
positive sign. Moreover, in the refined case, we see that the contributions of degenerate curves at higher arithmetic 
genus complete these Weyl orbit contributions to full representations of $E_8$. For instance, 
the  $N^{b=1,e=1}_{0,0}=248$  in  table \ref{refinedEstringb12} for the refined  E-String comes from 
$240$ rational curves in the Weyl orbit of $\frak{e}_8$ of roots of length $2$, supplemented by 
the contribution of 8 degenerate curves of arithmetic genus one, the highest spin 
representation of which in the Lefschetz decomposition of moduli space gives the contribution  
$N^{b=1,e=1}_{\frac{1}{2},\frac{1}{2}}=1$. Summing over the right 
spin quantum number $j_+$  according to (\ref{sumoverrightspin}), we get from this state 
$(-2)\left[\frac{1}{2}\right]_-=(-2) (T_1-2 T_0)$. Adding to this the contribution $248[0]_-=248 T_0$ at spin 0 yields the entries  
$I^{b=1,e=1}_0=252$ and $I^{b=1,e=1}_1=-2$  in table  \ref{estringb123}. This illustrates how representation 
theoretic aspects get obscured by the signs in the unrefined case. 

\begin{table}[h!]
	\centering
	{\footnotesize
		\begin{tabular}{|r|c|}\hline
			$N_{j_-j_+}^{(1,0)}$ &$2j_+=$0\\ \hline
			$2j_-=$0&1\\ 
			\hline \end{tabular}
		\begin{tabular}{|r|cc|}\hline
			$N_{j_-j_+}^{(1,1)}$ &$2j_+=$0&1\\ \hline
			$2j_-=$0&248&\\ 
			1 &&1\\ 
			\hline \end{tabular}
		\begin{tabular}{|r|ccc|}\hline
			$N_{j_-j_+}^{(1,2)}$ &$2j_+=$0&1&2\\ \hline
			$2j_-=$0&4125&&\\ 
			1 &&249&\\ 
			2 &&&1\\  
			\hline \end{tabular}
	}
%
%
%
	\centering
	{\footnotesize
		\begin{tabular}{|r|cccc|}\hline
			$N_{j_-j_+}^{(1,3)}$ &$2j_+=$0&1&2&3\\ \hline
			$2j_-=$0&35001&&1&\\ 
			1 &&4374&&\\
			2 &1&&249&\\
			3 &&&&1\\
			\hline \end{tabular}
		\begin{tabular}{|r|ccccc|}\hline
			$N_{j_-j_+}^{(1,4)}$ &$2j_+=$0&1&2&3&4\\ \hline
			$2j_-=$0&217501&&249&&\\ 
			1 & &39375&&1&\\ 
			2 &249&&4375&&\\ 
			3 &&1&&249&\\
			4 &&&&&1\\
			\hline \end{tabular}
		\begin{tabular}{|r|ccc|}\hline
			$N_{j_-j_+}^{(2,2)}$ &$2j_+=$0&1&2\\ \hline
			$2j_-=$0&3876&&\\ 
			1 &&248&\\ 
			2 &&&1\\  
			\hline \end{tabular}
		\begin{tabular}{|r|cccccc|}\hline
			$N_{j_-j_+}^{(2,3)}$ &$2j_+=$0&1&2&3&4&5\\ \hline
			$2j_-=$0&&186126&&249&&\\ 
			1 &4124 &&38877&&1&\\ 
			2 &&249&&4373&&\\ 
			3 &&&1&&249&\\
			4 &&&&&&1\\
			\hline \end{tabular}			
	}
	\caption{Refined BPS invariants of the massless E-string at base degrees $b = 1,2$ and $e \le 3$.}\label{refinedEstringb12}
\end{table}
We can use the classic projective description of the moduli space of smooth very high genus and high $\kappa$ curves also in the refined case, and obtain asymptotic formulae for the multiplicities of states. Let us introduce a degeneration parameter  $d=d_++d_-$ such as at the end of section  \ref{IntegerGeometric}. In a 
suitable range  $d\le d(\kappa)$, we can use the projective model of the moduli space  $P_n(M,\kappa)$ given by a 
$\mathbb{P}^{(C_\kappa^2- K \cdot C_\kappa)/2}$ bundle  over $S^{[d]}$, together with its $SU(2)_-\times SU(2)_+$ 
Lefschetz decomposition and the multiplicities of Weyl orbits to predict an infinite number of non-vanishing BPS 
numbers at arbitrary high base degree $b$: we start with any maximal left and maximal right spin (\ref{leftbound},\ref{rightbound}) 
and predict, using the methods of \cite{CKK}, the  lower spin BPS degeneracies  $N_{j^{max}_--\frac{d_-}{2},j^{max}_+-\frac{d_+}{2}}^{\kappa}$ bounded by $d_++d_-\le d(\kappa)=2 n+1$. 
Note that at the boundary for even $d$ the checkerboard pattern already predicts vanishing numbers. Applying this procedure to all $\kappa$ bounded by  $b<5$ and $e<13$, we find that the BPS numbers stabilize for 
\be 
e-b>n \ {\rm  if}\  b>1\  {\rm or} \  e-b>2n \ {\rm if}  \ \ b=1 \ .
\ee 
This suggests that in this range, no corrections of the type discussed at the end of section \ref{IntegerGeometric} occur and the numbers are indeed given by the projective model. 
Examples for such asymptotic BPS numbers are given in table  \ref{asymptoticE8} for $d\le 10$. 
\begin{table}[h!] 
	\centering
	{\footnotesize
		\begin{tabular}{|r|cccccccccc|}\hline
			$N_{j^{max}_--\frac{d_-}{2},j^{max}_+-\frac{d_+}{2}}^{(b>1,e>n+b)}$ &$d_+=$9&8&7&6&5&4&3&2&1&0\\ \hline
			$d_-=$9&-&&-&&-&&-&&0&\\
			8&&-&&-&&-&&0&&0\\
			7&-&&-&&-&&250&&0&\\
			6&&-&&-&&44250&&1&&0\\
			5&-&&-&&1398253&&4625&&0&\\
			4&&-&&44250&&261501&&250&&0\\
			3&-&&250&&4625&&39625&&1&\\ 
			2 &&0&&1&&250&&4375&&0\\ 
			1&0&&0&&0&&1&&249&\\
			0 &&0&&0&&0&&0&&1\\
			\hline \end{tabular}			
	}
	\caption{Asymptotics of refined  BPS invariants of the massless E-string for $d=2n+1$ with $n=4$. We indicate only those vanishings by $0$ that are not already implied by the 
	checkerboard pattern. }\label{asymptoticE8}
\end{table}
The fact that there is such a universal behavior of the BPS numbers for all $b$, which we have not used as boundary conditions for our computations at fixed $b$, suggests that there should be a more universal description of 
$\Ztop$ valid for all $b$. 

We have also solved the massive E-string up to base degree $b\le 3$. The results at the level of BPS numbers are 
available upon request. Some have already been discussed in~\cite{Huang:2013} based on computations using the refined holomorphic anomaly 
equations and imposing boundary conditions. In this work, we have provided the systematic underpinning of such calculations, and demonstrated the complete integrability of the massive E-string using our modular ansatz and the boundary conditions. Given a
solution of the massive E-string, \cite{Huang:2013} explains how to take limits to get solutions for the refined 
topological string on del Pezzo surfaces $B_k\mathbb{P}^2$ given by $\mathbb{P}^2$ blown up at $k$ points. For degree 
$9-k$ smaller than six, these are generically not toric and therefore not solvable by the topological vertex. Hence, our 
methods also provide complete solutions for these previously inaccessible cases.    

\subsection{The M-string}

The geometry of the M-string is less well understood  in the literature.  In the original 
paper~\cite{Haghighat:2013}, a `non-planar toric diagram' is considered with identifications 
of two  outer legs which indicate the sums with which the legs of  refined topological 
vertices have to be  connected to obtain the partition function. It is however not at all obvious 
how to compute the homology, the intersection numbers or the K\"ahler cone of the resulting local Calabi-Yau space $\check M$. Another description, most explicitly proposed in \cite{DelZotto:2014hpa,HKLV:2014}, is in terms of an elliptic fibration over a non-compact base containing 
a $(-2)$ curve. As explained in section \ref{EMstringgeometries}, this implies that $f_4$ and $g_6$ 
restrict to constants at the $(-2)$ curve and leads generically to $\check M$ as a non-compact threefold over a 
local surface $S=T^2\times \mathbb{P}^1$. This geometry appears too simple to allow for a 
mass deformation. One possibility to arrive at such a deformation is to tune  
complex structure moduli and  choose these constants so that $\Delta$ vanishes at the  
$(-2)$ curve, yielding an $I_1$ singularity over the entire curve, i.e. giving rise to a constant fibration with a nodal torus. 
This is reminiscent of the geometry described in \cite{DelZotto:2014hpa,HKLV:2014}, except for two additional
non-compact curves with $I_1$ singularities intersecting the compact $\mathbb{P}^1$ in those references. 
This  would allow the introduction of an additional  K\"ahler parameter as mass parameter by 
resolving the node over the compact $\mathbb{P}^1$.    

Below, we will analyze the BPS numbers of the M-string and their vanishing. We find BPS degeneracies  and vanishing 
conditions which roughly follow the pattern expected from a geometric description and are 
sufficient to solve the model completely. There are several interesting features 
of the BPS spectrum which might help clarify the mass deformation of the geometry, which we will elaborate on in the following.

The enhanced supersymmetry of the massless M-string is visible in the trivial fibration structure 
of the geometry described above (as the BPS numbers are independent of the complex structure, we can consider the generic geometry $T^2 \times S$ in the massless case). Consequently, the unrefined BPS invariants $I^{(b,e)}_g$ of the massless M-string vanish 
for $b\in \mathbb{N}$, $e\in \mathbb{N}_{0}$.\footnote{This 
can be seen in the BPS invariants of the  degree 24 hypersurface in the resolution of 
$\mathbb{P}^4(1,1,2,8,12)$ analyzed in~\cite{Hosono:1993qy}, which contains the geometry 
$S$ as a local surface. The ones at base degree zero are $I^{0,e}_1 = 1$, $I^{0,e}_g = 0$, $ \forall g\neq 1$, $e \in \mathbb{N} $~\cite{Huang:2013}.} 
For the refined invariants, this implies
\be 
\sum_{j_+\in \frac{1}{2} \mathbb{Z}}  (-1)^{2 j_+}(2 j_++1) N^{(b,e)}_{j_- j_+}=0 \ .
\ee
We also find that the sums over left spins vanish,
\be 
\sum_{j_-\in \frac{1}{2} \mathbb{Z}}  (-1)^{2 j_-}(2 j_-+1) N^{(b,e)}_{j_- j_+}=0 \ .
\ee
For the refined invariants $N_{j_- j_+}^{b,e}$, we find the vanishing conditions 
\be 
N_{j_-j_+}^{b,e}=0 \quad  \left\{\begin{array}{rl} & {\rm for}\  b>1 \ {\rm and} \    e=0 \ {\rm or} \ j_+< 5 - 4 b + b^2  \\  
	&{\rm for } \ \left\{\begin{array}{rl} 2 j_- &> 2 j_-^{max}(b,e)= b (e-1)+1  \\ & 
		{\rm or}  \\   
		2 j_+&> 2 j_+^{max}(b,e)=b(e+1)
	\end{array} \right. \ 
\end{array} \right. .
\label{Mcastelnouvorefined} 
\ee
We note by  comparing with \eqref{leftbound} and \eqref{rightbound} that  this would correspond to intersection numbers $C^2=0$, $C\cdot E=1$, $K\cdot C=-2$ and $K\cdot E=0$. Note 
in particular that this is not compatible with $C^2=-2$, our naive expectation. A possible explanation is a change in the normal bundle of the base curve $C$ required to allow for mass deformation, as discussed above. We list the BPS numbers for $b = 1,2$ and $e \le 3$ in the table \ref{masslessMstring}. 

Just as in the E-string case exemplified in table \ref{asymptoticE8}, the BPS numbers exhibit an asymptotic stability pattern for large $\kappa$ and  $j_\pm$, suggesting a simple projective description of the moduli space $P_n(\check M,\kappa)$. However, they do not follow a checkerboard pattern. 
This can be explained in part by the fact that in the mass deformed case, even and odd mass degrees exhibit a checkerboard pattern 
filling the ``white'' and ``black'' fields respectively. In the massless limit, these contributions are added.

\begin{table}[h!]
	\centering
	{\footnotesize
		\begin{tabular}{|r|cc|}\hline
			$N_{j_-j_+}^{(1,0)}$ &$2j_+=$0&1\\ \hline
			$2j_-=$0&2&1\\ 
			\hline \end{tabular}		
		\begin{tabular}{|r|ccc|}\hline
			$N_{j_-j_+}^{(1,1)}$ &$2j_+=$0&1&2\\ \hline
			$2j_-=$0&2&4&2\\ 
			1 &1&2&1\\ 
			\hline \end{tabular}
		\begin{tabular}{|r|cccc|}\hline
			$N_{j_-j_+}^{(1,2)}$ &$2j_+=$0&1&2&3\\ \hline
			$2j_-=$0&10&9&4&1\\ 
			1 &5&6&5&2\\ 
			2 &&1&2&1\\ 
			\hline \end{tabular}
		\begin{tabular}{|r|ccccc|}\hline
			$N_{j_-j_+}^{(1,3)}$ &$2j_+=$0&1&2&3&4\\ \hline
			$2j_-=$0&16&22&16&5&\\ 
			1 &14&20&15&6&1\\ 
			2 &4&6&6&5&2\\ 
			3 &&&1&2&1\\ 
			\hline \end{tabular}
		%
		%
		%
		\begin{tabular}{|r|ccccc|}\hline
			$N_{j_-j_+}^{(2,1)}$ &$2j_+=$0&1&2&3&4\\ \hline
			$2j_-=$0&&&2&4&2\\ 
			1 &&&1&2&1\\ 
			\hline \end{tabular}
		\begin{tabular}{|r|ccccccc|}\hline
			$N_{j_-j_+}^{(2,2)}$ &$2j_+=$0&1&2&3&4&5&6\\ \hline
			$2j_-=$0&2&9&16&18&14&5&\\ 
			1 &1&6&14&18&14&6&1\\ 
			2 &&1&4&6&6&5&2\\ 
			3 &&&&&1&2&1\\ 
			\hline \end{tabular}
		\begin{tabular}{|r|ccccccccc|}\hline
			$N_{j_-j_+}^{(2,3)}$ &$2j_+=$0&1&2&3&4&5&6&7&8\\ \hline
			$2j_-=$0&20&48&86&102&74&32&8&1&\\ 
			1 &20&52&90&108&90&52&20&4&\\ 
			2 &8&24&42&55&58&46&24&6&\\ 
			3 &1&4&8&14&22&24&16&6&1\\ 
			4 &&&&1&4&6&6&5&2\\ 
			5 &&&&&&&1&2&1\\ 
			\hline \end{tabular}
	}
	\caption{Refined BPS invariants of the massless M-string at base degrees $b = 1,2$ and $e \le 3$. }\label{masslessMstring}
\end{table}

From our ansatz (\ref{ansatz_z_nb}) for $Z_\beta$, in which the mass parameter $m$ enters via (\ref{JacobiringI}), and the Fourier expansion 
of $\cA_m$ (\ref{Aproduct}) and $\cB_m$ (\ref{ellipticgenusK3}), it follows that $Z_\beta$ is symmetric under $m \rightarrow -m$, i.e. coefficients 
of $Q_m^\mu$ and $Q_m^{-\mu}$, for $Q_m = \exp(m)$, coincide. This $\IZ_2$ symmetry is the analogue of the $E_8$ Weyl symmetry in the E-string case, and 
implies for the BPS numbers that  
\be 
I^{b,e,\mu}_{g}=I^{b,e,-\mu}_g \,, \quad N^{b,e,\mu}_{j_- j_+}=N^{b,e,-\mu}_{j_- j_+}\ .
\ee 
For this reason, we will only list BPS invariants for  $\mu\ge 0 $.   

The negative powers of $Q_m$ are at odds with the identification of $(t_b,t_e=\tau,t_m=m)$ as K\"ahler parameters in a large radius K\"ahler cone expansion of $\Ztop$. By the structure of the coefficients of weak Jacobi forms, we can however shift the  modular parameter $\tau$ as  
\be 
t_b=t_b,\quad \tilde \tau=\tau-t_m, \quad  t_m=m\  ,
\ee
thus absorbing all negative powers of $Q_m$, yielding a geometric large radius expansion interpretation. In the following, it will however be more convenient to continue working in unshifted classes.

The Castelnuovo bounds that we extract from our computations for the massive M-String depend on 
whether the degree $\mu$  in the mass parameter is even or odd. 
For even degree, we obtain
\be 
N_{j_-j_+}^{b,e, 2 \nu}=0 \quad  {\rm if } \quad  \left\{\begin{array}{rl} 2 j_- &> 2 j_-^{max}(b,e,\nu)=\frac{1}{2} (2 be - 2 \nu^2 - 2 b)+1  \\ & 
	{\rm or}  \\   
	2 j_+&> 2 j_+^{max}(b,e,\nu)= be - \nu^2+ b 
\end{array} \right. ,
\label{Mcastelnouvorefinedeven} 
\ee
whereas for odd degree,   
\be 
N_{j_-j_+}^{b,e, 2 \nu+1}=0 \ { \rm if} \ 
	 \left\{\begin{array}{rl} 2 j_- &> 2 j_-^{max}(b,e,\nu)=\frac{1}{2} (2 be - 2 \nu^2 - 2 b- 2\nu)+1  \\ & 
		{\rm or}  \\   
		2 j_+&> 2 j_+^{max}(b,e,\nu)=\left\{\begin{array}{ll} 0  \ \ {\rm if } \ \ b=1 \ {\rm and} \ e \le \nu(\nu+1)   \\  
		                                     \frac{1}{2} (2 be -2 \nu^2+ 2b-2\nu)  \,\,{\rm else.} \end{array} \right. 
	\end{array} \right.   
\label{Mcastelnouvorefinedodd} 
\ee
Note that both $2 j_\mp^{max}(b,e,\nu)$ must be non-negative in order for $N_{j_-j_+}^{b,e,\mu}$ not to vanish. 
Just as for the E-string, these vanishing conditions allow us to formulate a bound $\mu^{max}(e,b)$ at given $(b,e)$ on $\mu$ 
above which all invariants vanish. 

The massless M-string arises for $t_m=0$, i.e. $Q_m=1$. We can thus obtain the massless BPS invariants by summing  
$I_g^{b,e}=\sum_{\mu=-\mu_{max}(b,e)}^{\mu_{max}(b,e)} I_g^{b,e,\mu}$ or 
$N_{j_-,j_+}^{b,e}=\sum_{\mu=-\mu_{max}(b,e)}^{\mu_{max}(b,e)} N_{j_- j_+}^{b,e,\mu}$, just as 
for the E-string. Since  $\mu$ even and $\mu$ odd have an opposite checkerboard pattern,
\be
N^{(b,e)}_{j_-j_+}=0 \ \ {\rm if }\ \  2(  j_- +j_+) +\mu+1\neq 0\ {\rm mod} \ 2 , 
\label{checkerboardM}
\ee
the checkerboard pattern disappears in the massless case. However, it should  
still be true that black vs.\ white fields stem from the decomposition of different PT moduli spaces according to the parity 
of $\mu$. A hint for this can be seen in the asymptotic BPS numbers for high enough $j_\pm$, $(b,e)$ and small $\mu$, 
which  stabilize for even and odd $\mu$ to two classes of  patterns, shown in table  \ref{asymptoticMeven} 
and in table  \ref{asymptoticModd} for $d\le 10$  respectively. 

\begin{table}[h!] 
	\centering
	{\footnotesize
		\begin{tabular}{|r|cccccccccc|}\hline
			$N_{j^{max}_--\frac{d_-}{2},j^{max}_+-\frac{d_+}{2}}^{(b>1,e \gg b,\mu=0)}$ &$d_+=$9&8&7&6&5&4&3&2&1&0\\ \hline
			$d_-=$9&-&&-&&-&&-&&0&\\
			8&&-&&-&&-&&1&&0\\
			7&-&&-&&-&&21&&0&\\
			6&&-&&-&&107&&6&&0\\
			5&-&&-&&180&&40&&1&\\
			4&&-&&107&&74&&14&&0\\
			3&-&&21&&40&&28&&4&\\ 
			2 &&1&&6&&14&&10&&1\\ 
			1&0&&0&&1&&4&&3&\\
			0 &&0&&0&&0&&1&&1\\
			\hline \end{tabular}			
	}
	\caption{Asymptotics of refined  BPS invariants of the massive  M-string for $\mu$ even.}\label{asymptoticMeven}
\end{table}

\begin{table}[h!] 
	\centering
	{\footnotesize
		\begin{tabular}{|r|cccccccccc|}\hline
			$N_{j^{max}_--\frac{d_-}{2},j^{max}_+-\frac{d_+}{2}}^{(b>1,e \gg b,\mu=\pm 1)}$ &$d_+=$9&8&7&6&5&4&3&2&1&0\\ \hline
			$d_-=$9&&-&&-&&-&&0&&0\\
			8&-&&-&&-&&8&&0&\\
			7&&-&&-&&70&&2&&0\\
			6&-&&-&&196&&25&&0&\\
			5&&-&&196&&81&&7&&0\\
			4&-&&70&&81&&30&&2&\\
			3&&8&&25&&30&&11&&0\\ 
			2 &0&&2&&7&&11&&3&\\ 
			1&&0&&0&&2&&3&&1\\
			0 &0&&0&&0&&0&&1&\\
			\hline \end{tabular}			
	}
	\caption{Asymptotics of refined  BPS invariants of the massive  M-string for $\mu$ odd.}\label{asymptoticModd}
\end{table}

We have listed all non-vanishing refined BPS invariants for $b=1$ and $e\le 3$ in table \ref{mb1e3}. 
The BPS invariants for $b=2,3$ and $e\le 3$ are given in appendix \ref{appendixmassivemstring}. 
\begin{table}[h!]
	\centering
	{\footnotesize
		\begin{tabular}{|r|c|}\hline
			$N_{j_-j_+}^{(1,0,0)}$& 
			$2j_+=$1\\ \hline
			$2j_-=$0&1\\ \hline 
		\end{tabular}		
		\begin{tabular}{|r|c|}\hline
			$N_{j_-j_+}^{(1,0,1)}$& $2j_+=$0\\ \hline
			$2j_-=$0&1\\ \hline 
		\end{tabular}
		\begin{tabular}{|r|ccc|}\hline
			$N_{j_-j_+}^{(1,1,0)}$& 
			$2j_+=$0&1&2\\ \hline
			$2j_-=$0&&2&\\ 
			1 &1&&1\\ \hline 
		\end{tabular}		
		\begin{tabular}{|r|ccc|}\hline
			$N_{j_-j_+}^{(1,1,1)}$& 
			$2j_+=$0&1&2\\ \hline
			$2j_-=$0&1&&1\\ 
			1 &&1&\\ \hline 
		\end{tabular}
		\begin{tabular}{|r|c|}\hline
			$N_{j_-j_+}^{(1,1,2)}$& $2j_+=$1\\ \hline
			$2j_-=$0&1\\ \hline 
		\end{tabular}
		\begin{tabular}{|r|cccc|}\hline
			$N_{j_-j_+}^{(1,2,0)}$& $2j_+=$0&1&2&3\\ \hline
			$2j_-=$0&&5&&1\\ 
			1 &3&&3&\\ 
			2 &&1&&1\\ 
			\hline 
		\end{tabular}
		\begin{tabular}{|r|cccc|}\hline
			$N_{j_-j_+}^{(1,2,1)}$& 
			$2j_+=$0&1&2&3\\ \hline
			$2j_-=$0&4&&2&\\ 
			1 &&3&&1\\ 
			2 &&&1&\\ 
			\hline 
		\end{tabular}
		\begin{tabular}{|r|ccc|}\hline
			$N_{j_-j_+}^{(1,2,2)}$& $2j_+=$0&1&2\\ \hline
			$2j_-=$0&&2&\\ 
			1 &1&&1\\ 
			\hline 
		\end{tabular}		
		\begin{tabular}{|r|c|}\hline
			$N_{j_-j_+}^{(1,2,3)}$& $2j_+=$0\\  \hline
			$2j_-=$0&1\\ 
			\hline 
		\end{tabular}
		%
		%
		%
		\begin{tabular}{|r|ccccc|}\hline
			$N_{j_-j_+}^{(1,3,0)}$& $2j_+=$0&1&2&3&4\\ \hline
			$2j_-=$0&&12&&3&\\ 
			1 &8&&9&&1\\ 
			2 &&4&&3&\\ 
			3 &&&1&&1\\ 
			\hline 
		\end{tabular}
		\begin{tabular}{|r|ccccc|}\hline
			$N_{j_-j_+}^{(1,3,1)}$& $2j_+=$0&1&2&3&4\\ \hline
			$2j_-=$0&7&&7&&\\ 
			1 &&9&&3&\\ 
			2 &2&&3&&1\\ 
			3 &&&&1&\\ 
			\hline 
		\end{tabular}			
		\begin{tabular}{|r|cccc|}\hline
			$N_{j_-j_+}^{(1,3,2)}$& $2j_+=$0&1&2&3\\ \hline
			$2j_-=$0&&5&&1\\ 
			1 &3&&3&\\ 
			2 &&1&&1\\ 
			\hline
		\end{tabular}			
		\begin{tabular}{|r|ccc|}\hline
			$N_{j_-j_+}^{(1,3,3)}$& $2j_+=$0&1&2\\ \hline
			$2j_-=$0&1&&1\\ 
			1 &&1&\\ 
			\hline 
		\end{tabular}
	} 
	\caption{Refined BPS invariants of the massive M-string at base degree $b=1$ and $e \leq 3$.} \label{mb1e3}
	\end {table}

We can deduce from (\ref{Mcastelnouvorefinedeven},\ref{Mcastelnouvorefinedodd}) for every class $(b,e)$
a bound on $\mu$ which grows\footnote{In \cite{Haghighat:2013}, 
a bound $\mu_{max}(e,b)=e+b$ is proposed. This bound grows linearly in $e$ at fixed $b$, while the actual growth scales 
	only with $\sqrt{e}$. It is saturated for $(b,e)=\{(1,0),(1,1),(1,2)\}$, while our 
	bound is saturated for an infinite set of $(b,e)$.} to leading order in $e$ as
	\be 
\mu_{max}(b,e)=\sqrt{ 4 b e + c_1 b+c_0 }\ .
\label{mbound}
\ee
The precise values of the constants $c_0$ and $c_1$ appear to depend systematically on which spin 
representations occur at $\mu_{max}(b,e)$, i.e. for which $(j_-,j_+)$ the invariants $N^{b,e,\mu_{max}(b,e)}_{j_-,j_+}$ are non-vanishing

Most notably, when 
\be 
\mu_{\rm max}(b,e)=2\sqrt{b e -(b-1)} \in \mathbb{N} \ ,
\ee
we always find $N^{b,e,\mu_{\rm max}}_{0,\frac{2 b-1}{2}}=1$. 

In general, there are other spin representations at $\mu_{max}(e,b)$ for different choices of $c_1$ and $c_0$.
In particular, for $b=1$ we have $N^{b,e,\mu_{max}(b,e)}_{0,0}=1$ when $\mu_{max}(b,e)=\sqrt{4e + 1} \in \mathbb{N}$.
For $b=2$, we find spin representations that occur subleading for smaller values of $c_0$:
\be \label{roguetable}
{ \begin{tabular}{|r|ccccc|}\hline
		$N_{j_-j_+}^{(2,7,7)}$& $2j_+=$0&1&2&3&4\\  \hline
		$2j_-=$0&&&1&&1\\ 
		1 &&&&1&\\ 
		\hline \end{tabular} \atop  {\rm for} \ \mu_{max}(2,e)=\sqrt{8 e - 7}}  \ {\rm and} \
{\begin{tabular}{|r|ccccc|}\hline
		$N_{j_-j_+}^{(2,3,4)}$& $2j_+=$0&1&2&3&4\\ \hline
		$2j_-=$0&&&&2&\\ 
		1 &&&1&&1\\ \hline \end{tabular} \atop  {\rm for} \ \mu_{max}(2,e)=\sqrt{8 e - 8}}
\ , 
\ee
where we label the tables by those $(b,e,\mu)$  where these spin 
representations occur first. The bounds on $\mu$ can also be seen from the unrefined 
invariants listed in table \ref{massiveMinvariantsb1}, where $I^{(b,e)}_{g}=0$ in the massless limit can serve as a check.

\begin{table}[h!]
	\centering
	{\footnotesize
		\begin{tabular}{|r|cccccc|}\hline
			$I^{(1,e,\mu)}_{g=0}$& $\mu=$0&1&2&3&4&5\\ \hline
			$e=$0& -2 & 1 &&&&\\ 
			1 & -12 & 8 & -2 &&&\\ 
			2 & -56 & 39 & -12 & 1 &&\\ 
			3 & -208 & 152 & -56 & 8 &&\\ 
			4 & -684 & 513 & -208 & 39 & -2 &\\ 
			5 & -2032 & 1560 & -684 & 152 & -12 &\\ 
			6 & -5616 & 4382 & -2032 & 513 & -56 & 1\\ 
			\hline \end{tabular}} 
	\caption{Unrefined BPS invariants of the massive M-string at base degree $b=1$ and genus $g=0$. }\label{massiveMinvariantsb1}
\end{table}

\subsection{The E-M$^n$ string chain}
	
In this section, we report on Castelnuovo bounds for the E-M$^n$ string chain that we extract from our computations. According to the general theory, the bounds $j_-^{\rm max}({b_i}, e)$ and $j_+^{\rm max}({b_i}, e)$ grow quadratically in the base degrees ${b_i}$ and the elliptic fiber degree $e$. Furthermore, in the limit ${b_{i\geq 2}} = 0$ and the limit ${b_1} = 0$, one should recover the bounds for the E-string and the M-string respectively. The only freedom we are allowed are quadratic terms ${b_i} {b_j}(i<j)$  mixing base degrees in the bounds $j_-^{\rm max}({b_i}, e)$ and $j_+^{\rm max}({b_i}, e)$. These terms proved sufficient to deduce bounds for all cases considered. 
	
E.g., for the massive E-M string chain with base degrees $b_0, b_1>0$ and only the M-string mass turned on, we find the bounds\footnote{When either $b_1 = 0$ or $b_0 = 0$ we can use the E-string or the M-string bounds.}
\begin{equation}
N_{j_-j_+}^{b_0, b_1,e, \mu}=0 \quad  {\rm if } \quad  
	\left\{\begin{array}{rl} 
	2 j_- &> 2 j_-^{max}=b_0 e -\tfrac{1}{2}b_0(b_0+1)+b_1(e-1) +1+b_0b_1- \lfloor (\mu/2)^2\rfloor  \\ & 
	{\rm or}  \\   
	2 j_+ &> 2 j_+^{max}=b_0 e -\tfrac{1}{2}b_0(b_0-1)+ b_1(e+1) -b_0b_1 - \lfloor (\mu/2)^2\rfloor
	\end{array} \right. .
	\end{equation}

\section{Relating to the domain wall method} \label{section:domain_wall_method}

We now would like to compare our results on the E- and M-string with those of \cite{Haghighat:2014}. The philosophy underlying \cite{Haghighat:2014} is that as both the M- and the E-string descend from the M2 brane of M-theory, stretched between two M5 branes and an M5 and an M9 brane respectively, the elliptic genus capturing their BPS excitations should be constructed from the same ingredients. More precisely, \cite{Haghighat:2013} invokes the relation between the elliptic genus of the M-string and the topological string partition function to compute the former via the refined topological vertex formalism \cite{IKV}.\footnote{The geometries are not toric, but based on a representation first proposed in \cite{Hollowood:2003cv}, vertex methods apply.} The resulting expression is given the interpretation of the quantum mechanics of groundstates of M2 brane wrapped on $T^2$, labelled by Young tableaux (consistent with the findings of \cite{Kim:2010mr}). The computation is then organized in terms of certain matrices $D_{\nu^t \mu}$ which are interpreted in terms of M5 brane domain walls separating M2 brane states $\nu$ and $\mu$. In \cite{Haghighat:2014}, this interpretation is pushed further: an expressions $D_\mu$ for M9 domain walls is guessed, permitting the computation of both the Hilbert series for heterotic strings (M2 branes spanning between two M9 branes) and E-strings (spanning between an M5 and an M9 brane). Using this method, the elliptic genus of two E-strings was determined in \cite{Haghighat:2014}. The calculation was generalized to three E-strings in \cite{Cai:2014}. 
	
\subsection{The M-string via the topological vertex}
	The elliptic genus for the M-string obtained in \cite{Haghighat:2013} via a refined topological vertex calculation is
	\be \label{M_babak}
	Z = \sum_{\nu} Q^{|\nu|} \prod_{(i,j)\in \nu}\frac{\vartheta_1\left([\nu_i - j + \frac{1}{2}]\epsilon_1 + [-i + \frac{1}{2}] \epsilon_2 -m \right) \vartheta_1 \left( [-\nu_i +j - \frac{1}{2} ]\epsilon_1 + [i-\frac{1}{2}] \epsilon_2 - m \right)}{\vartheta_1\left( [\nu_i - j+1] \epsilon_1 +[\nu_j^t-i] \epsilon_2 \right) \vartheta_1\left( [\nu_i - j] \epsilon_1 + [\nu_j^t - i +1 ] \epsilon_2 \right) }\,.
	\ee
	Here, the first sum is taken over Young diagrams $\nu$, and the second over the integer coordinates of boxes of each such diagram. $|\nu|$ indicates the total number of boxes of the diagram $\nu$, and $\nu_i$ the number of boxes in the $i^{\rm th}$ row. Recall that the zeros of $\vartheta_1$, all of which are simple, lie at the origin and the lattice translates thereof. The expression (\ref{M_babak}) thus has an infinite number of poles that are unexpected from the vantage point of the Gopakumar-Vafa form of the refined topological string free energy \cite{Gopakumar:1998ii,Gopakumar:1998jq}. We will demonstrated here for the case $n_b=2$ that these poles are spurious.
	
	
	\paragraph{$\mathbf{n_b=1}$}
	
	The single M-string elliptic genus is given by (\ref{M_babak}) as
	\be
	Z_1 = \frac{\vartheta_1(\epsilon_+ - m) \vartheta_1(-\epsilon_+ - m)}{\vartheta_1(\epsilon_1) \vartheta_1(\epsilon_2)} \,.
	\ee
	This expression exhibits merely the expected poles at $\epsilon_{1,2}=0$. To identify this result with (\ref{Mnb1}), we use the sum of squares relation
	\be
	\vartheta_1(z) = \frac{\vartheta_2^2 \vartheta_3^2}{\vartheta_4^2} \left( \frac{\vartheta_3^2(z)}{\vartheta_3^2}-\frac{\vartheta_2^2(z)}{\vartheta_2^2} \right)
	\ee 
	and the addition formula
	\be
	\vartheta_1(w+z) \vartheta_1(w-z) = \frac{\vartheta_2^2 \vartheta_3^2}{\vartheta_4^2} \left(\frac{\vartheta_3^2(w)}{\vartheta_3^2} \frac{\vartheta_2^2(z)}{\vartheta_2^2} -\frac{\vartheta_3^2(z)}{\vartheta_3^2} \frac{\vartheta_2^2(w)}{\vartheta_2^2}  \right) \,.
	\ee
	Further identities among $\vartheta$-functions are collected in appendix \ref{JacobiIdentity}.

	\paragraph{$\mathbf{n_b=2}$}
	
	As the expressions and identities required rapidly grow lengthy, we here focus on the massless case. $Z_2$ in this case reduces to
	\begin{eqnarray}
	Z_2 = \frac{\vartheta_1^2(\frac{3}{2}\epsilon_1 + \frac{1}{2}\epsilon_2 ) \vartheta_1^2(\frac{\epsilon_1+\epsilon_2 }{2}  ) }{\vartheta_1(\epsilon_1) \vartheta_1(\epsilon_2)\vartheta_1(\epsilon_2-\epsilon_1)\vartheta_1(2\epsilon_1) } +
	(\epsilon_1\leftrightarrow\epsilon_2) . 
	\end{eqnarray}
	This expression is more compact than the result (\ref{Mnb2}), yet exhibits a pole on the diagonal $\epsilon_1 = \epsilon_2$. That this pole is spurious follows from the identity 
	\begin{eqnarray}\label{eq:idn}
	\lefteqn{\cA\left(\frac{3}{2}\epsilon_1 + \frac{1}{2}\epsilon_2 \right) \vartheta_1(2\epsilon_2)  - \cA \left(\frac{1}{2}\epsilon_1 + \frac{3}{2}\epsilon_2 \right) \vartheta_1(2\epsilon_1) } \nonumber \\
	&=& \frac{\vartheta_1(\epsilon_2-\epsilon_1) \cA_{-}}{1492992} 
	[\cB_{+}^6 - 15 \cA_{+}^2 \cB_{+}^4 E_4 - 45 \cA_{+}^4 \cB_{+}^2 E_4^2 + 40 \cA_{+}^3 \cB_{+}^3 E_6 + 
	24 \cA_{+}^5 \cB_{+} E_4 E_6  \nonumber \\  && ~~~~ + \cA_{+}^6 (27 E_4^3 - 32 E_6^2)] \,.
	\end{eqnarray}
	The easiest method to obtain such identities is by making a polynomial ansatz in $\cA_{\pm}, \cB_{\pm}, E_{4,6}$ with appropriate modular weight and elliptic indices in $\epsilon_{\pm}$, and then fixing the coefficients by comparing the small $\epsilon$ expansion. We relegate the proof of this identity to appendix \ref{appendix:proofs_identities}.
	
	\subsection{The E-string via the domain wall method}
	The formula (\ref{Enb1}) for the elliptic genus of a single E-string obtained in \cite{Klemm:1996}, which we rederived above with our methods, serves as input in \cite{Haghighat:2014} to determine the matrix element of the M9 domain wall operator between the states corresponding to $\nu = \cdot$ and $\nu={\tiny \yng(1)}$. For two E-strings, \cite{Haghighat:2014} obtain the result
	\begin{eqnarray} \label{Z2_3.17}
	Z_2 &=& \frac{q}{576 \eta^{12}} \frac{1}{\vartheta_1(\epsilon_1)\vartheta_1(\epsilon_2) \vartheta_1(\epsilon_2-\epsilon_1) \vartheta_1(2 \epsilon_1 )} [ (-4 A_1^2 E_4 +3A_2 E_4^2 +5B_2E_6) \cA(\epsilon_1)^2 \nonumber \\ && 
	-(3A_2 E_6 +5B_2E_4)\cA(\epsilon_1) \cB(\epsilon_1) +4A_1 \cB(\epsilon_1)^2 ] + (\epsilon_1\leftrightarrow\epsilon_2) . 
	\end{eqnarray}   
	The two terms are the contributions to a sum over Young diagrams as in (\ref{M_babak}), stemming from $\nu = {\tiny\yng(2)}$ and $\nu = {\tiny\yng(1,1)}$. The
	$A_i, B_i$ are the $E_8$ Jacobi forms with index $i$ and modular weights $4,6$ respectively already encountered above. Here again, a pole arises on the diagonal $\epsilon_1=\epsilon_2$. To show that it too is spurious, as required to match our result (\ref{Enb2}), we invoke the following identities to substitute for $\cA(\epsilon_1)^2$, $\cA(\epsilon_1) \cB(\epsilon_1)$, and $\cB(\epsilon_1)^2$ in the first, second, and third term in square brackets in (\ref{Z2_3.17}) respectively:

	\begin{eqnarray}  \label{identity3.18}
	\lefteqn{\cA(\epsilon_1)^2\vartheta_1(2\epsilon_2) - \cA(\epsilon_2)^2\vartheta_1(2\epsilon_1)}  \nonumber \\ 
	&=&   \frac{\vartheta_1(\epsilon_1-\epsilon_2)} {10368} 
	(\cA_{+}\cB_{-} -\cA_{-}\cB_{+}) (3\cA_{+}\cB_{-}\cB_{+}^2  + \cA_{-}\cB_{+}^3 -9\cA_{-}\cA_{+}^2\cB_{+}E_4 - 3\cA_{+}^3\cB_{-}E_4  \nonumber \\ &&   + 8\cA_{+}^3\cA_{-}E_6) \,,  \nonumber \\
	\lefteqn{\cA(\epsilon_1) \cB(\epsilon_1)\vartheta_1(2\epsilon_2) - \cA(\epsilon_2) \cB(\epsilon_2) \vartheta_1(2\epsilon_1)  }\\ 
	&=&   \frac{\vartheta_1(\epsilon_1-\epsilon_2)} {10368} (\cA_{+}\cB_{-} -\cA_{-}\cB_{+}) [ \cB_{-} (\cB_{+}^3  + 3 \cA_{+}^2 \cB_{+} E_4 - 4 \cA_{+}^3 E_6) + 
	3 \cA_{-} \cA_{+} (\cB_{+}^2 E_4  \nonumber \\ &&  + 3 \cA_{+}^2 E_4^2 - 4 \cA_{+} \cB_{+} E_6)] \,, 
	\nonumber \\
	\lefteqn{\cB(\epsilon_1)^2 \vartheta_1(2\epsilon_2) - \cB(\epsilon_2)^2 \vartheta_1(2\epsilon_1)} \nonumber \\  
	&=&   \frac{\vartheta_1(\epsilon_1-\epsilon_2)} {10368}  [-\cB_{-}^2 \cB_{+} (\cB_{+}^3 - 9 \cA_{+}^2 \cB_{+} E_4 + 8 \cA_{+}^3 E_6) + 
	6 \cA_{-} \cA_{+} \cB_{-} \cB_{+} (\cB_{+}^2 E_4 + 3 \cA_{+}^2 E_4^2  \nonumber \\ &&   - 4 \cA_{+} \cB_{+} E_6) +  \cA_{-}^2 \cA_{+} (27 \cA_{+} \cB_{+}^2 E_4^2 - 8 \cB_{+}^3 E_6 - 24 \cA_{+}^2 \cB_{+} E_4 E_6 +  \cA_{+}^3 (-27 E_4^3 + 32 E_6^2)) ] \,. 
	\nonumber 
	\end{eqnarray}    
	The validity of the identities (\ref{identity3.18}) can be shown directly via manipulations involving the $\vartheta$-function identities reviewed in appendix \ref{JacobiIdentity}. We do this for the first identity in (\ref{identity3.18}) in appendix \ref{appendix:proofs_identities}. The rest can be proved analogously.

	
	\section{The E-string and a generalizion of the blowup equation}
	\label{sc:blowup}
	
	Recently, there has been a lot of progress on the quantization of the mirror curve to a toric Calabi-Yau threefold $X$. One perspective on this problem is to view the mirror curve as the spectral curve of a quantum integrable system constructed by Goncharov and Kenyon \cite{Goncharov2011}. The spectrum of this system can be solved exactly by a quantization condition that is based on the refined free energy in the Nekrasov-Shatashvili limit of the topological string on $X$ \cite{Franco:2015rnr,Hatsuda2015,Wang:2015wdy}. This quantization condition is notably invariant under the transformation $\hbar \mapsto 4\pi^2/\hbar$ (with appropriate transformations of the K\"ahler moduli $t_i$, related to the Hamiltonians of the integrable system by quantum mirror maps).
	A second perspective, following \cite{Codesido:2016ixn,Codesido:2015dia,Gu:2015pda,Grassi:2014zfa}, extracts trace-class difference operators from the quantum mirror curve. The exact form of the Fredholm determinant of these operators, which allows for a Fermi gas and matrix model interpretation, has been obtained in \cite{Marino:2015ixa,Kashaev:2015wia}. The compatibility of these two perspectives on the quantum spectral curve implies an identity from which an infinite number of constraints on the BPS invariants of $X$ can be extracted \cite{Sun:2016obh}. We call this identity the compatibility formula. It can be formulated for any toric Calabi-Yau threefold.
	
	As we have discussed above, the massive E-string is associated to the local $\halfKthree$ geometry, which is not toric. However, generalizing work on Seiberg-Witten curves with $E$-type flavor symmetry \cite{Minahan:1996cj}, a mirror curve has also been formulated for this geometry \cite{Sakai:2011}. It is natural to ask whether a proper quantization of this curve leads to an equally rich story as in the toric case. If the two perspectives on the quantization of the mirror curve in the toric case generalize, then the BPS spectrum of the E-string should satisfy a generalization of the compatibility formula to the non-toric case.
	
	In \cite{Grassi:2016nnt}, it was shown that if the local Calabi-Yau manifold $X$ is the space $Y^{N,m}$, i.e. the resolution of the cone over a $Y^{N,m}$ singularity, the compatibility formula is the Nekrasov-Satashvili limit of the G\"{o}ttsche-Nakajima-Yoshioka $K$-theoretic blowup equation for Nekrasov partition functions \cite{Nakajima:2005fg,Gottsche:2006bm,Nakajima:2011}. We will here propose a generalization of this formula, pre-NS limit, which can be applied to all local Calabi-Yau manifolds $X$ which permit the refinement of the topological string, i.e. exhibit a $\mb C^*$ isometry. In particular, $X$ need not be toric.
	
	We will show that the BPS invariants of local $\halfKthree$ satisfy the constraints implied by this generalization of the blowup equation.

	\subsection{Generalizing the blowup equation}
	To formulate the blowup equation and its generalization, we separate the $n_X$ irreducible compact curve classes $C_i$ into two sets, based on their intersections with the $g_X$ irreducible compact divisor classes $D_j$ in the geometry. These are captured by the $n_X\times g_X$ intersection matrix~$-\md C$
	\begin{equation}\label{eq:C-matrix}
	- C_{ij} = C_i  \cdot  D_j \, .
	\end{equation}
	We distinguish between those $C_i$ which intersect at least one compact divisor, and those which do not. In the toric setting, $g_X$ corresponds to the genus of the mirror curve. When the geometry engineers a supersymmetric gauge theory, the K\"ahler parameters $t_i$ of the former set map to moduli of the gauge theory, while the parameters $m_i$ of the latter map to masses of hypermultiplets. We will retain this nomenclature and refer to the $m_i$ as mass parameters.
	
	We will also need to specify an $n_X$ dimensional integral vector $\md B$ such that non-vanishing BPS invariants $N^{\md d}_{j_-,j_+}$ occur only at
	\begin{equation} \label{condB}
	2j_- + 2j_+ + 1 \equiv \md B \cdot \md d \quad \mod 2 \, .
	\end{equation}		
	This condition specifies $\md B$ only mod 2. The existence of such a vector $\md B$ is guaranteed by the fact that the non-vanishing BPS invariants follow a so-called checkerboard pattern, as first observed in \cite{CKK}.
	
	We define the twisted refined free energies $\widehat{F}_{\rm ref}(\md t, \epsilon_1,\epsilon_2)$ via
	\begin{equation}\label{eq:twist}
	\widehat{F}_{\rm ref}(\md t;\epsilon_1,\epsilon_2) = F^{\rm pert}_{\rm ref}(\md t;\epsilon_1,\epsilon_2) + F^{\rm inst}_{\rm ref}(\md t+\pi\ri \md B; \epsilon_1,\epsilon_2) \ .
	\end{equation}	
	Here, the perturbative contributions are given by
	\begin{equation}\label{eq:F-pert}
	F_{\rm ref}^{\rm pert}(\md t;\epsilon_1, \epsilon_2) = \frac{1}{\epsilon_1\epsilon_2}\( \frac{1}{6} \sum_{i,j,k=1}^{n_X} a_{ijk} t_i t_j t_k + 4\pi^2 \sum_{i=1}^{n_X} b_i^{\rm NS}t_i\)  + \sum_{i=1}^{n_X} b_i t_i - \frac{(\epsilon_1+\epsilon_2)^2}{\epsilon_1\epsilon_2} \sum_{i=1}^{n_X} b_i^{\rm NS} t_i \ ,
	\end{equation}
	where $a_{ijk}$ and $b_i$ are related to the topological intersection numbers in $X$, and $b_i^{\rm NS}$ can be obtained from the refined genus one holomorphic anomaly equation \cite{Huang:2010,Krefl:2010}. The instanton contributions are given by the refined Gopakumar-Vafa formula (\ref{defrefinedinv}).

	In terms of these quantities, we formulate the following generalization of the blowup equation to arbitrary local Calabi-Yau manifolds with $\IC^*$ isometry:
	\begin{equation}\label{eq:blowup}
	\begin{aligned}
	\sum_{\md n \in \mb Z^{g_X}}  (-1)^{\sum_{i=1}^{g_X} n_i} \exp&\(\widehat{F}_{\rm ref}\( \md t+ \epsilon_1(\md C\cdot \md n + \tfrac{1}{2}\md r), \epsilon_1, \epsilon_2 - \epsilon_1 \) \right.\\
	&\phantom{===}\left.+ \widehat{F}_{\rm ref}\( \md t+ \epsilon_2(\md C\cdot \md n+\tfrac{1}{2} \md r), \epsilon_1 - \epsilon_2, \epsilon_2 \) \) = 0 \ .
	\end{aligned}
	\end{equation}
	Here $\md r$ is chosen among an appropriate subset of integer vectors satisfying (\ref{condB}), as (\ref{eq:blowup-exp}) depends on the explicit representative of the mod 2 class. Two different vectors $\md r$, $\md r'$ however do yield the same constraints if there exists a $g_X$ dimensional integral vector $\md m$ such that
	\begin{equation}\label{eq:r-equivalence}
	\md r' = \md r + 2\md C \cdot \md m \ ;
	\end{equation} 
	changing from $\md r$ to $\md r'$ only amounts to a shift in the index vector $\md n$ in the blowup equation. It is conjectured in \cite{Sun:2016obh} that at least $g_X$ inequivalent choices for the vector $r$ exist for toric Calabi-Yau geometries, such that the equation \eqref{eq:blowup} is true. For the $Y^{N,m}$ geometry with $g_X = N-1$, at least $(N-1)^2$ such vectors exist \cite{Grassi:2016nnt}. In the next section, we will find one appropriate $\md r$ vector for local $\halfKthree$, which has $g_X=1$.
	
	We can separate the perturbative and non-perturbative contributions to \eqref{eq:blowup} to obtain a form of this equation more amenable to computation. Defining the invariants $n^{\md d}_{g,n}$ via
	\begin{equation}\label{eq:F-inst-exp}
	F^{\rm inst}_{\rm ref}(\md t,\epsilon_1,\epsilon_2) = \sum_{g,n=0}^\infty (\epsilon_1\epsilon_2)^{g-1}(\epsilon_1+\epsilon_2)^{2n} n^{\md d}_{g,n} \ie^{-\md d \cdot \md t} 
	\end{equation}
	and introducing
	\begin{equation}
	f(\md d,\md R,\epsilon_1,\epsilon_2) = \sum_{g,n\geq 0} (-1)^{\md d\cdot \md B} n^{\md d}_{g,n}\( (\epsilon_1(\epsilon_2 - \epsilon_1))^{g-1}\epsilon_2^{2n} \ie^{-\epsilon_1\md d\cdot \md R} + (\epsilon_2(\epsilon_1 - \epsilon_2))^{g-1} \epsilon_1^{2n} \ie^{-\epsilon_2 \md d\cdot \md R} \) \,,
	\end{equation}	
	we arrive at
	\begin{equation}\label{eq:blowup-exp}
	\begin{aligned}
	0=\sum_{\md n \in \mb Z^{g_X}} (-1)^{\sum_{i=1}^{g_X}n_i}  &\exp\( \(\sum_{i=1}^{n_X} (b_i - b_i^{\rm NS}) R_i - \frac{1}{6} \sum_{i,j,k=1}^{n_X} a_{ijk} R_i R_j R_k\)(\epsilon_1+\epsilon_2) \)  \\
	&\times \ie^{-\tfrac{1}{2} \sum_{i=1}^{n_X} a_{ijk} t_i R_j R_k} \exp\(\sum_{\md d} \ie^{-\md d \cdot \md t} f(\md d,\md R,\epsilon_1,\epsilon_2) \) \ ,
	\end{aligned}
	\end{equation}
	where we have used the notation 
	\begin{equation}
	R_i = \sum_{j=1}^{n_X} C_{ij} n_j + r_j/2 \ .
	\end{equation}	
	
	Thus, in order to extract constraints on BPS invariants one needs, in addition to the BPS invariants, the following perturbative data 
	\begin{equation}\label{eq:pert-data}
	\md{C}, \; \md{B}, \; a_{ijk}, \; b_i, \;  b_i^{\rm NS} \ .
	\end{equation}

	\subsection{Constraints on the BPS invariants of $\halfKthree$}
	
	In this section, we verify that the constraints implied by \eqref{eq:blowup} are satisfied by the BPS invariants of local $\halfKthree$, with an appropriately chosen integer vector $\md r$. We begin by determining the perturbative data \eqref{eq:pert-data} for this geometry.
	
	
	We have discussed the $\halfKthree$ surface in section~\ref{EMstringgeometries}, and in particular noted that it exhibits an elliptic fibration. We will denote the K\"ahler moduli associated to the base $C_1$ and  elliptic fiber $E$ of this fibration by $t_b$ and $t_f$ respectively.

	Local $\halfKthree$, the total space of the canonical line bundle of $\halfKthree$, has a single compact divisor, given by the zero section $\mc S = \halfKthree$ of the line bundle. Using the adjunction formula, we find that 
	\begin{equation}\label{eq:X-ints}
	\mc S\cdot C_1 = 1 \ ,\quad \mc S \cdot E = \mc S \cdot e_i = 0 \,,\quad i=1, \ldots,8 \ .
	\end{equation}
	Therefore, following the nomenclature introduced above, $t_f$ as well as all the $m_i$ are mass parameters. The $\md C$ matrix reads
	\begin{equation}
	\md C = (1,0,0,0,0,0,0,0,0,0 )^t \ .
	\end{equation}
	By checking the pattern of the BPS numbers we have computed in the previous sections, we can constrain the integer vector $\md B$ via
	\begin{equation}
	\md B \equiv (1,0,0,0,0,0,0,0,0,0) \mod 2 \ .
	\end{equation}
	Due to the $E_8$ Weyl symmetry acting on the mass parameters $m_i$, it is convenient to keep track of only $t_b, t_f$ in the free energy \eqref{eq:F-inst-exp}, and hide the dependence on the masses $m_i$ in $n^{d_b, d_f}_{g,n}$ which decompose into $E_8$ characters or $E_8$ Weyl orbits. We can thus work with a reduced intersection matrix and integer vector, which we continue to denote by the same symbols,
	\begin{equation}\label{eq:CB}
	\md C = (1,0)^t \ , \quad \md B \equiv (1,0) \mod 2 \ .
	\end{equation}	

	The coefficients $a_{ijk}, b_i$ of the perturbative free energy can in principle be computed from the topological intersection numbers of divisors in the local $\halfKthree$. The coefficients $b_i^{\rm NS}$ however cannot be computed from the topology of $X$; we will compute them by applying the refined holomorphic anomaly equation to the mirror curve of $X$ constructed by Sakai \cite{Sakai:2011}. Indeed, we will extract all of the remaining perturbative data $a_{ijk}, b_i, b_i^{\rm NS}$ from this curve.

	For this calculation, we express Sakai's mirror curve in the Weierstrass form
	\begin{equation}\label{eq:Sakai}
	y^2 = 4 x^3 - g_2(u, \tau, \md m) x - g_3 (u, \tau, \md m) \ .
	\end{equation}
	The coefficient functions $g_2(u, \tau, \md m)$ and $g_3(u, \tau, \md m)$ are polynomials in $u$ of degree $4$ and $6$ respectively. Their explicit forms are reproduced in appendix \ref{sc:jacobi_weyl}. The modulus $u$ is chosen such that $u\rightarrow 0$ is the large volume limit. $\tau$ is the elliptic modulus of the elliptic fiber inside $\halfKthree$ in the limit $u\rightarrow 0$. These parameters are related to the K\"ahler moduli $t_b, t_f$ of $X$ by
	\begin{equation}
	- t_b = \log u +\mc O(u) \ , \quad  q = \ie^{2\pi \ri\tau} = \ie^{-t_f} \ .
	\end{equation}
	The coefficients $a_{ijk}$ can now be extracted from the prepotential which can be computed through the special geometry relations following the procedure in \cite{Huang:2013}. We find
	\begin{equation}\label{eq:F0-pert}
	F_0(t_b, t_f, \md m) = \frac{1}{2} t_b^2 t_f  + \frac{1}{2} t_b t_f^2 + \frac{1}{6}t_f^3+ \mc O(\ie^{-\md{t}}) \ .
	\end{equation}
	The coefficients of $t_b t_f^2$ and $t_f^3$ are integral constants, and they are fixed by the requirement that $F_0(t_b, t_f, \md m)$ splits and reproduces the prepotential of the local del Pezzo $E_8$ surface in the blow down limit of the base $e_9$:
	\begin{equation}
	t_b \rightarrow -\infty, \quad t_b+t_f =: t \;\textrm{ finite} \ .
	\end{equation}
	Next, we fix the genus one refined holomorphic anomaly equations \cite{Huang:2010,Krefl:2010} by the known genus one BPS invariants and obtain\footnote{The coefficients of $t_f$ can also be fixed by going to the blowdown limit. But they in fact do not contribute to the blowup equation.}
	\begin{equation}\label{eq:F1-pert}
	\begin{aligned}
	F_1^{\rm ST}(t_b,t_f,\md m) &= -\frac{1}{2}t_b + \mc O(\ie^{-\md t})\ ,\\
	F_1^{\rm NS}(t_b,t_f,\md m) &= -\frac{1}{2}t_b + \mc O(\ie^{-\md t}) \ .
	\end{aligned}
	\end{equation}

	We have thus computed all of the perturbative data needed in order to extract constraints on the BPS invariants of local $\halfKthree$ from the generalization \eqref{eq:blowup} of the blowup equation. Substituting \eqref{eq:CB} as well as $a_{ijk}, b_i, b_i^{\rm NS}$ taken from \eqref{eq:F0-pert}, \eqref{eq:F1-pert} into equation \eqref{eq:blowup-exp}, we find that for $\md r = (1,0)$, the highly non-trivial identities for the refined Gromov-Witten invariants extracted from equation \eqref{eq:blowup-exp} are satisfied. For instance, some of these identities are
	\begin{equation}\label{eq:GW-constraints}
	\begin{aligned}
	0=&\, 24 - n^{1,1}_{0,0} + 24 n^{1,1}_{0,1} + 24 n^{1,1}_{1,0} \ , \\
	0=&\, 27 n_{0, 0}^{1, 0} - n_{0, 0}^{1, 1} - 72 n_{0, 1}^{1, 0} + 
	24 n_{0, 1}^{1, 1} - 72 n_{1, 0}^{1, 0} + 24 n_{1, 0}^{1, 1}  \ ,\\
	0=&\, 27 n_{0, 0}^{1, 1} - n_{0, 0}^{1, 2} - 72 n_{0, 1}^{1, 1} + 
	24 n_{0, 1}^{1, 2} - 72 n_{1, 0}^{1, 1} + 24 n_{1, 0}^{1, 2} \ , \\
	& \cdots \cdots \cdots
	\end{aligned}
	\end{equation}
	Note that we have here absorbed the dependence on the mass parameters $m_i$ into the invariants $n^{\md d}_{g,n}$ introduced in \eqref{eq:F-inst-exp}, so that the latter are linear combinations of $W_i(\md m)$ defined by
	\begin{equation}
	W_i(\md m) = \sum_{\md w \in \mc O_i} \ie^{2\pi\ri \md m \cdot \md w} \ ,
	\end{equation}
	where on the right hand side we sum over all the weights $\md w$ in certain Weyl orbit $\mc O_i$ of $\mf e_8$. The identities \eqref{eq:GW-constraints} can be translated to constraints on the BPS invariants. Taking the vanishing conditions on these invariants into account, we find from \eqref{eq:GW-constraints}
	\begin{equation}
	\begin{aligned}
	0 =& -1 + N_{1/2,1/2}^{1,1} \ ,\\
	0 =& N_{0,0}^{1,0} - N_{1/2,1/2}^{1,1} \ , \\
	0 =& N_{0,0}^{1,1} + 7 N_{1/2,1/2}^{1,1} - N_{1/2,1/2}^{1,2} - 6 N_{1,1}^{1,2} \ ,\\
	& \cdots \cdots \cdots
	\end{aligned}
	\end{equation}
	
	We have expanded equation \eqref{eq:blowup-exp} up to degree 5 in terms of $\ie^{-t_b}, \ie^{-t_f}$ and total genus $g+n \leqslant 4$; all the constraints thus obtained are satisfied.

	\section{Conclusions}
	\label{Conclusions}
	
	We have demonstrated that our approach to computing $\Ztop$ is sufficiently powerful to completely solve $\Ztop$ on all local elliptically fibered Calabi-Yau 3-folds leading to superconformal 6d field theories without gauge symmetry. In forthcoming work \cite{AGHKZ}, we will extend this analysis to singularities in the elliptic fibration that lead to gauge symmetry.
	
	While we have justified the general form of our ansatz based on the holomorphic anomaly equations, we obtained the exact form of the index polynomial based on anomaly considerations. To further our understanding of the refined topological string, it would be important to derive the index purely from topological string considerations.
	
	Regarding our proposed generalization of the blowup equation, much remains to be done. Its range of validity should be determined, and a derivation provided within this class of geometries. More ambitiously still, one should study to what extent the Grassi-Hatsuda-Mari\~no conjecture or variants thereof can be extended to this class of geometries.

	One aspect of our approach that we find intriguing is that its justification lies in the Witten 
	form \cite{Witten:1993ed} of the holomorphic anomaly equations \cite{HKK2}, 
	which is formulated for the partition function directly. The holomorphic anomaly 
	equations in their BCOV incarnation \cite{BCOV} on the other hand are formulated 
	in terms of the coefficients $F_{n,g}$ of the coupling constant expansion of the 
	topological string amplitudes. The two formulations yield complementary results for $\Ztop$; the former yields closed expressions for $Z_\beta$, coefficients of an expansion in base classes, while the latter yields exact results for $F_{n,g}$, coefficients in coupling constant expansions. Understanding the precise relation between the two 
	formulations might thus be a stepping stone towards understanding $\Ztop$ pre-expansion in any variables.

	\section*{Acknowledgments}
	We would like to thank Francesco Benini, Michele Del Zotto, Valery Gritsenko, Sheldon Katz, Kimyeong Lee and Guglielmo Lockhart for helpful conversations and correspondence. MH thanks Kaiwen Sun and Xin Wang for discussions on related projects. JG and AKKP acknowledge support from the grant ANR-13-BS05-0001. MH is supported by the ``Young Thousand People" plan by the Central Organization Department in China,  Natural Science Foundation of China and CAS Center for Excellence in Particle Physics (CCEPP). AK thanks the IHES for hospitality during the time when this work was initiated.

	\appendix
	

	\section{Definitions and properties of modular forms and weak Jacobi forms}
	\label{Appendixmod}   
	In this appendix, we collect some facts regarding modular forms and weak Jacobi forms. We follow 
	closely the treatment of~\cite{Zagierbook} for the former and~\cite{EZ} and~\cite{Dabholkar:2012}
	for the latter.      
	\subsection{Modular forms} 
	The Eisenstein series are defined as
	\be 
	E_{2m}(\tau)=1 +\frac{2}{\zeta(1-2m)}\sum_{n=1}^\infty\frac{n^{2m-1} q^n}{1- q^n} \quad\quad\quad\quad m \geq 1\ ,
	\label{defEisenstein}
	\ee
	where
	\be 
	q=\exp(2 \pi i \tau)  \ , 
	\ee
	with $\tau$ taking values in the complex upper half plane $\IH=\{ \tau \in \mathbb{C}| {\rm Im}(\tau)>0\}$.   
	For integer $m>1$, the Eisenstein series are modular forms of weight $k=2m$. 
	By definition a weight $k$ modular form $f_k(\tau)$ transforms under an ${\rm SL}(2,\mathbb{Z})$ action
	\be 
	\tau\mapsto \tau_\gamma=\frac{a\tau +b}{c\tau + d}, \qquad {\rm for}  \qquad 
	\gamma=\left(\begin{array}{cc} a& b \\c& d \end{array}\right)\in {\rm SL}(2,\mathbb{Z})
	\label{PSL2tau} 
	\ee
	as 
	\be 
	f_k(\tau_\gamma)=(c \tau +d)^k f_k(\tau)\ . 
	\ee
	The two modular forms 
	\be 
	Q:=E_4, \quad {\rm  and}\quad R:=E_6 
	\ee
	generate the ring of holomorphic modular forms $M_*=\oplus_{k\ge 0} M_k(\Gamma_1)$
	of the modular group  $\Gamma_1={\rm SL}(2,\mathbb{Z})$. 
	
	The Dedekind $\eta$ function is defined as 
	\be
	\eta(\tau)=q^{\frac{1}{24}} \prod_{n=1}^\infty (1-q^n) \ . 
	\ee
	Its $24^{th}$ power,
	\begin{equation}
	\Delta(\tau)=\eta^{24}(\tau)= \frac{E_4^3-E_6^2}{1728}  \ ,
	\end{equation}
	is a weight $k=12$ modular form called the discriminant function. The modular invariant $j$-function is defined as
	\be
	j(\tau)=\frac{E_4^3(\tau)}{\Delta(\tau)}=\frac{1}{q} + 744 + 196884 q + 21493760 q^2 + 864299970 q^3+{\cal O}(q^4)\ .
	\label{j-invariant}
	\ee
	Any modular function can be written as a rational function of $j(\tau)$.
	
	The second Eisenstein series $E_2$ is an interesting special case  
	as it transforms under an ${\rm SL}(2,\mathbb{Z})$ action with a shift 
	\be 
	E_2(\tau_\gamma)=(c \tau +d)^2 E_2(\tau) -\frac{6 i}{\pi} c(c\tau+d)\ .
	\label{quasimodulartransformation} 
	\ee
	This Eisenstein series is an example of a quasimodular form (of weight 
	$k=2$). More generally $P:=E_2$, $Q$ and $R$ generate the ring of 
	quasimodular forms $\tilde M_*(\Gamma_1)=M_*(\Gamma_1)[E_2]$ of even weight. 
	The ring $\tilde M_*(\Gamma_1)$ is closed under differentiation with respect to 
	$\tau$. Alternatively, one can define the almost holomorphic second Eisenstein series     
	\be 
	\hat E_2(\tau)=E_2(\tau)-\frac{3}{\pi {\rm Im}(\tau)}\ ,
	\label{anholomorphicE2}
	\ee
	which does transform as a modular function of weight $k=2$.  The ring 
	of almost holomorphic modular forms is defined  as $\hat M_*(\Gamma_1)=
	M_*(\Gamma)[\hat E_2]$.  Quasimodular forms are the holomorphic parts
	of almost holomorphic forms, but they are not necessarily modular themselves.
	
	\subsection{Jacobi forms} 
	Jacobi forms~\cite{EZ} are functions $\phi:\mathbb{H}\times \mathbb{C}\rightarrow \mathbb{C}$ that depend on a modular 
	parameter $\tau\in \mathbb{H}$ and an elliptic parameter $z\in \mathbb{C}$. They transform 
	under the action of the modular group on $\IH \times \IC$, given by
	\be
	\tau \mapsto \tau_\gamma=\frac{a \tau+ b}{c \tau + d}, \quad z \mapsto z_\gamma=\frac{z}{c \tau + d}\quad  {\rm with} \quad  
	\left(\begin{array}{cc}a&b\\ c& d \end{array}\right) \in {\rm SL}(2; \mathbb{Z})  \ ,
	\label{PSL2} 
	\ee 
	as  
	\be
	\phi\left(\tau_\gamma, z_\gamma\right)= (c \tau + d)^k e^{\frac{2 \pi i m c z^2}{c \tau + d}} \phi(\tau,z)        \ .
	\label{mod}
	\ee
	Furthermore, they enjoy the property of quasi-periodicity,
	\be
	\phi(\tau,z +\lambda \tau+ \mu)=e^{- 2 \pi i m (\lambda^2 \tau+ 2 \lambda z)}\phi(\tau,z) \quad \forall  \lambda, \mu \in \mathbb{Z} \ .
	\label{shiftJacobi} 
	\ee
	$k\in \mathbb{Z}$ is called the \emph{weight} and $m\in \IN$ the \emph{index} of the Jacobi form. 
	
	Due to the periodicity under $\tau\mapsto\tau+1, z\mapsto z+1$, the Jacobi forms enjoy a double Fourier expansion
	\be 
	\phi(\tau,z)=\sum_{n,r} c(n,r) q^n y^r, \qquad {\rm where} \ \ q=e^{2 \pi i \tau},\ \  y=e^{2 \pi i z} \ .
	\ee
	It is in fact more appropriate to write the coefficients as $c(n,r)=C(4 n m-r^2,r)$ as the combination $4n m-r^2$ is invariant under the transformation \eqref{shiftJacobi} and $C(4 n m-r^2,r)$ has a periodicity of $2m$ in $r$. Holomorphic Jacobi forms satisfy the constraint $c(n,r)=0$ unless $4 mn \ge r^2$, cusp forms satisfy $c(n,r)=0$ unless $4 mn > r^2$, while for weak Jacobi forms, one imposes the condition $c(n,r)=0$ unless $n\ge 0$. 
	
	According to~\cite{EZ}, the ring of weak Jacobi forms of integer index is freely 
	generated over the ring of modular forms by the two generators $\phi_{-2,1} (\tau,z)$
	and $\phi_{0,1} (\tau,z)$ of index 1. Introducing the notation
	\be
	A(\tau,z)=\phi_{-2,1} (\tau,z) \quad {\rm and }\quad B(\tau,z)=\phi_{0,1} (\tau,z) \ ,  
	\ee
	we see that the vector space of weak Jacobi forms of weight $k$ and index $m$
	is equal to
	\be 
	J^{\rm weak}_{k,m}=\bigoplus_{j=0}^m M_{k+2j} (\Gamma_1) A^j B^{m-j}\ . 
	\ee
	The generators $A$ and $B$  are of index $m=1$ and weight $-2$ and $0$ respectively. 
	They can be defined as 
	\begin{eqnarray}  \label{conventions4.205}
	A(\tau,z) &=& -\frac{ \theta_1(\tau,z)^2} {\eta^6(\tau)} \,  \nonumber \\ 
	B(\tau,z) &=& 4\(\frac{ \theta_2(\tau,z)^2} {  \theta_2(\tau, 0)^2 } +\frac{ \theta_3(\tau,z)^2} {\theta_3(\tau, 0)^2 } +\frac{ \theta_4(\tau,z)^2} {  \theta_4(\tau, 0)^2 } \). 
	\end{eqnarray}
	Our  conventions for the theta functions associated to the spin structure on the torus are  
	\be 
	\Theta\left[a \atop b\right](\tau,z)=\sum_{n\in \mathbb{Z}} e^{\pi i (n + a)^2 \tau + 2 \pi i z (n+a) + 2 \pi i bn}\ . 
	\ee
	The  Jacobi theta functions are then $\theta_1=i\Theta\left[\frac{1}{2} \atop \frac{1}{2}\right]$, 
	$\theta_2=\Theta\left[\frac{1}{2} \atop 0\right]$, $\theta_3=\Theta\left[0 \atop 0\right]$ and 
	$\theta_4=\Theta\left[0 \atop \frac{1}{2}  \right]$. In particular, we have
	\begin{equation}
	\theta_1(\tau,z) = z \cdot \eta(\tau)^3   \exp\( \sum_{k=1}^{\infty} \frac{ B_{2k}}{2k(2k)!} (iz)^{2k} E_{2k} (\tau) \) \ .
	\end{equation}
	In order to accommodate our convention for the normalization of $\epsilon_i$, we will also use the notation
	\begin{equation}
	\cA(\tau,2\pi z):= A(\tau, z)\ ,\; \cB(\tau,2\pi z) := B(\tau, z) \ ,\; \vartheta_i(\tau, 2\pi z):= \theta_i(\tau,z),\;  i =1,\ldots,4\ .
	\end{equation}

	Using the Jacobi triple product for $\theta_1$ and the notation
	\begin{equation}
	x_m =\left(2 \sin \pi m z \right)^2=-(y^\frac{m}{2}-y^{-\frac{m}{2}})^2 \ ,\quad y=\exp(2 \pi i z) \ ,
	\end{equation}
	one finds that the weak Jacobi form $A$ has a simple product form
	\ba 
	A(\tau,z) &=&  (y^{\frac{1}{2}} - y^{-\frac{1}{2}})^2 \prod_{n=1}^\infty \frac{(1 - q^n y)^2 (1 - \frac{q^n}{y})^2}{(1-q^n)^4} \label{Aproduct_shift} \\
	&=& - x_1 \prod_{n=1}^\infty \frac{( 1 + x_1 q^n - 2 q^n +q^{2n})^2}{(1-q^n)^4}\ . 
	\label{Aproduct}
	\ea

	The weight zero index one weak Jacobi form $B$ is one half of the elliptic genus of K3,
	\be
	\label{ellipticgenusK3}
	\chi({\mathrm K3};q,y)=2 B(\tau,z)=
	\left(2 y+20+\frac{2}{y}\right)+\left(\frac{20}{y^2}-\frac{128}{y} +216 -128 y+20 y^2\right)q+  {\cal O}(q^2)  \ ,
	\ee
	and it enjoys the following expansion in $x_1$
	\be
	B(\tau,z)=- x_1 (1- 10 x_1 q + x_1^2 q^2) +\sum_{n=0}^\infty q^n g_n(x_1)\ , 
	\label{Bform}
	\ee
	with $g_n(x_1)$ a polynomial in $x_1$ of order $n$. Note that $A(\tau,z)$ vanishes when $z = 0$, while $B(\tau,0)=12$, as can be seen from the expansion of these Jacobi forms in $z$ with quasi modular coefficients     
	\begin{eqnarray}
	A(\tau,z) &=& -z^2 + \frac{E_2 }{12}z^4 + \frac{-5 E_2^2 + E_4}{1440} z^6 + \frac{35 E_2^3 - 21 E_2 E_4 + 4 E_6}{362880}z^8  + \mathcal{O}(z^{10}), \nonumber \\ 
	B(\tau,z) &=& 12 - E_2 z^2 + 
	\frac{E_2^2 + E_4}{24} z^4 + \frac{-5 E_2^3 - 15 E_2 E_4 + 8 E_6}{4320} z^6  + \mathcal{O}(z^8) . 
	\label{ABquasimodular}
	\end{eqnarray}    
	
	The real zeros of $A$ coincide with the zeros of $x_1$. All complex zeros are obtained as translates of these zeros by multiples of 1 and $\tau$. 
	
	The weak Jacobi forms $A(\tau, n z)$ of index $n^2$ will play an important role in this paper. $A(\tau, n_1 z)$ is divisible by $A(\tau, n_2 z)$ if the integer $n_1$ is divisible by $n_2$. Based on this observation, it is convenient to define a more primitive weak Jacobi form, due to Zagier,
	\be
	P_d(\tau,z) = \prod_{k|d} A(\tau, kz)^{\mu(d/k)},   \label{P4.4}
	\ee
	where $\mu(n)$ is the M\"{o}bius function. The first few $P_d$ are given by
	\be
	P_1 = A_1 \,, \quad P_2 = \frac{A_2}{A_1} \,, \quad P_3 = \frac{A_3}{A_1} \, \ldots ,
	\ee
	where we have defined $A_k = A(\tau,kz)$. For any $d$, one can show that $P_d(\tau, z)$ has no poles and vanishes only at primitive $d$-torsion points, i.e. at $z=2\pi (n_1+\tau n_2)/d$ for integers $n_1$, $n_2$  with the greatest common divisor $\textrm{gcd}(n_1,n_2,d)=1$. So $P_d(\tau, z)$ is a weak Jacobi form and can be written as a polynomial in $A(\tau, z), B(\tau, z), E_4, E_6$. These polynomials $P_d$ will feature prominently in a forthcoming paper \cite{HKK2}. We can express $A(\tau,nz)$ in terms of these more basic building blocks via   
	\be 
	A(\tau, nz) =  \prod_{k|n} P_k(\tau,z) .   \label{A4.5}
	\ee
	
	\subsection{Some useful identities for Jacobi theta functions} 
	\label{JacobiIdentity}
	We review some basic identities for Jacobi theta functions in this appendix, which can be found e.g. in \cite{WW}. For simplicity of notation, we will only indicate the elliptic argument of theta functions, and use the notation $\vartheta_i := \vartheta_i (\tau,0)$. Apart from the odd function $\vartheta_1(z)$, all other Jacobi functions $\vartheta_i(z), ~ i=2,3,4$ are even functions of the elliptic argument. When the elliptic argument is set to zero, the Jacobi theta functions satisfy the famous abstruse identity,
	\be
	\vartheta_4^4 =\vartheta_3^4 -\vartheta_2^4 \,.
	\ee 
	Only two of the four theta functions are algebraically independent, as they satisfy the relations
	\begin{equation} \label{A.27}
	\vartheta_2^2(z)\vartheta_4^2  =   \vartheta_4^2(z)\vartheta_2^2-\vartheta_1^2(z)\vartheta_3^2,  ~~~~
	\vartheta_3^2(z)\vartheta_4^2 =	  \vartheta_4^2(z)\vartheta_3^2-\vartheta_1^2(z)\vartheta_2^2. 
	\end{equation}
	Denoting $\epsilon_{\pm} \equiv \frac{1}{2} (\epsilon_1\pm \epsilon_2)$ and $\vartheta_{i\pm}\equiv \vartheta_i( \epsilon_{\pm})$, the following formulae express theta functions with elliptic arguments $\epsilon_{1,2}$ in terms of theta functions with elliptic arguments $\epsilon_{\pm}$: 
	\begin{equation} \label{A.28}
	\begin{aligned}
	&& \vartheta_1(\epsilon_1)\vartheta_1(\epsilon_2)\vartheta_4^2	=	\vartheta_{3+}^2 \vartheta_{2-}^2-\vartheta_{2+}^2 \vartheta_{3-}^2  = \vartheta_{1+}^2 \vartheta_{4-}^2 - \vartheta_{4+}^2 \vartheta_{1-}^2 ,	
	\\ 
	&&\vartheta_2(\epsilon_1)\vartheta_2(\epsilon_2)\vartheta_4^2	=	\vartheta_{4+}^2\vartheta_{2-}^2-\vartheta_{1+}^2\vartheta_{3-}^2	
	=	\vartheta_{2+}^2\vartheta_{4-}^2-\vartheta_{3+}^2\vartheta_{1-}^2,	
	\\ 
	&&\vartheta_3(\epsilon_1)\vartheta_3(\epsilon_2)\vartheta_4^2	=	\vartheta_{4+}^2\vartheta_{3-}^2-\vartheta_{1+}^2\vartheta_{2-}^2	
	=	\vartheta_{3+}^2\vartheta_{4-}^2 - \vartheta_{2+}^2 \vartheta_{1-}^2,
	\\ 
	&&\vartheta_4(\epsilon_1)\vartheta_4(\epsilon_2)\vartheta_4^2	=	\vartheta_{3+}^2 \vartheta_{3-}^2-\vartheta_{2+}^2\vartheta_{2-}^2	
	=	\vartheta_{4+}^2 \vartheta_{4-}^2-\vartheta_{1+}^2 \vartheta_{1-}^2,  \\ 
	&&\vartheta_1(\epsilon_1)\vartheta_2(\epsilon_2)\vartheta_3\vartheta_4 
	=\vartheta_{1+}\vartheta_{2+}\vartheta_{3-}\vartheta_{4-} + \vartheta_{3+}\vartheta_{4+}\vartheta_{1-}\vartheta_{2-} ,  \\ 
	&&\vartheta_1(\epsilon_1)\vartheta_3(\epsilon_2)\vartheta_2\vartheta_4 
	=\vartheta_{1+}\vartheta_{3+}\vartheta_{2-}\vartheta_{4-} + \vartheta_{2+}\vartheta_{4+}\vartheta_{1-}\vartheta_{3-}  ,  \\ 
	&&\vartheta_1(\epsilon_1)\vartheta_4(\epsilon_2)\vartheta_2\vartheta_3 
	=\vartheta_{1+}\vartheta_{4+}\vartheta_{2-}\vartheta_{3-} + \vartheta_{2+}\vartheta_{3+}\vartheta_{1-}\vartheta_{4-}  ,  \\ 
	&&\vartheta_2(\epsilon_1)\vartheta_3(\epsilon_2)\vartheta_2\vartheta_3 
	=\vartheta_{2+}\vartheta_{3+}\vartheta_{2-}\vartheta_{3-}  - \vartheta_{1+}\vartheta_{4+}\vartheta_{1-}\vartheta_{4-} ,  \\ 
	&&\vartheta_2(\epsilon_1)\vartheta_4(\epsilon_2)\vartheta_2\vartheta_4 
	=\vartheta_{2+}\vartheta_{4+}\vartheta_{2-}\vartheta_{4-}  - \vartheta_{1+}\vartheta_{3+}\vartheta_{1-}\vartheta_{3-} ,  \\ 
	&& \vartheta_3(\epsilon_1)\vartheta_4(\epsilon_2)\vartheta_3\vartheta_4 
	=\vartheta_{3+}\vartheta_{4+}\vartheta_{3-}\vartheta_{4-} - \vartheta_{1+}\vartheta_{2+}\vartheta_{1-}\vartheta_{2-} .
	\end{aligned}
	\end{equation}
	These identities are not entirely independent. For example, we can transform the product 
	$\vartheta_1(\epsilon_1)\vartheta_1(\epsilon_2)\vartheta_2(\epsilon_1)\vartheta_2(\epsilon_2)$ with either the first two formulae or the fifth formula above, and obtain the the same result due to (\ref{A.27}). We can also derive the duplication formulas from the above (\ref{A.28}). For example, set $\epsilon_1=2z, \epsilon_2=0$ in the 5th formula we obtain
	\be \label{duplication_formula}
	\vartheta_1(2z)= \frac{2\vartheta_1(z)\vartheta_2(z)\vartheta_3(z)\vartheta_4(z)}{\vartheta_2\vartheta_3\vartheta_4} \,.
	\ee
	The Eisenstein series can be expressed in terms of the the Jacobi theta functions. We have
	\be \label{Eisen}
	E_4 = \frac{1}{2} ( \vartheta_2^4 + \vartheta_3^4 +\vartheta_4^4) \,, \quad
	E_6 = \frac{1}{2}(- 3 \vartheta_2^8 (\vartheta_3^4+\vartheta_4^4) + \vartheta_3^{12}+\vartheta_4^{12}) \,.
	\ee
	Finally, the product of the three even theta functions yields twice the Dedekind eta function,
	\be \label{eta}
	2 \eta = \vartheta_2 \vartheta_3 \vartheta_4 \,.
	\ee

	\subsection{$E_8$ Weyl invariant Jacobi forms} \label{sc:jacobi_weyl}
	$E_8$ Weyl invariant Jacobi forms $\phi_{k,m} : \IH \times \IC^8 \rightarrow \IC$ of weight $k$ and index $m$ depend on a modular parameter $\tau \in \IH$ and an element $\md m \in \Lambda_{\mathfrak{e}_8}\otimes \IC = \IC^8$ of the complexified root lattice of $\mathfrak{e}_8$. They are invariant under the action of the Weyl group of $E_8$ on $\md m$, and satisfy the modular and quasi-periodicity relations
	\be
	\phi_{k,m} \left( \frac{a \tau +b }{c \tau + d} , \frac{\md m}{c \tau + d} \right) = (c\tau + d) \exp \left( m \pi i \frac{c}{c \tau +d} (\md m, \md m)_{\mathfrak{e}_8} \right) \phi_{k,m}(\tau,\md m) \,,
	\ee
	and
	\be
	\phi_{k,m}(\tau, \md m + \boldsymbol{\mu} + \tau \boldsymbol{\nu}) = \exp \Big(-m \pi i \big(\tau (\boldsymbol{\nu},\boldsymbol{\nu})_{\eeight} + 2 (\boldsymbol{\mu},\boldsymbol{\nu})_{\eeight} \big)\Big) \phi_{k,m}(\tau,\md m) \,, 
	\ee
	for $\boldsymbol{\mu},\boldsymbol{\nu} \in  \Lambda_{\eeight}$. Recall that the inner product $(\cdot,\cdot)_{\eeight}$ on $\Lambda_{\eeight}$ is normalized such that the roots have length squared 2.
	
	The ring of $E_8$ Weyl invariant Jacobi forms is generated by the following forms \cite{Sakai:2011}:
			\begin{align}
	& A_1 = \Theta(\tau, \md m) = \frac{1}{2}\sum_{k=1}^4 \prod_{j=1}^8 \theta_4(\tau,m_j) \ ,\quad A_4 = \Theta(\tau,2\md m) \ ,\nn\\
	& A_n = \frac{n^3}{n^3+1} \( \Theta(n\tau, n\md m)  + \frac{1}{n^4}\sum_{k=0}^{n-1}\Theta(\tfrac{\tau+k}{n}, \md m)\) \ ,  \quad n = 2,3,5\ ,\nn\\
	& B_2 = \frac{8}{15}\((\theta_3^4+\theta_4^4) \Theta(2\tau, 2\md m) + \frac{1}{2^4}(-\theta_2^4-\theta_3^4)\Theta(\tfrac{\tau}{2}, \md m) + \frac{1}{2^4} (\theta_2^4 - \theta_4^4) \Theta(\tfrac{\tau+1}{2}, \md m)\) \ , \nn\\
	& B_3 = \frac{81}{80}\( h(\tau)^2 \Theta(3\tau,3\md m)-\frac{1}{3^5}\sum_{k=0}^2 h(\tfrac{\tau+k}{3})^2 \Theta(\tfrac{\tau+k}{3},\md m) \) \ , \nn\\
	& B_4 = \frac{16}{15}\( \theta_4(2\tau)^4 \Theta(4\tau,4\md m)-\frac{1}{2^4}\theta_4(2\tau)^4 \Theta(\tau+\tfrac{1}{2}, 2\md m)- \frac{1}{4^5}\sum_{k=0}^3\theta_2(\tfrac{\tau+k}{2})^4\Theta(\tfrac{\tau+k}{4}, \md m) \) , \nn\\
	& B_6 = \frac{9}{10}\( h(\tau)^2\Theta(6\tau,6\md m)+\frac{h(\tau)^2}{2^4}\sum_{k=0}^1 \Theta(\tfrac{3\tau+3k}{2}, 3\md m) -\frac{1}{3^5}\sum_{k=0}^2 h(\tfrac{\tau+k}{3})^2 \Theta(\tfrac{2\tau+2k}{3}, 2\md m) \right. \nn\\
	& \phantom{=====}\left.- \frac{1}{3\cdot 6^4}\sum_{k=0}^5 h(\tfrac{\tau+k}{3})^2 \Theta(\tfrac{\tau+k}{6},\md m) \)\ ,
	\end{align}
	where we have set $\md m = \sum_{i=1}^8 m_i e_i$ and
	\begin{equation}
	h(\tau) = \theta_2(2\tau)\theta_2(6\tau) + \theta_3(2\tau)\theta_3(6\tau) \ .
	\end{equation}
	$A_n$ and $B_n$ have index $n$ and weight 4 and 6 respectively.
	
	\section{Proof of some identities involving $\vartheta$-functions and Jacobi forms} 
	\label{appendix:proofs_identities}
	
	In this appendix, we prove two identities invoked in section \ref{section:domain_wall_method}.
	
	To prove the identity \eqref{eq:idn}, which we reproduce in the following
	\begin{eqnarray} \label{app:id1}
	\lefteqn{\cA\left(\tfrac{3}{2}\epsilon_1 + \tfrac{1}{2}\epsilon_2 \right) \vartheta_1(2\epsilon_2)  - \cA \left(\tfrac{1}{2}\epsilon_1 + \tfrac{3}{2}\epsilon_2 \right) \vartheta_1(2\epsilon_1) } \nonumber \\
	&=& \frac{\vartheta_1(\epsilon_2-\epsilon_1) \cA_{-}}{1492992} 
	[\cB_{+}^6 - 15 \cA_{+}^2 \cB_{+}^4 E_4 - 45 \cA_{+}^4 \cB_{+}^2 E_4^2 + 40 \cA_{+}^3 \cB_{+}^3 E_6 + 
	24 \cA_{+}^5 \cB_{+} E_4 E_6  \nonumber \\  && ~~~~ + \cA_{+}^6 (27 E_4^3 - 32 E_6^2)] \,,
	\end{eqnarray}
	we begin with the first term: we multiply the numerator and the denominator by $\vartheta_{1-}^2$, pair this factor with $\vartheta_1(\tfrac{3}{2}\epsilon_1 +\tfrac{1}{2}\epsilon_2)^2$ via \eqref{A.28}, and replace $\vartheta_1(2\epsilon_2)$ using the duplication formula \eqref{duplication_formula}. We find
	\begin{equation}
	\begin{aligned}
	& \cA(\tfrac{3}{2} \epsilon_1 + \tfrac{1}{2}\epsilon_2) \vartheta_1(2\epsilon_2)  \\
	=& -\frac{2\vartheta_1(\epsilon_2)\vartheta_2(\epsilon_2)\vartheta_3(\epsilon_2)\vartheta_4(\epsilon_2)}{\eta^6\vartheta_{1-}^2\vartheta_2\vartheta_3\vartheta_4^5}\(\vartheta_3(\epsilon_1)^4\vartheta_{2+}^4 + \vartheta_2(\epsilon_1)^4\vartheta_{3+}^4 - 2\vartheta_2(\epsilon_1)^2 \vartheta_3(\epsilon_1)^2 \vartheta_{2+}^2 \vartheta_{3+}^2\) \ .
	\end{aligned}
	\end{equation}
	Now we can pair each of the four factors $\vartheta_i(\epsilon_2)$ in the prefactor with the four $\vartheta_j(\epsilon_1)$ in each term inside the parentheses and apply \eqref{A.28}. The result involves theta functions of elliptic argument $\epsilon_-$, $\epsilon_+$ only, and we denote it by $P[\vartheta_{i}(\epsilon_+),\vartheta_{i}(\epsilon_-)]$.
	The second term on the LHS of (\ref{app:id1}) can be transformed in exactly the same manner, and the result is $P[\vartheta_{i}(\epsilon_+),\vartheta_{i}(-\epsilon_-)]$. Since $\vartheta_1(z)$ is an odd function, while $\vartheta_2(z), \vartheta_3(z), \vartheta_4(z)$ are even functions, all terms with even powers of $\vartheta_{1-}$ in $P[\vartheta_{i}(\epsilon_+),\vartheta_{i}(\epsilon_-)]$ drop out of the difference $P[\vartheta_{i}(\epsilon_+),\vartheta_{i}(\epsilon_-)] - P[\vartheta_{i}(\epsilon_+),\vartheta_{i}(-\epsilon_-)]$. The surviving terms take the form
	\begin{equation}
	\begin{aligned}
	&-\frac{4}{\eta^6 \vartheta_{1-}^2 \vartheta_2^3\vartheta_3^3\vartheta_4^9}\(\vartheta_{1-}^5 \vartheta_{2-} \vartheta_{3-} \vartheta_{4-}(\ldots)+ \vartheta_{1-}^3 \vartheta_{2-} \vartheta_{3-} \vartheta_{4-}^3(\ldots)+ \right.\\
	&\left. \vartheta_{1-}^3 \vartheta_{2-}^3 \vartheta_{3-} \vartheta_{4-}(\ldots)+ \vartheta_{1-}^3 \vartheta_{2-} \vartheta_{3-}^3 \vartheta_{4-}(\ldots)+\vartheta_{1-} \vartheta_{2-}^3 \vartheta_{3-} \vartheta_{4-}^3(\ldots)+\vartheta_{1-} \vartheta_{2-} \vartheta_{3-}^3 \vartheta_{4-}^3(\ldots) \) \ ,
	\end{aligned}
	\end{equation}
	where $(\ldots)$ are polynomials in $\vartheta_i,\vartheta_{i+}$. Next we can reduce the number of products of $\vartheta_{i-}$ by using \eqref{A.27} to transfer the powers of $\vartheta_{2-}, \vartheta_{3-}$ to  $\vartheta_{1-}, \vartheta_{4-}$. 
	The result is
	\begin{equation}
	\begin{aligned}
	-\frac{4}{\eta^6 \vartheta_{1-}^2 \vartheta_2^3\vartheta_3^3\vartheta_4^{11}}\(\vartheta_{1-}^5 \vartheta_{2-} \vartheta_{3-} \vartheta_{4-}(\ldots)+ \vartheta_{1-}^3 \vartheta_{2-} \vartheta_{3-} \vartheta_{4-}^3(\ldots)+\vartheta_{1-} \vartheta_{2-} \vartheta_{3-} \vartheta_{4-}^5(\ldots) \) \ .
	\end{aligned}
	\end{equation}
	Using \eqref{A.27} and similar relations 
	\begin{equation}
	\vartheta_1(z)^2\vartheta_4^2 = \vartheta_3(z)^2\vartheta_2^2 - \vartheta_2(z)^2\vartheta_3^2 \ ,\quad \vartheta_4(z)^2\vartheta_4^2 = \vartheta_3(z)^2\vartheta_3^2 - \vartheta_2(z)^2\vartheta_2^2  \ ,
	\end{equation}
	it is easy to see the coefficients of $\vartheta_{1-} \vartheta_{2-} \vartheta_{3-} \vartheta_{4-}^5$ and $\vartheta_{1-}^3 \vartheta_{2-} \vartheta_{3-} \vartheta_{4-}^3$ vanish. For instance, the coefficient of $\vartheta_{1-} \vartheta_{2-} \vartheta_{3-} \vartheta_{4-}^5$ reads
	\begin{equation}
	\begin{aligned}
	\vartheta_2^2 \vartheta_{1+}^2 \vartheta_{2+}^2 - \vartheta_3^2 \vartheta_{1+}^2 \vartheta_{3+}^2 - \vartheta_3^2 \vartheta_{2+}^2 \vartheta_{4+}^2 + \vartheta_2^2 \vartheta_{3+}^2 \vartheta_{4+}^2 
	= -\vartheta_{1+}^2\vartheta_{4+}^2\vartheta_4^2 + \vartheta_{4+}^2\vartheta_{1+}^2\vartheta_4^2 = 0 \ .
	\end{aligned}
	\end{equation}
	Now all the theta functions of $\epsilon_-$ can be factored out. We are left with the following coefficient of the factor $\vartheta_1(\epsilon_2 - \epsilon_1) \cA_{-}$:
	\begin{align} \label{app:after_massage1}
	-\frac{2}{\vartheta_2^2\vartheta_3^2\vartheta_4^{10}}&(-\vartheta_3^2 \vartheta_{1+}^2 \vartheta_{2+}^8 \vartheta_{3+}^2 
	+ 2 \vartheta_3^2 \vartheta_{1+}^2 \vartheta_{2+}^4 \vartheta_{3+}^6
	- \vartheta_2^2 \vartheta_{1+}^2 \vartheta_{2+}^2 \vartheta_{3+}^8 
	- \vartheta_4^2 \vartheta_{1+}^2 \vartheta_{2+}^8 \vartheta_{4+}^2 
	- \vartheta_2^2 \vartheta_{2+}^8 \vartheta_{3+}^2 \vartheta_{4+}^2 \nn\\
	&- 2 \vartheta_4^2 \vartheta_{1+}^2 \vartheta_{2+}^4 \vartheta_{3+}^4 \vartheta_{4+}^2 
	+ \vartheta_4^2 \vartheta_{1+}^2 \vartheta_{3+}^8 \vartheta_{4+}^2 
	+ \vartheta_3^2 \vartheta_{2+}^2 \vartheta_{3+}^8 \vartheta_{4+}^2 
	- 2 \vartheta_4^2 \vartheta_{2+}^2 \vartheta_{3+}^6 \vartheta_{4+}^4) \ .
	\end{align}
	Equality with
	\begin{align} \label{app:preim1}
	\frac{1}{1492992}&(\cB_{+}^6 - 15 \cA_{+}^2 \cB_{+}^4 E_4 - 45 \cA_{+}^4 \cB_{+}^2 E_4^2 + 40 \cA_{+}^3 \cB_{+}^3 E_6 + 
	24 \cA_{+}^5 \cB_{+} E_4 E_6  \nn \\
	& + \cA_{+}^6 (27 E_4^3 - 32 E_6^2))
	\end{align}
	now can either be seen directly by expressing both (\ref{app:after_massage1}) and (\ref{app:preim1}) in terms of algebraically independent $\vartheta$ functions, or by noting that both expressions define weak Jacobi forms with the same weight and index, and fixing coefficients by comparing their small $q$ expansion to sufficiently high order.\footnote{Note that we cannot prove the identity \eqref{eq:idn} using the same argument because both sides depend on two elliptic parameters.}.
	
	We next prove the first formula in (\ref{identity3.18}), which we also reproduce here
	\begin{eqnarray}  
	\lefteqn{\cA(\epsilon_1)^2\vartheta_1(2\epsilon_2) - \cA(\epsilon_2)^2\vartheta_1(2\epsilon_1)} \nonumber \\ 
	&=&   \frac{\vartheta_1(\epsilon_1-\epsilon_2)} {10368} 
	(\cA_{+}\cB_{-} -\cA_{-}\cB_{+}) (3\cA_{+}\cB_{-}\cB_{+}^2  + \cA_{-}\cB_{+}^3 -9\cA_{-}\cA_{+}^2\cB_{+}E_4 - 3\cA_{+}^3\cB_{-}E_4  \nonumber \\ &&   + 8\cA_{+}^3\cA_{-}E_6) \,. 
	\end{eqnarray}
	First, we use the duplication formula (\ref{duplication_formula}) to obtain 
	\begin{eqnarray}
	\lefteqn{ \cA(\epsilon_1)^2\vartheta_1(2\epsilon_2) - \cA(\epsilon_2)^2\vartheta_1(2\epsilon_1)} \nonumber \\ &=&\frac{2\vartheta_1(\epsilon_1)\vartheta_1(\epsilon_2)}{\vartheta_2\vartheta_3\vartheta_4 \eta^{12}} 
	[\vartheta_2(\epsilon_2)\vartheta_3(\epsilon_2)\vartheta_4(\epsilon_2)\vartheta_1^3(\epsilon_1) - \vartheta_2(\epsilon_1)\vartheta_3(\epsilon_1)\vartheta_4(\epsilon_1)\vartheta_1^3(\epsilon_2)] \,.
	\end{eqnarray} 
	Using the formulae in (\ref{A.28}), we obtain the identity  
	\begin{eqnarray}
	\vartheta_1(\epsilon_1)\vartheta_1(\epsilon_2) = -\frac{\eta^6}{12} (\cA_+\cB_- - \cA_{-}\cB_+) \,.
	\end{eqnarray} 
	Applying the formulae in (\ref{A.28}) to the pairs $\vartheta_1(\epsilon_1) \vartheta_i(\epsilon_2)$, $i=2,3,4$ in the first term, and likewise with $\epsilon_1$ and $\epsilon_2$ exchanged in the second term, we find 
	\begin{eqnarray} \label{afterallthat}
	\lefteqn{\vartheta_2(\epsilon_2)\vartheta_3(\epsilon_2)\vartheta_4(\epsilon_2)\vartheta_1^3(\epsilon_1) - \vartheta_2(\epsilon_1)\vartheta_3(\epsilon_1)\vartheta_4(\epsilon_1)\vartheta_1^3(\epsilon_2) }  \\ \nonumber 
	&=&  \frac{\vartheta_{1-}\vartheta_{2-}\vartheta_{3-}\vartheta_{4-}}{2\eta^6}[ \vartheta_{1+}^2\vartheta_{2+}^2\vartheta_{3+}^2\vartheta_{4-}^2+ 
	\vartheta_{1+}^2\vartheta_{2+}^2\vartheta_{3-}^2\vartheta_{4+}^2+\vartheta_{1+}^2\vartheta_{2-}^2\vartheta_{3+}^2\vartheta_{4+}^2
	+\vartheta_{1-}^2\vartheta_{2+}^2\vartheta_{3+}^2\vartheta_{4+}^2] \,.
	\end{eqnarray} 
	The four factors of $\vartheta_{i-}$ in front are proportional via the duplication identity (\ref{duplication_formula}) to the expected factor $\vartheta_1(\epsilon_1-\epsilon_2)$ required to cancel the pole in (\ref{Z2_3.17}). It is straightforward to work out the rest by re-expressing the weak Jacobi forms in (\ref{identity3.18}) and the RHS of (\ref{afterallthat}) in terms of an independent set of theta functions, and invoking the identities (\ref{Eisen}) and (\ref{eta}). This then completes the proof of the first identity in (\ref{identity3.18}). The proof of the others follows the same pattern.

	\section{BPS invariants from the modular approach}
	\label{Modularandenumerative}
	
	\subsection{Massless E-string $n_b=4,5$}
	
	Here in Tabs.~\ref{tb:E-b4}, \ref{tb:E-b5} we give the unrefined BPS indices for massless E-string at base degree $n_b=4,5$ respectively.

	\begin{table}[h!]
		\centering
		{\footnotesize
			\begin{tabular}{|r|ccc|}\hline
				$I^{4,e}_g$& $e=$4&5&6\\ \hline
				$g=$0&-114265008&-23064530112&-1972983690880\\ 
				1 &76413833&27863327760&3478600320920\\ 
				2 &-26631112&-18669096840&-3493725635712\\ 
				3 &5889840&8744913564&2548788575530\\ 
				4 &-835236&-3051708946&-1455980703978\\ 
				5 &69587&804336322&669294682633\\ 
				6 &-2642&-158420138&-249630702534\\ 
				7 &11&22611609&75407691994\\ 
				8 &&-2209196&-18284982166\\ 
				9 &&132731&3503417329\\ 
				10 &&-3828&-517711576\\ 
				11 &&15&56863333\\ 
				12 &&&-4374392\\ 
				13 &&&211796\\ 
				14 &&&-5068\\ 
				15 &&&19\\ 
				\hline 
			\end{tabular}
			\caption{Unrefined BPS indices $I^{n_b,n_e}_{g}$ for massless E-string at $n_b=4$.}\label{tb:E-b4}
			\begin{tabular}{|r|ccc|}\hline
				$I^{5,e}_g$& $e=$5&6&7\\ \hline
				$g=$0&18958064400&5105167984850&594537323257800\\ 
				1 &-23436186176&-9930641443350&-1585090167772500\\ 
				2 &16150498760&11074858711765&2457788116576020\\ 
				3 &-7785768630&-8996745286730&-2835031032258700\\ 
				4 &2795423986&5741344855169&2636754649061672\\ 
				5 &-757807700&-2964403762286&-2045465858241700\\ 
				6 &153448015&1252200262512&1344639656186269\\ 
				7 &-22499836&-433368495468&-754544640140708\\ 
				8 &2255437&122266349837&362345939803835\\ 
				9 &-138768&-27823676688&-148816360041566\\ 
				10 &4085&5022627207&52113548906686\\ 
				11 &-16&-701630204&-15479922392000\\ 
				12 &&73076867&3871431860954\\ 
				13 &&-5347102&-806989428116\\ 
				14 &&247076&138311499769\\ 
				15 &&-5670&-19134877318\\ 
				16 &&21&2082390360\\ 
				17 &&&-171656442\\ 
				18 &&&10099677\\ 
				19 &&&-381760\\ 
				20 &&&7340\\ 
				21 &&&-26\\ 
				\hline 
			\end{tabular}
			\caption{Unrefined BPS indices $I^{n_b,n_e}_{g}$ for massless E-string at $n_b=5$.}\label{tb:E-b5}
		}
	\end{table}
	
	\subsection{Massless M-string $n_b=3,4$ and $n_e\leq 3$}
	
	We give here the BPS numbers for the massless M-string at base degree 3, 4 and fiber degree up to 3 in Tabs.~\ref{tb:Mmassless-BPS-b3}, \ref{tb:Mmassless-BPS-b4} respectively. In the head of each table, $(n_b, n_e)$ means base degree $n_b$ and fiber degree $n_e$. Conspicuously, the checkerboard pattern disappears in these tables.
	
	\begin{table}[h!]
		\centering
		{\tiny
			\begin{tabular}{|r|ccccccc|}\hline
				$(3,1)$ &$2j_+=$0&1&2&3&4&5&6\\ \hline
				$2j_-=$0&&&&&2&4&2\\ 
				1 &&&&&1&2&1\\ 
				\hline \end{tabular}
			\begin{tabular}{|r|cccccccccc|}\hline
				$(3,2)$ &$2j_+=$0&1&2&3&4&5&6&7&8&9\\ \hline
				$2j_-=$0&10&9&8&19&36&40&26&10&2&\\ 
				1 &5&6&10&20&31&36&31&18&5&\\ 
				2 &&1&4&7&10&16&20&15&6&1\\ 
				3 &&&&&1&4&6&6&5&2\\ 
				4 &&&&&&&&1&2&1\\ 
				\hline \end{tabular}
			\begin{tabular}{|r|ccccccccccccc|}\hline
				$(3,3)$ &$2j_+=$0&1&2&3&4&5&6&7&8&9&10&11&12\\ \hline
				$2j_-=$0&70&128&176&243&322&345&270&147&58&19&4&&\\ 
				1 &75&138&200&286&384&428&370&242&118&40&7&&\\ 
				2 &36&69&112&171&236&285&292&240&144&56&12&1&\\ 
				3 &7&16&33&56&84&120&152&154&115&60&21&4&\\ 
				4 &&1&4&8&16&32&50&62&62&47&24&6&\\ 
				5 &&&&&1&4&8&14&22&24&16&6&1\\ 
				6 &&&&&&&&1&4&6&6&5&2\\ 
				7 &&&&&&&&&&&1&2&1\\ 
				\hline \end{tabular}
		}
		\caption{BPS numbers of massless M-string at base degree 3 and fiber degree up to 3.}\label{tb:Mmassless-BPS-b3}
	\end{table}
	
	\begin{table}[h!]
		\centering
		{\tiny
			\begin{tabular}{|r|ccccccccc|}\hline
				$(4,1)$ &$2j_+=$0&1&2&3&4&5&6&7&8\\ \hline
				$2j_-=$0&&&&&&&2&4&2\\ 
				1 &&&&&&&1&2&1\\ 
				\hline \end{tabular}
			\begin{tabular}{|r|ccccccccccccc|}\hline
				$(4,2)$ &$2j_+=$0&1&2&3&4&5&6&7&8&9&10&11&12\\ \hline
				$2j_-=$0&2&9&16&19&28&45&58&61&48&22&4&&\\ 
				1 &1&6&14&20&27&40&57&70&62&34&11&2&\\ 
				2 &&1&4&7&10&17&28&38&40&32&18&5&\\ 
				3 &&&&&1&4&7&10&16&20&15&6&1\\ 
				4 &&&&&&&&1&4&6&6&5&2\\ 
				5 &&&&&&&&&&&1&2&1\\ 
				\hline \end{tabular}
			\begin{tabular}{|r|ccccccccccccccccc|}\hline
				$(4,3)$ &$2j_+=$0&1&2&3&4&5&6&7&8&9&10&11&12&13&14&15&16\\ \hline
				$2j_-=$0&106&243&406&556&680&818&964&987&794&491&244&103&34&6&&&\\ 
				1 &124&284&475&664&848&1056&1262&1336&1175&836&471&202&62&12&1&&\\ 
				2 &74&167&282&414&570&753&934&1058&1060&890&584&283&100&26&4&&\\ 
				3 &25&54&94&154&239&344&459&580&673&662&519&314&142&44&7&&\\ 
				4 &4&8&16&34&62&100&154&227&302&347&338&265&152&57&12&1&\\ 
				5 &&&1&4&8&16&34&60&91&128&159&158&116&60&21&4&\\ 
				6 &&&&&&1&4&8&16&32&50&62&62&47&24&6&\\ 
				7 &&&&&&&&&1&4&8&14&22&24&16&6&1\\ 
				8 &&&&&&&&&&&&1&4&6&6&5&2\\ 
				9 &&&&&&&&&&&&&&&1&2&1\\ 
				\hline \end{tabular}
		}
		\caption{BPS numbers of massless M-string at base degree 4 and fiber degree up to 3.}\label{tb:Mmassless-BPS-b4}
	\end{table}

	\subsection{Massive M-string $n_b=2,3$ and $n_e\le 3$}
	\label{appendixmassivemstring}  
	
	Here we display the BPS numbers for massive M-string with base degree 2 and 3 and fiber degree up to 3 in Tabs.~\ref{tb:Mmassive-BPS-b2-1}, \ref{tb:Mmassive-BPS-b2-2}, \ref{tb:Mmassive-BPS-b3-1}, \ref{tb:Mmassive-BPS-b3-2}. In the head of each table $(n_b,n_e,n_m)$ means base degree $n_b$, fiber degree $n_e$, and degree $n_m$ in the mass parameter. Noticeably, the checkerboard pattern is recovered in every table.
	
	\begin{table}[h!]
		\centering
		{\tiny
			\begin{tabular}{|r|ccc|}\hline
				$(2,1,0)$& 
				$2j_+=$2&3&4\\ \hline
				$2j_-=$0&&2&\\ 
				1 &1&&1\\ 
				\hline \end{tabular}
			\begin{tabular}{|r|ccc|}\hline
				$(2,1,1)$& 
				$2j_+=$2&3&4\\ \hline
				$2j_-=$0&1&&1\\ 
				1 &&1&\\ 
				\hline \end{tabular}		
			\begin{tabular}{|r|c|}\hline
				$(2,1,2)$& $2j_+=$3\\ \hline
				$2j_-=$0&1\\ 
				\hline \end{tabular}
			
			\begin{tabular}{|r|ccccccc|}\hline
				$(2,2,0)$& $2j_+=$0&1&2&3&4&5&6\\ \hline
				$2j_-=$0&&5&&10&&3&\\ 
				1 &1&&8&&8&&1\\ 
				2 &&1&&4&&3&\\ 
				3 &&&&&1&&1\\ 
				\hline \end{tabular}
			\begin{tabular}{|r|ccccccc|}\hline
				$(2,2,1)$& $2j_+=$0&1&2&3&4&5&6\\ \hline
				$2j_-=$0&1&&7&&6&&\\ 
				1 &&3&&8&&3&\\ 
				2 &&&2&&3&&1\\ 
				3 &&&&&&1&\\ 
				\hline \end{tabular}
			\begin{tabular}{|r|ccccc|}\hline
				$(2,2,2)$& $2j_+=$1&2&3&4&5\\ \hline
				$2j_-=$0&2&&4&&1\\ 
				1 &&3&&3&\\ 
				2 &&&1&&1\\ 
				\hline \end{tabular}
			\begin{tabular}{|r|ccc|}\hline
				$(2,2,3)$& $2j_+=$2&3&4\\ \hline
				$2j_-=$0&1&&1\\ 
				1 &&1&\\ 
				\hline \end{tabular}
		} 
		\caption{BPS numbers of massive M-string with base degree 2 and fiber degree 1 or 2. }\label{tb:Mmassive-BPS-b2-1}
		\end {table}
		
		\begin{table}[h!]
			\centering
			{\tiny
				\begin{tabular}{|r|ccccccccc|}\hline
					$(2,3,0)$& $2j_+=$0&1&2&3&4&5&6&7&8\\ \hline
					$2j_-=$0&&26&&50&&18&&1&\\ 
					1 &12&&48&&48&&12&&\\ 
					2 &&14&&31&&26&&4&\\ 
					3 &1&&6&&14&&10&&1\\ 
					4 &&&&1&&4&&3&\\ 
					5 &&&&&&&1&&1\\ 
					\hline \end{tabular}
				\begin{tabular}{|r|ccccccccc|}\hline
					$(2,3,1)$& $2j_+=$0&1&2&3&4&5&6&7&8\\ \hline
					$2j_-=$0&9&&36&&31&&4&&\\ 
					1 &&23&&46&&23&&2&\\ 
					2 &4&&19&&26&&11&&\\ 
					3 &&2&&7&&11&&3&\\ 
					4 &&&&&2&&3&&1\\ 
					5 &&&&&&&&1&\\ 
					\hline \end{tabular}
				\begin{tabular}{|r|cccccccc|}\hline
					$(2,3,2)$& $2j_+=$0&1&2&3&4&5&6&7\\ \hline
					$2j_-=$0&&11&&24&&7&&\\ 
					1 &4&&20&&20&&4&\\ 
					2 &&5&&12&&10&&1\\ 
					3 &&&1&&4&&3&\\ 
					4 &&&&&&1&&1\\ 
					\hline \end{tabular}
				\begin{tabular}{|r|ccccccc|}\hline
					$(2,3,3)$& $2j_+=$0&1&2&3&4&5&6\\ \hline
					$2j_-=$0&1&&7&&6&&\\ 
					1 &&3&&8&&3&\\ 
					2 &&&2&&3&&1\\ 
					3 &&&&&&1&\\ 
					\hline \end{tabular}							
				\begin{tabular}{|r|ccc|}\hline
					$(2,3,4)$& $2j_+=$2&3&4\\ \hline
					$2j_-=$0&&2&\\ 
					1 &1&&1\\ 
					\hline \end{tabular}
			} 
			\caption{BPS numbers of massive M-string with base degree 2 and fiber degree 3. }\label{tb:Mmassive-BPS-b2-2}
			\end {table}

			\begin{table}[h!]
				\centering
				{\tiny
					\begin{tabular}{|r|ccc|}\hline
						$(3,1,0)$& $2j_+=$4&5&6\\ \hline
						$2j_-=$0&&2&\\ 
						1 &1&&1\\ 
						\hline 
					\end{tabular}
					\begin{tabular}{|r|ccc|}\hline
						$(3,1,1)$& $2j_+=$4&5&6\\ \hline
						$2j_-=$0&1&&1\\ 
						1 &&1&\\ 
						\hline 
					\end{tabular}			
					\begin{tabular}{|r|c|}\hline
						$(3,1,2)$& $2j_+=$5\\ \hline
						$2j_-=$0&1\\ 
						\hline 
					\end{tabular}
					\begin{tabular}{|r|cccccccccc|}\hline
						$(3,2,0)$& $2j_+=$0&1&2&3&4&5&6&7&8&9\\ \hline
						$2j_-=$0&&5&&11&&20&&6&&\\ 
						1 &3&&6&&17&&17&&3&\\ 
						2 &&1&&5&&10&&9&&1\\ 
						3 &&&&&1&&4&&3&\\ 
						4 &&&&&&&&1&&1\\ 
						\hline 
					\end{tabular}
					\begin{tabular}{|r|cccccccccc|}\hline
						$(3,2,1)$& $2j_+=$0&1&2&3&4&5&6&7&8&9\\ \hline
						$2j_-=$0&4&&4&&15&&11&&1&\\ 
						1 &&3&&9&&16&&8&&\\ 
						2 &&&2&&5&&9&&3&\\ 
						3 &&&&&&2&&3&&1\\ 
						4 &&&&&&&&&1&\\ 
						\hline 
					\end{tabular}
					\begin{tabular}{|r|ccccccccc|}\hline
						$(3,2,2)$& $2j_+=$0&1&2&3&4&5&6&7&8\\ \hline
						$2j_-=$0&&2&&4&&9&&2&\\ 
						1 &1&&2&&7&&7&&1\\ 
						2 &&&&1&&3&&3&\\ 
						3 &&&&&&&1&&1\\ 
						\hline 
					\end{tabular}
					\begin{tabular}{|r|cccccccc|}\hline
						$(3,2,3)$& $2j_+=$0&1&2&3&4&5&6&7\\ \hline
						$2j_-=$0&1&&&&3&&2&\\ 
						1 &&&&1&&2&&1\\ 
						2 &&&&&&&1&\\ 
						\hline 
					\end{tabular}												
					\begin{tabular}{|r|c|}\hline
						$(3,2,4)$& $2j_+=$5\\ \hline
						$2j_-=$0&1\\ 
						\hline 
					\end{tabular}
				} 
				\caption{BPS numbers of massive M-string at base degree 3 and fiber degree 1 or 2. }\label{tb:Mmassive-BPS-b3-1}
				\end {table}
				
				\begin{table}[h!]
					\centering
					{\tiny
						\begin{tabular}{|r|ccccccccccccc|}\hline
							$(3,3,0)$& $2j_+=$0&1&2&3&4&5&6&7&8&9&10&11&12\\\hline 
							$2j_-=$0&&64&&121&&163&&75&&11&&&\\ 
							1 &39&&106&&192&&184&&64&&5&&\\ 
							2 &&39&&93&&149&&124&&32&&1&\\ 
							3 &5&&21&&50&&84&&63&&13&&\\ 
							4 &&1&&6&&20&&36&&27&&4&\\ 
							5 &&&&&1&&6&&14&&10&&1\\ 
							6 &&&&&&&&1&&4&&3&\\ 
							7 &&&&&&&&&&&1&&1\\ 
							\hline 
						\end{tabular}
						\begin{tabular}{|r|ccccccccccccc|}\hline
							$(3,3,1)$& $2j_+=$0&1&2&3&4&5&6&7&8&9&10&11&12\\ \hline
							$2j_-=$0&29&&74&&130&&109&&26&&2&&\\ 
							1 &&59&&121&&175&&102&&18&&&\\ 
							2 &16&&50&&102&&124&&62&&6&&\\ 
							3 &&8&&26&&54&&67&&27&&2&\\ 
							4 &&&2&&8&&23&&28&&11&&\\ 
							5 &&&&&&2&&7&&11&&3&\\ 
							6 &&&&&&&&&2&&3&&1\\ 
							7 &&&&&&&&&&&&1&\\ 
							\hline 
						\end{tabular}
						\begin{tabular}{|r|cccccccccccc|}\hline
							$(3,3,2)$& $2j_+=$0&1&2&3&4&5&6&7&8&9&10&11\\ \hline
							$2j_-=$0&&30&&57&&82&&34&&4&&\\ 
							1 &17&&45&&89&&86&&26&&1&\\ 
							2 &&15&&38&&65&&55&&12&&\\ 
							3 &1&&6&&17&&33&&25&&4&\\ 
							4 &&&&1&&6&&13&&10&&1\\ 
							5 &&&&&&&1&&4&&3&\\ 
							6 &&&&&&&&&&1&&1\\ 
							\hline 
						\end{tabular}
						\begin{tabular}{|r|ccccccccccc|}\hline
							$(3,3,3)$& $2j_+=$0&1&2&3&4&5&6&7&8&9&10\\ \hline
							$2j_-=$0&6&&14&&30&&25&&3&&\\ 
							1 &&10&&22&&38&&19&&2&\\ 
							2 &2&&6&&16&&22&&10&&\\ 
							3 &&&&2&&6&&10&&3&\\ 
							4 &&&&&&&2&&3&&1\\ 
							5 &&&&&&&&&&1&\\ 
							\hline 
						\end{tabular}
						\begin{tabular}{|r|ccccccccc|}\hline
							$(3,3,4)$& $2j_+=$0&1&2&3&4&5&6&7&8\\ \hline
							$2j_-=$0&&2&&4&&9&&2&\\ 
							1 &1&&2&&7&&7&&1\\ 
							2 &&&&1&&3&&3&\\ 
							3 &&&&&&&1&&1\\ 
							\hline 
						\end{tabular}
						\begin{tabular}{|r|ccc|}\hline
							$(3,3,5)$& $2j_+=$4&5&6\\ \hline
							$2j_-=$0&1&&1\\ 
							1 &&1&\\ 
							\hline 
						\end{tabular}
					} 
					\caption{BPS numbers of massive M-string at base degree 3 and fiber degree 3. }\label{tb:Mmassive-BPS-b3-2}
				\end{table}
				
				\subsection{$N_{j_- j_+}^\kappa$ for a mass deformation of $T^2 \times$ the resolved $D_4$ singularity}
				In table \ref{D4}, we report on numbers $N_{j_- j_+}^\kappa$ for the massive $D_4$ case as extracted from our computations based on the methods explained in section \ref{Anomaly}. They satisfy basic consistency requirements: they are positive integers, and they follow a checkerboard pattern.
				
				\begin{table}[h!]
					\centering
					{\tiny
						\begin{tabular}{|r|cc|}\hline
							$N_{j_-j_+}^{(1,1,1,1;0,0)}$ &$2j_+=$0&1\\ \hline
							$2j_-=$0 &0&1\\ 
							\hline 
						\end{tabular}
						\begin{tabular}{|r|c|}\hline
							$N_{j_-j_+}^{(1,1,1,1;0,1)}$ &$2j_+=$0\\ \hline
							$2j_-=$0&1\\ 
							\hline 
						\end{tabular}
						\begin{tabular}{|r|ccc|}\hline
							$N_{j_-j_+}^{(1,1,1,1;1,0)}$ &$2j_+=$0 & 1 & 2\\ \hline
							$2j_-=$ 0 && 14 &\\ 
							1 & 10 &  & 4 \\ 
							\hline 
						\end{tabular}
						%
						%
						%
						\begin{tabular}{|r|ccc|}\hline
							$N_{j_-j_+}^{(1,1,1,1;1,1)}$ &$2j_+=$0 & 1 & 2\\ \hline
							$2j_-=$ 0 & 13 & & 4\\ 
							1 & &  7&  \\ 
							\hline 
						\end{tabular}
						\begin{tabular}{|r|cc|}\hline
							$N_{j_-j_+}^{(1,1,1,1;1,2)}$ &$2j_+=$0 & 1\\ \hline
							$2j_-=$ 0 & & 7\\ 
							1 & 3& \\ 
							\hline 
						\end{tabular}
						\begin{tabular}{|r|c|}\hline
							$N_{j_-j_+}^{(1,1,1,1;1,3)}$ &$2j_+=$0 \\ \hline
							$2j_-=$ 0 & 3\\ 
							\hline 
						\end{tabular}
						\begin{tabular}{|r|cccc|}\hline
							$N_{j_-j_+}^{(1,1,1,1;2,0)}$ &$2j_+=$0 & 1 & 2 & 3\\ \hline
							$2j_-=$ 0 &  & 173 & & 22\\ 
							1 & 129 &  & 81 & \\
							2 & & 49 & & 10 \\ 
							\hline 
						\end{tabular}
						\begin{tabular}{|r|cccc|}\hline
							$N_{j_-j_+}^{(1,1,1,1;2,1)}$ &$2j_+=$0 & 1 & 2 & 3\\ \hline
							$2j_-=$ 0 & 133 & & 74 &\\ 
							1 & & 126 &  & 16 \\
							2 & 30 & & 22 & \\ 
							\hline 
						\end{tabular}
						\begin{tabular}{|r|cccc|}\hline
							$N_{j_-j_+}^{(1,1,1,1;2,2)}$ &$2j_+=$0 & 1 & 2 & 3\\ \hline
							$2j_-=$ 0 & & 89 & & 6\\ 
							1 & 67 & & 37 & \\
							2 & & 15 & & \\ 
							\hline 
						\end{tabular}
						\begin{tabular}{|r|ccc|}\hline
							$N_{j_-j_+}^{(1,1,1,1;2,3)}$ &$2j_+=$0 & 1 & 2 \\ \hline
							$2j_-=$ 0 & 40 & & 15 \\ 
							1 & & 27 &\\
							2 & 3 & &\\ 
							\hline 
						\end{tabular}
						\begin{tabular}{|r|cc|}\hline
							$N_{j_-j_+}^{(1,1,1,1;2,4)}$ &$2j_+=$0 & 1  \\ \hline
							$2j_-=$ 0 & & 12  \\ 
							1 & 6 &\\
							\hline 
						\end{tabular}
						\begin{tabular}{|r|c|}\hline
							$N_{j_-j_+}^{(1,1,1,1;2,5)}$ &$2j_+=$0   \\ \hline
							$2j_-=$ 0 & 3 \\ 
							\hline 
						\end{tabular}				
					}			
					\caption{Refined BPS invariants of the massive $D_4$ chain at base degrees $\beta = (1,1,1,1)$ and fiber degree $e\leq 2$.}\label{D4}
				\end{table}

				\section{Sakai's mirror curve for the massive E-string}
				\label{sc:Sakai}
				
				The Sakai mirror curve for the massive E-string is given by \eqref{eq:Sakai}, with coefficient functions $g_2(u,\tau,\md m), g_3 (u,\tau,\md m)$  \cite{Sakai:2011}
				\begin{equation}
				g_2(u,\tau, \md m) = \sum_{j=0}^4 a_j u^j \ ,\quad g_3 (u,\tau,\md m) = \sum_{j=0}^{6} b_j u^j \ .
				\end{equation}
				The coefficients are
				\begin{gather}
				a_0 = \frac{1}{12}E_4 \ , \quad b_0 = \frac{1}{216} E_6 \ ,\nn \\
				a_1 = 0 \ ,\quad b_1 = -\frac{4}{E_4} A_1 \ ,\nn \\
				a_2 = \frac{6\cdot 1728}{E_4(E_4^3 - E_6^2)}(- E_4 A_2 + A_1^2) \ ,\nn \\
				b_2 = \frac{5\cdot 1728}{6E_4^2(E_4^3 - E_6^2)}( E_4^2 B_2 - E_6 A_1^2) \ , \\
				\cdots \cdots \cdots \ . \nonumber
				\end{gather}
				where we have used the nine generators of the ring of $E_8$ Weyl invariant Jacobi forms introduced in subsection \ref{sc:jacobi_weyl}.
				 The other coefficients $a_3, a_4, b_3, \ldots, b_6$ of the Sakai curve are not needed in our computations and can be found in \cite{Sakai:2011}. Note that compared to Sakai's original notation, we have inverted $u$ so that $u\rightarrow 0$ is the large volume limit.

				\bibliographystyle{utcaps}
				\bibliography{draft_refEM}
				
			\end{document}